\documentclass[10pt]{article}
\RequirePackage{fix-cm}

\usepackage{amssymb}
\usepackage{amsmath}
\usepackage{graphicx,color}
\usepackage{amsfonts}
\usepackage{times}
\usepackage[sort&compress,numbers]{natbib}
\usepackage{setspace}
\usepackage{booktabs,bm,multirow, pifont,multirow, placeins}
\usepackage{wrapfig}
\usepackage{slashbox}
\usepackage[labelfont=bf]{caption}
\usepackage{pifont}
\usepackage{color}
\usepackage{ulem}
\usepackage{hyperref}
\usepackage{mhchem}

\newcommand{\xmark}{\ding{55}}%

\newif\ifmoditem
\newcommand{\setupmodenumerate}{%
  \global\moditemfalse
  \let\origmakelabel\makelabel
  \def\moditem##1{\global\moditemtrue\def\mesymbol{##1}\item}%
  \def\makelabel##1{%
    \origmakelabel{##1\ifmoditem\rlap{\mesymbol}\fi\enspace}%
    \global\moditemfalse}%
}

\begin{document}

\begin{center}
\setstretch{1}
 \large {\textbf{Progress and Prospects in Two-Dimensional Magnetism of van der Waals Materials}}\\
 \normalsize
\vspace{0.1in}
Youngjun \textsc{Ahn$^{a}$*}, Xiaoyu \textsc{Guo$^{a}$} Suhan \textsc{Son$^{a}$}, Zeliang \textsc{Sun$^{a}$} \& Liuyan \textsc{Zhao$^{a}$*} \\
$^{a}$Department of Physics, University of Michigan, Ann Arbor, MI 48109, USA \\ 
Corresponding to: yoahn@umich.edu \& lyzhao@umich.edu
\end{center}
 
\setstretch{1}

\noindent\Large\textbf{Abstract}
\vspace{0.15in}

\normalsize
Two-dimensional (2D) magnetism in van der Waals (vdW) atomic crystals and moir\'e superlattices has emerged as a topic of tremendous interest in the fields of condensed matter physics and materials science within the past half-decade since its first experimental discovery in 2016 -- 2017. It has not only served as a powerful platform for investigating phase transitions in the 2D limit and exploring new phases of matter, but also provided new opportunities for applications in microelectronics, spintronics, magnonics, optomagnetics, and so on. Despite the flourishing developments in 2D magnetism over this short period of time, further efforts are welcome in multiple forefronts of 2D magnetism research for achieving the ultimate goal of routinely implementing 2D magnets as quantum electronic components. In this review article, we will start with basic concepts and properties of 2D magnetism, followed by a brief overview of historical efforts in 2D magnetism research and then a comprehensive review of vdW material-based 2D magnetism. We will conclude with discussions on potential future research directions for this growing field of 2D vdW magnetism.

\section{Introduction to two-dimensional magnetism}
Two-dimensional (2D) magnetism concerns the magnetic moment arrangements and their collective magnetic excitations in atomically thin materials and structures, as a consequence of magnetic interactions and spin fluctuations in the 2D limit. This research topic was initially investigated as theoretical models of phase transitions in the reduced dimensions between 1940s -- 1970s \cite{onsager1944crystal,mermin1966absence,berezinskii1971destruction,kosterlitz1973ordering,kosterlitz1974critical}. Afterwards, it was experimentally pursued first in quasi-2D bulk materials \cite{de1990magnetic}, then in magnetic elemental metal thin films \cite{venables1983nucleation,davey1968epitaxial,farrow2013thin}, and further in defected graphene films \cite{han2014graphene} mostly during 1980s -- 2010s, and eventually blossomed in van der Waals (vdW) magnetic atomic crystals \cite{huang2017layer,gong2017discovery} and twisted moir\'e superlattice structures \cite{sharpe2019emergent,tang2020simulation} after 2016 -- 2017. 

\subsection{Theoretical descriptions of interacting spins in two dimensions}

The understanding of 2D magnetism roots in a solid theoretical foundation \cite{onsager1944crystal,mermin1966absence,berezinskii1971destruction,kosterlitz1973ordering,kosterlitz1974critical}. Here in this subsection, the early theoretical studies of 2D magnetism are reviewed based on their spin Hamiltonian. For the sake of simplicity, the discussions below are based on the spin Hamiltonian including only the nearest-neighbor interactions:

$$\mathcal{H} = -\sum_{i<j} J(\hat{S}_i^x\hat{S}_j^x+\hat{S}_i^y\hat{S}_j^y+\alpha \hat{S}_i^z\hat{S}_j^z),$$

\noindent where $J$ is the nearest-neighbor exchange coupling, being positive for ferromagnetic (FM) coupling and negative for antiferromagnetic (AFM) coupling; $\hat{S}_{i,j}^{x,y,z}$ are the $x-$, $y-$, and $z-$components of the spin operators at the $i$ and $j$ sites; and $\alpha$ is a dimensionless parameter to tune the magnetic exchange anisotropy between the in-plane $xy-$component and the out-of-plane $z-$component. Aside from exchange anisotropy, magnetic anisotropy can also arise from the single ion anisotropy that is caused by the interactions between the magnetic ion and its surrounding crystalline field, also referred to as magnetocrystalline anisotropy. Its role in defining the spin Hamiltonian classes and determining the magnon band dispersion is similar to the exchange anisotropy. More sophisticated spin Hamiltonian of 2D magnetic systems can be found in other review articles and textbooks \cite{de1990magnetic,gibertini2019magnetic,skomski2008simple,wang2022magnetic}. 

Depending on $\alpha$, this simple spin Hamiltonian can be classified into three cases, easy-axis ($\alpha>1$), easy-plane ($\alpha<1$), and isotropic ($\alpha=1$). Taking the extreme conditions, $\alpha\gg1$ for the easy-axis case can be considered as the Ising-type spin Hamiltonian with the spin dimension, $n$, of $n=$1, $\alpha \sim 0$ for the easy-plane case as the XY-type with $n=$2, and $\alpha =1$ for the isotropic case as the isotropic Heisenberg-type with $n=$3. From a symmetry perspective, the Ising-, XY-, and Heisenberg-type spin Hamiltonian has the discrete $Z_2$, the continuous $U(1)$ (or classically, the continuous $O_2$), and the continuous $SU(2)$ (classically, the continuous $O_3$) symmetry, respectively.

\begin{figure}[th]
\begin{center}
\includegraphics[width=1.0\textwidth]{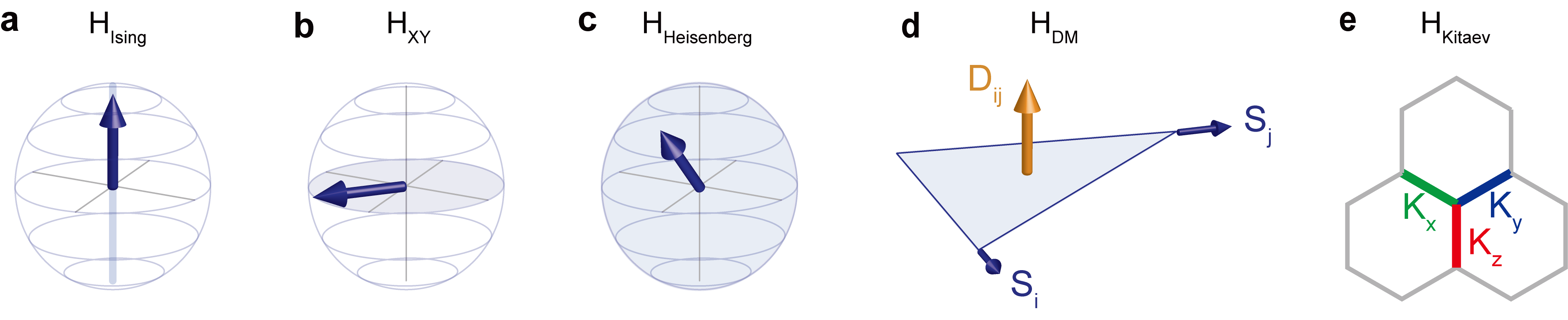}\caption{\small Sketch to highlight five types of spin interactions. (a) Ising-type; (b) XY-type; (c) isotropic Heisenberg-type; shaded areas correspond to the preferred spin orientations; (d) DM interaction; (e) Kitaev interaction.}
\label{magnetic exchange}
\end{center}
\end{figure}

\subsubsection{The 2D Ising model -- Onsager theorem}
The spin dimension, $n=1$, corresponds to the Ising model with only one spin component $S^z$, which shares the same spin Hamiltonian symmetry as the case of $\alpha>1$ in $\mathcal{H}$ above. The Ising-type spin Hamiltonian can be noted as:

$$\mathcal{H}_\mathrm{Ising} = -\sum_{i<j} J_z \hat{S}_i^z\hat{S}_j^z.$$

In this case, the spin at each site is restricted to only two spin states, up or down along the $z-$axis (Figure \ref{magnetic exchange}a). The Ising model was exactly solved in 2D by L. Onsager in 1944 \cite{onsager1944crystal}, whose result predicts the survival of long-range magnetic orders at finite temperatures for 2D Ising-type magnets, or equivalently, 2D magnets with the easy-axis magnetic anisotropy. This result can also be understood in such a way that due to the discrete $Z_2$ symmetry, it always causes finite energy to tilt the spin off from the most favored direction and therefore a finite temperature is required to provide the energy for destroying such Ising-type, or equivalently, easy-axis spin orders in 2D.

\subsubsection{The 2D XY model -- Berezinskii–Kosterlitz–Thouless transition}
The spin dimension, $n=2$, yields the XY model with two spin components $S^x$ and $S^y$, which belongs to the same spin symmetry class as the case of $\alpha<1$ in  $\mathcal{H}$ above. The XY-type spin Hamiltonian is described below:

$$\mathcal{H}_\mathrm{XY} = -\sum_{i<j} J_{x-y}(\hat{S}_i^x\hat{S}_j^x+\hat{S}_i^y\hat{S}_j^y).$$

In this model, the spin at each site can point to any direction within the $xy-$plane without any energy penalty (Figure \ref{magnetic exchange}b). Therefore, one can imagine that in the 2D limit where the thermal and quantum fluctuations are much stronger than in the three-dimensional (3D) counterpart, it is easy to rotate the spin orientation within the $xy-$plane even at zero temperature and thus hinders the formation of long-range magnetic orders. Indeed, Berenzinskii, Kosterlitz, and Thouless theoretically demonstrated in the early 1970s \cite{berezinskii1971destruction,kosterlitz1973ordering, kosterlitz1974critical} that only quasi-long-range orders can exist in 2D XY-type magnets, or equivalently, 2D magnets with the easy-plane magnetic anisotropy. The phase transition into such a quasi-long-range order, i.e., Berenzinskii-Kosterlitz-Thouless (BKT) phase transition, happens at finite temperatures with possible formations of spin vortex and anti-vortex pairs.

\subsubsection{The 2D isotropic Heisenberg model -- Mermin-Wagner Theorem}
The spin dimension, $n=3$, results in the isotropic Heisenberg model with all three spin components $S^x$, $S^y$, and $S^z$ involved, corresponding to the case of $\alpha=1$ in $\mathcal{H}$ above. One can write the isotropic Heisenberg Hamiltonian as below:

$$\mathcal{H}_\mathrm{Heisenberg} = -\sum_{i<j} J(\hat{S}_i^x\hat{S}_j^x+\hat{S}_i^y\hat{S}_j^y+\hat{S}_i^z\hat{S}_j^z).$$

In this case, the spin at each site is allowed to direct to any orientation within the 3D space without costing any extra energy (Figure \ref{magnetic exchange}c). In a more profound manner than 2D XY magnets, the enhanced spin fluctuations in the 2D limit necessarily destroy any long-range or quasi-long-range orders even at zero temperature in 2D isotropic Heisenberg magnets. This consequence was rigorously proved by Mermin and Wagner in 1966 \cite{mermin1966absence}. 

\begin{table}[th]
    \center
    \begin{tabular}{|c||*{3}{c|}}\hline
    \backslashbox[48mm]{spin dimension $n$}{lattice dimension $d$}&\makebox[3em]{$d=1$}&\makebox[3em]{$d=2$}&\makebox[3em]{$d=3$}\\\hline\hline
    $n=1:$ Ising-type $\Longleftrightarrow \alpha>1$: easy-axis & \xmark & \checkmark & \checkmark \\\hline
    $n=2:$ XY-type $\Longleftrightarrow \alpha<1$: easy-plane & \xmark & \xmark \checkmark & \checkmark \\\hline
    $n=3:$ Heisenberg-type $\Longleftrightarrow \alpha=1$: isotropic & \xmark & \xmark & \checkmark\\\hline
    \end{tabular}
    \caption{\small Summary of the presence (\checkmark) and the absence (\xmark) of conventional long-range magnetic orders at finite temperatures, depending on the material lattice dimension, $d$, and the spin dimension, $n$. Note that in the cell of $d=2$ and $n=2$, the symbol of \xmark \checkmark represents the presence of quasi-long-range order.}
    \label{spin-lattice dimensions}
\end{table}

Table \ref{spin-lattice dimensions} summarizes whether or not a long-range magnetic order can form depends on the material lattice dimension, $d$, and the spin dimension, $n$. The discussion in this subsection on theoretical models is dedicated to the case of $d=2$, magnetism in the 2D lattices.  

\subsubsection{Complementary perspective via magnon density of states}

\begin{figure}[th]
\begin{center}
\includegraphics[width=0.7\textwidth]{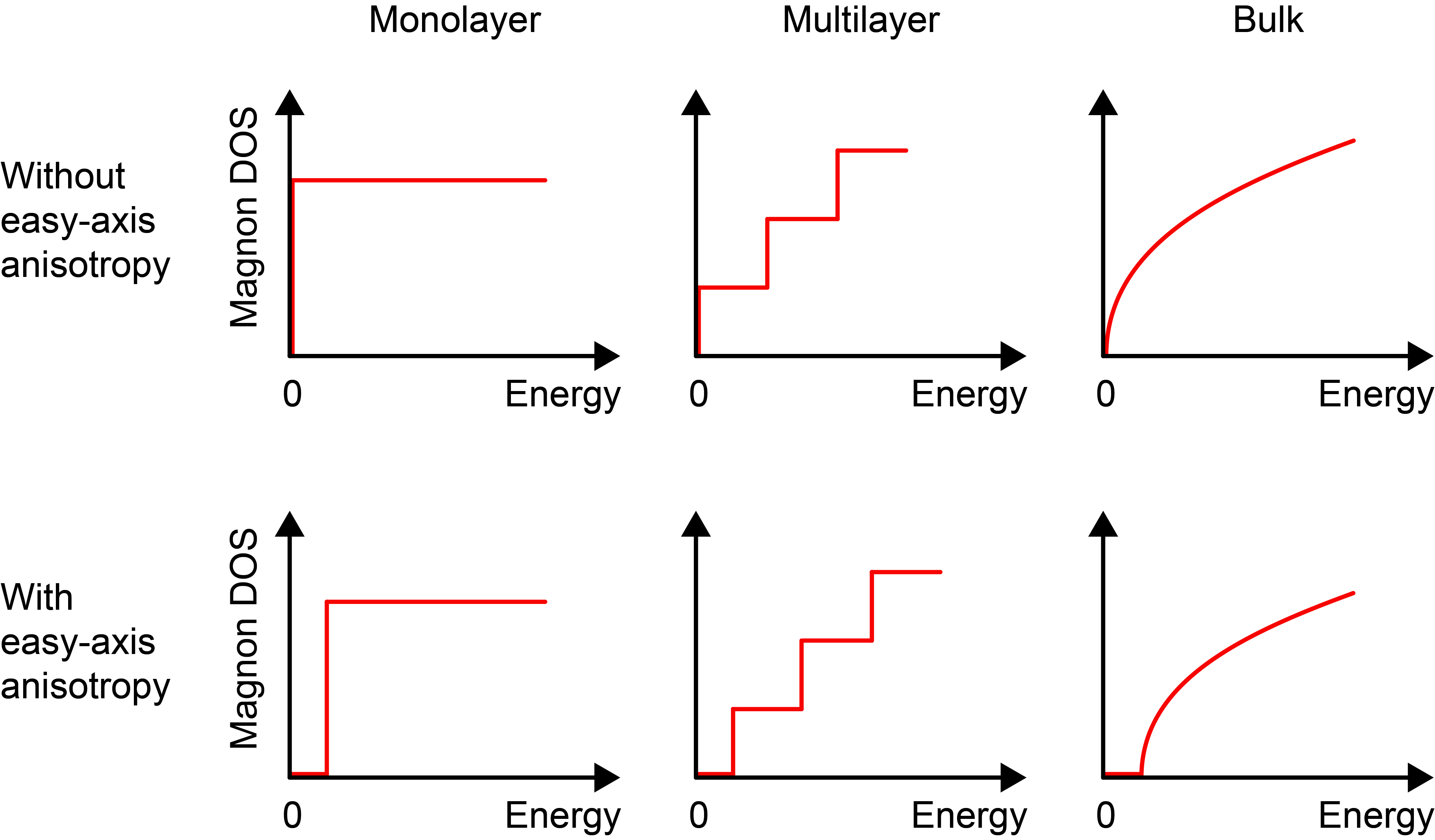}\caption{\small Schematic illustration of the magnon DOS spectra for a simple (i) monolayer, (ii) multilayer, and (iii) bulk FM, for cases with and without the easy-axis anisotropy. Figure is adapted and modified from Ref. \cite{gong2019two}}
\label{magnon dos}
\end{center}
\end{figure}

In addition to seeing the magnetic order and/or disorder through the spin and lattice dimensions above, magnon density of states (DOS) can provide a complementary and intuitive perspective. As discussed more in details later on in Section 4.2, magnons are collective spin excitations of magnetic orders and show well-defined momentum-energy dispersion (i.e., magnon bands) for magnetic crystals. A piece of key established knowledge is that the magnon band gap scales with the easy-axis magnetic anisotropy: the magnon band is gapped and the magnon DOS onsets at finite nonzero energies, only when the easy-axis magnetic anisotropy exists.

Taking the simple FM as an example, its low-energy magnon band follows the parabolic dispersion, i.e., $\mathbf{\mathrm{E}}(\Vec{k})\sim k^2$, and therefore, the magnon DOS shows as a step function for a 2D monolayer and scales to $\sqrt{\mathrm{E}}$ for a 3D bulk. Note that the magnon DOS onsets at the threshold of magnon excitations (i.e., magnon gap energy). In between the monolayer and the bulk, the magnon DOS spectrum evolves as a gradual building-up process, and for example, for a multilayer case, its magnon DOS features the sum of multiple step functions showing the increasing trend (see Figure \ref{magnon dos}). As a result, for the same amount of thermal excitations, their impact in the 2D magnetism is much greater than in their 3D counterpart, because of the lower magnon DOS in the 2D than in the 3D \cite{gong2017discovery, gong2019two}.

In the 2D limit, without an easy-axis magnetic anisotropy ($d=2,n=2,3$), a combination of gapless magnon band, low magnon DOS, and diverging Bose-Einstein statistics at zero energy leads to massive magnon excitations at any nonzero temperatures and therefore absence of long-range magnetic orders. On the other hand, with an easy-axis magnetic anisotropy in 2D magnets ($d=2, n=1$), the presence of magnon gaps counteracts thermal excitations and makes it possible to have long-range magnetic orders at finite temperatures.

\subsection{Other magnetic exchange coupling terms in two-dimensional magnetism}
In Section 1.1, we have primarily focused on a simplified spin Hamiltonian with three standard classes. It is, however, worth highlighting two more magnetic exchange coupling terms that are known to play important roles in the recent studies on quantum magnetism and can potentially have profound manifestations in the 2D limit, the Dzyaloshinskii–Moriya (DM) interaction and the Kitaev interaction. 

\subsubsection{The Dzyaloshinskii-Moriya interaction}
The antisymmetric exchange coupling between two neighboring spins, $\Vec{S}_i$ and $\Vec{S}_j$, is described by the DM interaction, whose Hamiltonian is:

$$\mathcal{H}_\mathrm{DM} = -\Vec{D}_{ij}\cdot(\hat{\Vec{\mathrm{S}}}_i \times \hat{\Vec{\mathrm{S}}}_j),$$

\noindent where $\Vec{D}_{ij}$ is the Dzyaloshinskii vector pointing along a high-symmetry direction of the magnetic crystals (Figure \ref{magnetic exchange}d). 

For a magnetic ground state, the effect of this Hamiltonian is to provide an additional energy gain if the two spins are aligned perpendicular to each other. For systems where the DM interaction takes up a substantial weight in the total spin Hamiltonian, the spin alignments often deviate from those of the simple collinear FMs and AFMs. For example, the manifestation of DM interaction includes weak FM \cite{dzyaloshinsky1958thermodynamic,moriya1960anisotropic}, magnetic Skyrmion \cite{skyrme1962unified,mühlbauer2009skyrmion,nagaosa2013topological}, magnetoelectric effect in multiferroic material \cite{fiebig2005revival,cheong2007multiferroics} and others. For magnetic excitation, the DM interaction to the magnon dispersion spectra is analogous to the spin-orbit coupling to the electronic band structure. For systems like honeycomb and kagome magnets which support the Dirac-like magnon dispersion near the Brillouin zone corners (i.e., K points), the presence of DM interaction can open up magnon gaps at the Dirac points and introduce topologically nontrivial spin wave edge states \cite{chisnell2015topological,pershoguba2018dirac,chen2018topological,chen2021magnetic}. 

\subsubsection{The Kitaev interaction}
The unique bond-directional Ising-type exchange coupling amongst the neighboring spin sites, $i$ and $j$, is described by the Kitaev interaction, and it is quantitatively described by the Kitaev spin Hamiltonian \cite{kitaev2006anyons,winter2017models,takagi2019concept}:

$$\mathcal{H}_\mathrm{Kitaev} = -\sum_{\gamma-bonds} {K_{\gamma}\hat{S}_i^\gamma \hat{S}_j^\gamma},$$

\noindent where the easy-axis $\gamma$ depends on the spatial orientation of an exchange bond, i.e., $\gamma$-bond. This is better illustrated by an example of the honeycomb lattice which has three nearest neighbors for any spin site, i.e., three nearest-neighbor exchange bonds. The Kitaev Hamiltonian specifies that one of the bonds has an easy axis along the $x-$direction, another along the $y-$direction, and the last along the $z-$direction (shown in Figure \ref{magnetic exchange}e).

The attraction of the Kitaev spin model lies in the fact that it is an exactly solvable spin model on the 2D honeycomb lattice and supports quantum spin liquid as its ground state. Over the past decades, tremendous efforts have been put into searching for 2D and 3D material candidates where the Kitaev exchange coupling term dominates the rest of spin interactions \cite{winter2017models,takagi2019concept,trebst2022kitaev}. On the other hand, in materials where the Kitaev exchange coupling term is at a weaker energy scale than other spin interaction terms, it has been shown recently that the Kitaev term can still impact the magnon excitation spectra significantly \cite{xu2018interplay,lee2020fundamental}.

\subsection{Historical overview of experimental efforts in two-dimensional magnetism}
\begin{figure}[th]
\begin{center}
\includegraphics[width=1.0\textwidth]{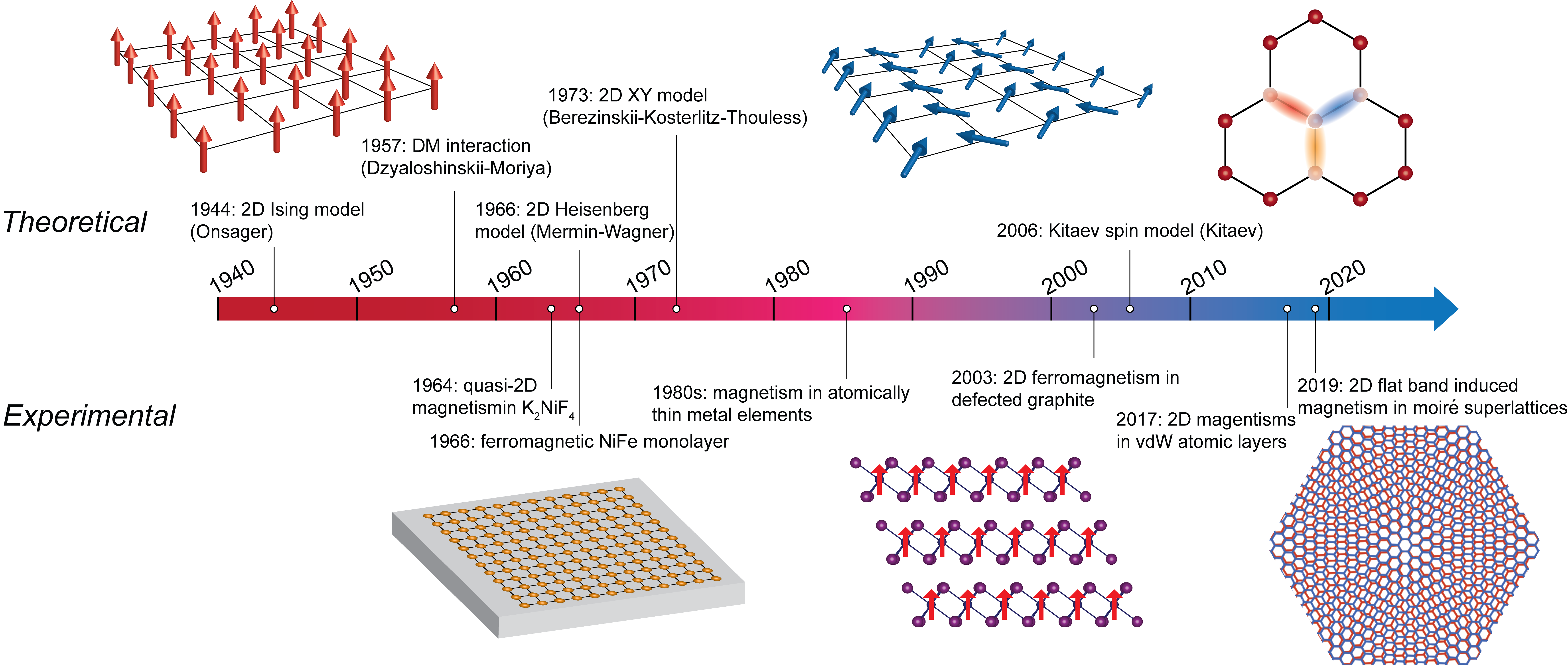}\caption{\small Timeline of the historical development in the 2D magnetism research. Top and bottom panels highlight the theoretical and experimental milestones, respectively.}
\label{historical overview}
\end{center}
\end{figure}

The theoretical foundation of 2D magnetism sparked much experimental interest in the early years dating back to the 1960s in 3D bulk materials with quasi-2D magnetic structures \cite{plumier1964neutron}. Such experimental enthusiasm has continued since then and, indeed, has achieved substantive progress before and after the discovery of 2D vdW magnets \cite{cortie2020two}. Figure \ref{historical overview} illustrates the timeline for the theoretical and experimental milestones in the research of 2D magnetism. 

An overview of four distinct stages of experimental pursuit in 2D magnetism is provided in this subsection. It consists of the studies on quasi-2D bulk magnetic materials in the 1960s, the investigations of magnetic elemental thin films in the 1980s, then the attempts of defected graphene monolayers in the 2000s, and finally the success of intrinsic 2D vdW magnetic atomic crystals and moir\'e superlattices since the late 2010s. 

\subsubsection{Quasi-two-dimensional magnetism in three-dimensional bulk magnets}
Magnetic exchange interactions are typically short-ranged in nature \cite{pajda2001ab}. As a result, it is not so rare to achieve magnetic materials whose magnetic interaction between neighboring spin sites is negligible if the site separation is big (i.e., $> 1$ nm). In such cases, even though the host material is a 3D bulk, the effective magnetic interactions and, hence, the magnetic properties are, in fact, of lower dimensionality. Quasi-2D magnets are those magnetic materials where the magnetic interactions are strong within the plane, but extremely weak along the out-of-plane direction. 

The first quasi-2D magnetism was demonstrated in the 1960s, using K$_2$NiF$_4$ as a material realization platform \cite{plumier1964neutron,lines1969magnetism}. A key experimental finding was achieved by neutron diffraction measurements in which the long-range magnetic order establishes within the 2D plane but remains incoherent between planes (i.e., magnetic diffraction peaks appear in the 2D momentum space, while magnetic ridges show up along the out-of-plane momentum direction). Over the years since the 1960s, more and more magnetic material families were found to be or close to quasi-2D magnets. Examples include but are not limited to vdW layered magnets \cite{de1990magnetic}, Ruddlesden-Popper perovskite magnets \cite{phan2007review}, honeycomb, hexagonal, and kagome magnets \cite{oitmaa1992quantum,yin2022topological}, etc., whose interlayer coupling is reduced either by large separations or by competing exchange pathways \cite{cortie2020two}. We further note that the excitement in quasi-2D magnetism extends beyond the investigation of 2D magnetic ordering behaviors and expands to the potential of mediating other novel phases of matter, such as quantum spin liquid \cite{takagi2019concept}, unconventional superconductivity \cite{anderson1987resonating}, etc. 

\subsubsection{Two-dimensional magnetism in atomically thin elemental metal films}
The epitaxial growth technology opened up the pathway of synthesizing atomically thin magnetic films on top of non-magnetic substrates. As compared to the quasi-2D magnets, such epitaxially grown magnetic films, first, allowed for the investigation of the thickness dependence of magnetism -- a knob to tune the lattice dimensionality and its interplay with the spin dimensionality, and second, enabled many interesting phenomena absent in 3D bulk crystals -- a platform stimulating both scientific and application leaps.

The first realization of 2D magnetism in epitaxial thin films was made in the late 1960s, in which 1.8 monolayer-thick NiFe films were grown on the Cu(111) substrate, and the Curie temperature was measured to be around 220 K \cite{gradmann1966struktur,gradmann1968very}. In the 1980s, the field expanded to the growth of magnetic elemental films of Fe, Ni, Co, and Cr on top of substrates of Au, Cu, Pd, and Ag, and moreover, synthesis of heterostructures made up of metal substrate/magnetic elemental film/metal film. During the 1980s -- 2000s, experimental investigations examined the critical behaviors of 2D magnetism in the elemental metal films and compared them with the theoretical classifications of 2D Ising, XY, and isotropic Heisenberg models, resulting in the major findings of long-range magnetic orders in both 2D Ising and 2D XY magnets \cite{qiu1994two,bander1988ferromagnetism} and the discovery of spin reorientation transitions \cite{chappert1986ferromagnetic,lugert1989evidence,qiu1993asymmetry,pappas1990reversible,fritzsche1994angular}. At the same time, studies of this family of elemental metal films also raised concerns about how intrinsic the 2D magnetism is, because of the coupling between films and substrates, the structural defects present in the film, and the finite size effects in the lateral dimensions. Nonetheless, the research on such thin magnetic films enabled tremendous technological advancement through giant magneto-resistance \cite{binasch1989enhanced}, magneto-tunneling junction \cite{zhu2006magnetic}, spin transfer torque \cite{ralph2008spin}, etc.

\subsubsection{Atomic defect-induced two-dimensional ferromagnetism in graphene monolayers}
Graphene, a 2D honeycomb lattice made of carbon atoms, was discovered through the mechanical exfoliation of graphite back in 2004 \cite{novoselov2004electric,zhang2005experimental,novoselov2005two}, and ever since, opened the era of 2D vdW atomic crystals and structures \cite{novoselov2005two}. As the first 2D atomic crystals, graphene hosts a wealth of exceptional physical properties. Structurally, it has ultra-low atomic defect concentration in exfoliated graphene monolayers \cite{yang2018structure}. Mechanically, it is the strongest material ever found \cite{papageorgiou2017mechanical}. Electronically, it has ultra-high carrier mobility and massless Dirac fermions \cite{zhang2005experimental,novoselov2005two}. Yet, because graphene is made of non-magnetic carbon atoms with the \textit{sp} bonding scheme, it is diamagnetic intrinsically \cite{sepioni2010limits}. 

A comprehensive review article on introducing magnetism in graphene is published recently \cite{tuvcek2018emerging}. Theoretically, it was predicted that the selected atomic defects and adatoms are possible to introduce finite magnetic moment into graphene monolayers \cite{yazyev2007defect}. Experimentally, it was shown that weak proton irradiation of highly ordered pyrolytic graphite (HOPG) introduces atomic defects and also generates weak FM in HOPG \cite{esquinazi2003induced}. Furthermore, it was also suggested that the atomic defects and the light adatoms introduce paramagnetism in atomically thin graphene layers \cite{nair2012spin}. In addition, many intriguing physical properties together with the flexible gate-tunability also attracted research interests in using ``magnetized" graphene for spintronic applications \cite{han2014graphene}. 

\subsubsection{Two-dimensional magnetism in van der Waals magnetic atomic crystals and moir\'e superlattices}
The rapid expansion of 2D materials research in the 2010s identified that vdW magnetic materials can be a viable platform for realizing 2D magnetism. The very first discoveries of 2D magnetism in atomically thin vdW magnets mainly concern the Ising-type of magnets with the easy-axis anisotropy, including 2D AFM in FePS$_3$ in 2016 \cite{lee2016ising} and 2D FM in CrI$_3$ \cite{huang2017layer} and Cr$_2$Ge$_2$Te$_6$ \cite{gong2017discovery} in 2017. These findings soon triggered the blossom of research in 2D vdW magnets \cite{burch2018magnetism, gibertini2019magnetic}, expanding into numerous research directions such as searching for more 2D magnetism candidates, exploring 2D magnetism of different spin Hamiltonian classes, investigating impact of enhanced fluctuations on 2D magnetism, implementing 2D magnetism-based spintronic devices, integrating with other 2D materials for new functionalities, and so on. As of now, the field of 2D vdW magnetism is at a steep growth speed that generates new scientific insights into phase transitions in the 2D limit and stimulates novel application ideas for future micro-spintronics.

Built upon the maturing of 2D vdW magnetism research and further inspired by the moir\'e electronic physics in twisted graphene and twisted transition metal dichalcogenides (TMDCs), very recent efforts have been made to realize moir\'e magnetism in twisted 2D vdW magnets. The idea of moir\'e magnetism is to introduce a spatially modulated magnetic exchange coupling at the moir\'e wavelength and harvest the results from the competition between the modulating moir\'e interlayer exchange coupling and the uniform intralayer exchange coupling. Theoretical predictions include noncollinear spin textures \cite{hejazi2020noncollinear}, skrymion lattices \cite{tong2018skyrmions,hejazi2021heterobilayer,akram2021skyrmions,akram2021moire,ghader2022whirling,zheng2023magnetic}, moir\'e magnons \cite{li2020moire}, topological magnons \cite{kim2022theory}, etc. Experimental efforts have been mainly focused on the twisted CrI$_3$ system with the major findings of coexisting layered AFM and FM phases \cite{song2021direct,xu2022coexisting, xie2022twist,cheng2023electrically} and emerging noncollinear spins \cite{xie2023evidence}.  

\subsubsection{Correlation-induced two-dimensional magnetism from the moir\'e flat bands}
Nearly concurrent with the discovery of 2D vdW magnets, flat electronic bands, i.e., electronic bands with no or little energy dispersion, were realized by the formation of moir\'e superlattices in twisted graphene bilayers first \cite{cao2018correlated,cao2018unconventional} and then twisted TMDCs \cite{tang2020simulation,regan2020mott}. As a result, twisted graphene and twisted TMDCs moir\'e superlattices become not only quantum simulators for many strongly correlated physics models \cite{kennes2021moire}, but also promising platforms to realize new quantum phases in the 2D limit. Examples of emergent electronic states include Mott insulator \cite{cao2018correlated}, Wigner crystal \cite{regan2020mott}, charge order insulator \cite{xu2020correlated}, unconventional superconductivity \cite{cao2018unconventional}, nematic order \cite{rubio2022moire}, heavy fermions \cite{zhao2023gate}, etc. 

Magnetism could be one manifestation of orbital and spin degrees of freedom from the strong electronic correlations, and therefore, twisted graphene and twisted TMDC moir\'e superlattices are another two platforms, in addition to vdW magnets, for realizing correlation-induced magnetism in the 2D limit. So far, orbital FM has been reported in ABC trilayer graphene on hexagonal boron nitride (hBN) \cite{sharpe2019emergent}, twisted graphene bilayers on hBN \cite{nuckolls2020strongly}, and twisted TMDC heterostructures \cite{li2021continuous}. In addition, the FM-to-AFM phase transition is also achieved by applying a vertical displacement field in twisted rhombohedral stacked MoTe$_2$ homobilayers \cite{anderson2023programming}. Although the AFM ordered phase has not been directly measured, its presence in the Mott insulating state has been suggested in twisted TMDC heterostructures \cite{tang2020simulation,tang2023evidence}. As compared to vdW atomic and moir\'e magnets with intrinsic magnetic elements, the correlation-induced 2D magnetism in twisted graphene and TMDCs typically emerges at lower temperatures. At the same time, they provide unique opportunities to investigate the interplay among magnetism, correlations, and topology.

\section{vdW materials platform for realizing two-dimensional magnetism}

The intrinsic 2D vdW magnets mainly develop in two families of systems - vdW atomic crystals with magnetic elements and twisted moir\'e superlattices with flat bands. The mechanisms of 2D magnetism for these two families are distinct. The first family is primarily based on the spin degree of freedom, although the spin-orbit coupling in heavy magnetic elements calls for the participation of the orbital degree of freedom. The second family is, however, mostly dominated by the orbital degree of freedom, even though sometimes there are unpaired electrons per moir\'e supercell. In recent years, for the first family, the material pool is rapidly expanding, and for the second family, more tuning knobs and device geometries have been developed, although the material platforms are mainly twisted graphene and twisted TMDC moir\'e superlattices.

\subsection{Two-dimensional magnetic atomic crystals and moir\'e superlattices}

We start with the first family, vdW magnetic atomic crystals where elements with partially filled $d$ or $f$ orbitals provide the intrinsic magnetic moments. Since the discovery of 2D magnetic atomic crystals in 2016 -- 2017 \cite{lee2016ising, huang2017layer, gong2017discovery}, the rapidly expanding 2D magnet database primarily includes the following classes \cite{jiang2021recent}: (i) binary $d-$, $f-$electron metal halides, MX$_3$, MX$_2$, MX, and M$_3$X$_8$, with M = 3$d$, 4$d$, 5$d$ transition metals, and 4$f$ and 5$f$ rare earth elements, and X = F, Cl, Br, and I; (ii) binary $d-$electron chalcogenides, MX$_2$, M$_5$X$_8$, M$_2$X$_3$, M$_3$X$_4$, and MX, with M = 3$d$ transition metals, and X = S, Se, and Te; (iii) ternary $d-$electron compounds, MXY$_3$ with M = 3d transition metals, X = P and Y = S, Se, and Te, or X = Si, Ge, and Sn, and Y = Te, MXY with M = 3$d$ transition metals, X = O, S, Se, and Te, and Y = Cl, Br, and I, Fe-Ge-Te compounds, and Mn-Bi-Te compounds. 

\begin{figure}[th]
\begin{center}
\includegraphics[width=0.95\textwidth]{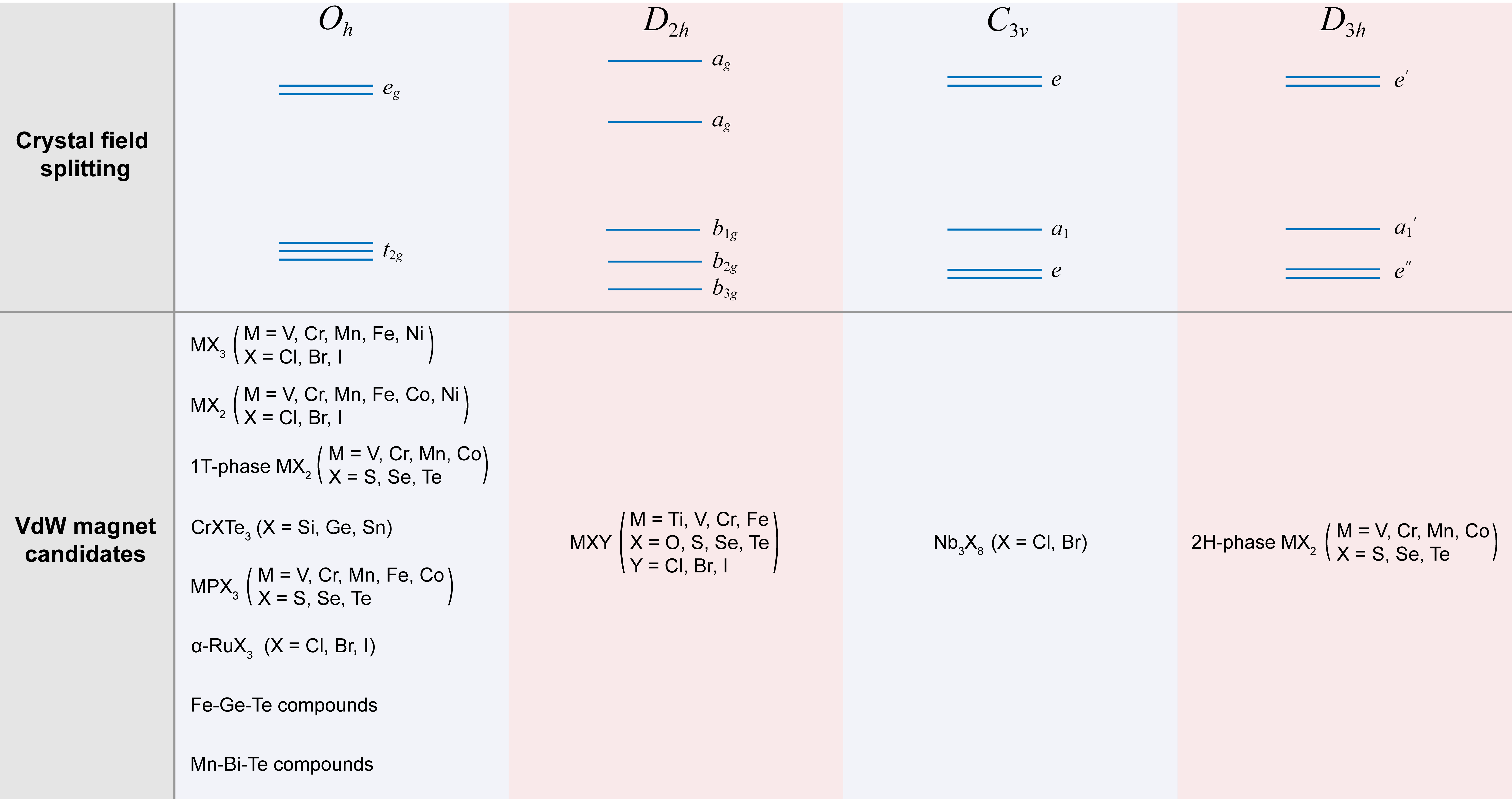}\caption{\small Summary of representative crystal field effects on lifting the $d$-orbital degeneracy. (Top row) Sketch of orbital degeneracy diagrams under four different crystal fields, $O_h$, $D_{2h}$, $C_{3v}$, and $D_{3h}$. (Bottom row) Material candidates of 2D vdW magnets listed for the corresponding crystal fields.}
\label{crystal field}
\end{center}
\end{figure}

For all the vdW magnetic atomic crystals above, the magnetic ions are subject to the crystal field from the surrounding ligand ions. The crystal field effect is classified by its symmetry, for which the common ones include octahedral ($O_h$), distorted octahedral ($D_{2h}$), trigonal prismatic ($D_{3h}$), triangular prismatic ($C_{3v}$), hexagonal ($C_{6v}$), tetragonal ($T_d$), etc. Taking the $d$-orbital as an example, the orbital degeneracy lifting is summarized in Figure \ref{crystal field} with the corresponding vdW magnet candidates listed as well. The ligand field theory provides an intuitive model to predict the magnetic behavior, based on the coordination environment and the number of $d$ or $f$ electrons for the magnetic ions. For example, for the Cr$^{3+}$ ion inside an $O_h$ crystal field, based on Hund's rule, the three remaining 3$d$ electrons half-fill the lower t$_{\mathrm{2g}}$ manifold, and therefore, give a local magnetic moment of 3$\mu_\mathrm{B}$. In this subsection, we will group 2D vdW magnetic atomic crystals by the involved magnetic ions. 

\subsubsection{3$d$ electron-based two-dimensional vdW magnets}

The first and most explored class of vdW magnets concerns 3$d$ transition metal compounds. Most 3$d$ transition metal ions can have a partially filled 3$d$ shell to host localized magnetic moment, including Ti, V, Cr, Mn, Fe, Co, and Ni. Some of them, in fact, can have different oxidation states depending on the ligand environment, for example, Fe$^{3+}$ in FeI$_3$ \cite{guan2021strain}, Fe$^{2+}$ in FePS$_3$ \cite{lanccon2016magnetic}, and both Fe$^{3+}$ and Fe$^{2+}$ in Fe$_3$GeTe$_2$ \cite{fei2018two,deng2018gate}. In addition, 3$d$ transition metal-based vdW magnets cover a large variety of crystal classes, including (i) binary transition metal halides, (ii) binary transition metal chalcogenides, and (iii) ternary transition metal compounds. 

\textbf{3$d$ transition metal halides} have gained tremendous research interest since the discovery of 2D FM in monolayer CrI$_3$ in 2017, which opened the era of 2D vdW magnetism. As of now, this family of vdW magnets includes 3$d$ transition metal trihalides (MX$_3$) and 3$d$ transition metal dihalides (MX$_2$). For both MX$_3$ and MX$_2$, the transition metal ion M is situated in the center of an octahedral cage made up of halide ligand ions X = F, Cl, Br, and I.  

\begin{figure}[th]
\begin{center}
\includegraphics[width=1.0\textwidth]{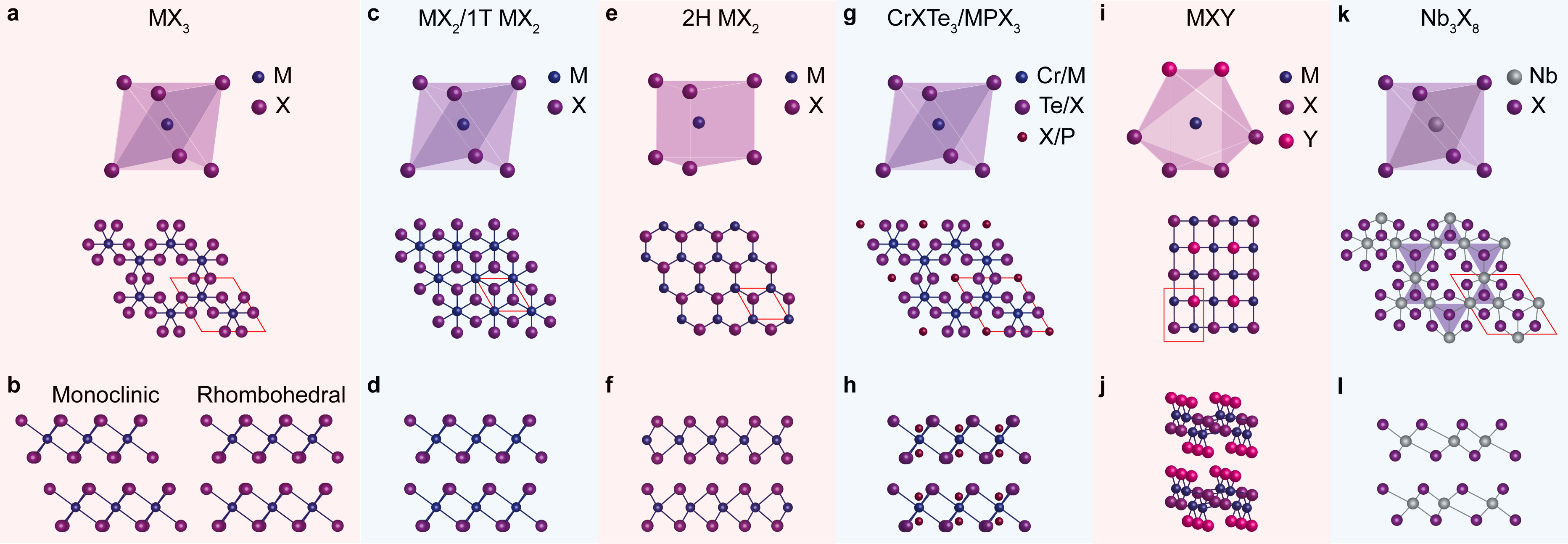}\caption{\small Summary of crystal structures of $d$ electron-based 2D vdW magnets. In-plane lattice structure and interlayer stacking geometry for (a-b) 3$d$ transition metal trihalides, MX$_3$; (c-d) 3$d$ transition metal dihalides MX$_2$ and 1T-phase of 3$d$ transition metal dichalcogenides 1T MX$_2$; (e-f) 2H-phase of 3$d$ transition metal dichalcogenides 2H MX$_2$;(g-h) ternary 3$d$ transition metal compound CrXTe$_3$ and MPX$_3$; (i-j) ternary 3$d$ transition metal compound MXY; and (k-l) Nb$_3$X$_8$.} 
\label{material structure1}
\end{center}
\end{figure}

The known MX$_3$ identified by experiment and theory includes M = V, Cr, Mn, Fe, and Ni. The trivalence and the octahedral coordination sets V, Cr, Mn, Fe, and Ni cations to be partially filled with 3$d^2$, 3$d^3$, 3$d^4$, 3$d^5$, and 3$d^7$ electronic configurations with $S=1, 3/2, 2, 5/2$, and $3/2$, respectively. Within a monolayer MX$_3$, the MX$_6$ octahedral cages are arranged in an edge-sharing manner into a honeycomb lattice, with two magnetic sites per primitive cell (Figure \ref{material structure1}a). The intralayer exchange coupling is FM for M = V, Cr, Mn, and Ni and AFM for M = Fe \cite{tomar2019intrinsic,mcguire2017crystal}. The spin anisotropy type is tunable by changing the ligand halide ions. For example, CrX$_3$ evolves from easy-plane to isotropic to easy-axis anisotropy as X switches from Cl to Br to I \cite{kim2019evolution}. For the vertical stacking, two stacking geometries have been reported for CrX$_3$, the rhombohedral stacking and the monoclinic stacking (Figure \ref{material structure1}b), with very similar elastic energies \cite{sivadas2018stacking}. Very interestingly, the interlayer exchange coupling can depend on the interlayer stacking geometries, that is, in CrX$_3$, interlayer AFM for the monoclinic stacking and interlayer FM for the rhombohedral stacking \cite{sivadas2018stacking}. From comparative density functional theory (DFT) calculations, MX$_3$ monolayers are semiconductors for M = V, Cr, and Fe, and half-metals for M = Mn and Ni.

MX$_2$ candidates are mostly identified via theory, including M = V, Cr, Mn, Fe, Co, and Ni \cite{mcguire2017crystal}. Similar to MX$_3$, the electronic and spin configurations are determined by the divalence and the octahedral coordination of M cations, with 3$d^3$, 3$d^4$, 3$d^5$, 3$d^6$, 3$d^7$, and 3$d^8$ electronic configurations and $S = 3/2, 2, 5/2, 2, 3/2$, and $1$ spin configurations, for M = V, Cr, Mn, Fe, Co, and Ni, respectively. Within the layer of MX$_2$, the MX$_6$ octahedra are arranged into a hexagonal lattice with only one magnetic site per structural primitive cell (Figure \ref{material structure1}c), in contrast to two per cell for MX$_3$. The intralayer coupling is predicted to be FM for FeX$_2$, CoCl$_2$, CoBr$_2$, and NiX$_2$, and AFM for VX$_2$, CrX$_2$, MnX$_2$, and CoI$_2$, where X = Cl, Br, and I. Along the out-of-plane direction of MX$_2$, the stacking geometry (mostly overlaying stacking, Figure \ref{material structure1}d),  the interlayer exchange coupling, and their relationship are much less explored than those in MX$_3$.

\textbf{3$d$ transition metal chalcogenides} are another popular class of vdW magnets, including the dominant transition metal dichalcogenides (MX$_2$) and the other less investigated candidates with higher stochiometry (e.g., M$_5$X$_8$, M$_2$X$_3$, M$_3$X$_4$, and MX). The widely studied MX$_2$ magnetic monolayers form the X-M-X sandwich structure where the transition metal ion M sits in between the chalcogenide ligand ions X. There are two major types of structural phases reported for the transition metal dichalcogenide magnets MX$_2$, trigonal prismatic H-phase and octahedral T-phase, with distinct crystal field splitting shown in Figure \ref{crystal field}.

For the T-phase magnetic MX$_2$, the identified materials candidates include M = V, Cr, Mn, and Co and X = S, Se, and Te. Same as transition metal dihalides, the MX$_6$ octahedra here also arrange into a hexagonal lattice within the plane and stack on top of each other along the out-of-plane direction (Figures \ref{material structure1}c and \ref{material structure1}d). T-phase VSe$_2$, VTe$_2$, MnSe$_2$, and CrTe$_2$ are of particular interest from both theoretical and experimental perspectives, because they were predicted and demonstrated to achieve magnetic critical temperatures above the room temperature \cite{bonilla2018strong,tang2022strain,o2018room,zhang2021room} and can host interplay between magnetism and charge density wave (CDW) orders \cite{feng2018electronic,wang2019evidence,otero2020controlled}. T-VSe$_2$ and T-VTe$_2$ monolayers are among the first room-temperature 2D FMs, with a 3$d^1$ electronic configuration and an $S=1/2$ spin configuration that support both their metallic and magnetic properties \cite{bonilla2018strong}. CDW orders have been reported in both T-VSe$_2$ \cite{feng2018electronic} and T-VTe$_2$ \cite{wang2019evidence}. The competitive relationship between FM and CDW expected from a theoretical perspective remains debated through experimental data.

For the H-phase magnetic MX$_2$, the theoretically and experimentally investigated materials include M = V, Cr, Mn, and Co and X = S, Se, and Te, very similar to those in the T-phase. Within the plane of layers, the MX$_6$ trigonal prisms form a honeycomb lattice with one of its two sublattices occupied by a transition metal ion M and the other by chalcogen ligand ions X (Figure \ref{material structure1}e). When stacking the layers along the out-of-plane direction, there are two stacking phases, the 2H phase where the two layers within one unit cell are rotated by 180$^{\mathrm{o}}$ and stacked vertically on top of each other (Figure \ref{material structure1}f), and the 3R phase where the three layers within the unit cell are aligned at 0$^{\mathrm{o}}$ and stacked in the rhombohedral ABC sequence. According to LDA+U calculations \cite{ataca2012stable}, H-phase MX$_2$ are predicted to be FM metals. Despite being metallic, no predictions or observations of CDW orders have been made in H-phase magnetic MX$_2$. Furthermore, little discussions have been devoted in literature on the comparison of the interlayer exchange coupling between the 2H and 3R phases.

\textbf{Ternary 3$d$ transition metal compounds} include a rich variety of chemical compositions and a diverse wealth of magnetic properties. Generally speaking, ternary transition metal compounds comprise the M-X$'$-X$''$ structure where M is the 3$d$ transition metal ions, and X$'$/X$''$ are the non-magnetic ligand ions from main groups IV, V, VI, or VII in the periodic table. Five classes of materials within this family have been identified experimentally and theoretically, namely, CrXTe$_3$ (X = Si, Ge, and Sn), MPX$_3$ (M = V, Cr, Mn, Fe, Co, and Ni; X = S, Se, and Te), MXY (M = Ti, V, Cr, and Fe; X = O, S, Se, and Te; Y = Cl, Br, and I), Fe-Ge-Te compound, and Mn-Bi-Te compound.

CrXTe$_3$ (X = Si, Ge, Sn) has received much research interest since 2017 when the FM long-range order was discovered in CrGeTe$_3$ \cite{gong2017discovery}, one of the first 2D FM semiconductors. Cr ion is in the trivalence state and sits in the center of the Te octahedra, consistent with the 3$d^3$ electronic configuration and the $S=3/2$ spin configuration. Within the layers, the CrTe$_6$ octahedra arrange in a honeycomb lattice with two magnetic sites per primitive cell, and the X ions situate in the hollow center of this honeycomb lattice, forming the double pyramid skeleton [X$_2$Te$_6$]$^{6-}$ (Figure \ref{material structure1}g). Between layers, the CrXTe$_3$ planes follow the rhombohedral ABC stacking sequence \cite{gong2017discovery} (Figure \ref{material structure1}h). CrSiTe$_3$ and CrGeTe$_3$ are experimentally verified to be FM semiconductors \cite{lin2016ultrathin,lin2017tricritical} while CrSnTe$_3$ is theoretically predicted to be an FM semiconductor as well \cite{zhuang2015computational}.  Using the hybrid density functional method, the key magnetic properties of CrXTe$_3$ are analyzed in a comparative and unified manner. As X is changed from Si to Ge to Sn, the magnetic critical temperature increases from 90 K theoretically (32 K experimentally) to 130 K (63 K experimentally) to 170 K (N.A. experimentally) due to the enhanced superexchange coupling from 2.10 meV to 3.07 meV to 3.92 meV for the nearest neighboring Cr sites \cite{zhuang2015computational}.

 MPX$_3$ (M = V, Cr, Mn, Fe, Co, and Ni; X = S, Se, and Te) is the class of vdW magnets from which the first 2D AFM, monolayer FePS$_3$ \cite{lee2016ising}, was discovered in 2016 and therefore has attracted highly focused efforts ever since then. The crystalline structure of MPX$_3$ monolayers is very similar to that of CrXTe$_3$ monolayers, except replacing the group IV X ions in CrXTe$_3$ by the group V P ions (Figure \ref{material structure1}g). Between layers, MPX$_3$ crystals have the monoclinic or rhombohedral stacking geometry (Figure \ref{material structure1}h). The change of transition metal ions can tune the magnetic anisotropy and the magnetic ground states. Comparing across the three members of 3D MnPS$_3$, FePS$_3$, and NiPS$_3$ with the same [P$_2$S$_6$]$^{4-}$ skeleton \cite{joy1992magnetism}, MnPS$_3$ was shown to have nearly isotropic Heisenberg-type spin interactions and host an AFM order with spins aligned along the out-of-plane direction and pointing in opposite directions between two honeycomb sublattices; FePS$_3$ exhibits the Ising-type spin Hamiltonian and realizes a zigzag AFM order with spins aligned in the same out-of-plane direction in the zigzag chains and the opposite directions between neighboring zigzag chains; NiPS$_3$ hosts a XY-type spin coupling and supports a zigzag AFM order with spins aligned along the zigzag chain direction. Furthermore, it has been shown that the zigzag chains are locked to the monoclinic stacking, for which three degenerate domain states are present \cite{joy1992magnetism}. Down to the monolayer limit, the zigzag out-of-plane AFM order in FePS$_3$ survives \cite{lee2016ising}, the zigzig in-plane AFM order in NiPS$_3$ is suppressed \cite{kim2019suppression}, and the out-of-plane N\'eel AFM order in MnPS$_3$ remains debated \cite{long2020persistence}. Compared to MnPS$_3$, MnPSe$_3$ is presumed to have an XY-type spin interaction and to hold an in-plane N\'eel AFM order with the spins aligned along the zigzag chain direction \cite{jeevanandam1999magnetism}. Despite the XY-nature, this in-plane N\'eel AFM order survives down to the monolayer limit attributed to the small strain built in during the mechanical exfoliation process \cite{ni2021imaging}.   

 MXY (M = Ti, V, Cr, and Fe; X = O, S, Se, and Te; Y = Cl, Br, and I) is a new class of vdW magnets that have received interest due to its orthorhombic magnetic lattice distinct from the popular honeycomb magnetic lattice discussed above. Within each layer, MXY consists of M$_2$X$_2$ sandwiched by the halogen Y layers where the transition metal M ion locates at the center of distorted octahedra ($D_{\mathrm{2h}}$ symmetry) made up of X and Y ligands (Figure \ref{material structure1}i). Between layers, MXY layers aligned on top of each other (Figure \ref{material structure1}j) to form the orthorhombic $Pmmn$ space group. Among the large class of MXY magnets, CrSBr is the first experimentally identified 2D orthorhombic magnet \cite{lee2021magnetic}. CrSBr is an Ising-type magnet with the easy-axis along the $b$-axis and realizes a layered A-type AFM order \cite{liu2022three}, which survives down to the 2D limit. Of particular interest to CrSBr is the highly anisotropic electronic band structure that leads to linearly polarized exciton emissions. More interestingly, this unique exciton couples efficiently to both the A-type AFM order \cite{wilson2021interlayer} and the spin waves \cite{diederich2023tunable}. In addition to CrSBr, a few theoretical studies predict that CrXY (X = S, Se, Te; Y = Cl, Br, I) monolayers are FM semiconductors with high magnetic transition temperatures \cite{jiang2018screening,guo2018chromium,wang2019family,han2020prediction} with similar electronic, photonic, and magnetic properties as CrSBr, albeit the difference in spin anisotropy. Furthermore, MXY (X = S, Se, Te; Y = Cl, Br, I) with 3$d$ transition metals other than Cr were computed using the hybrid functional theory, which predicts more FM semiconductors, such as TiTeI, VSI, and VSeI \cite{pan2020quantum}. Finally, transition metal oxyhalides MOY (M = V, Cr, and Fe; Y = F, Cl, Br, and I) monolayers were predicted to support 2D magnetism, including FM insulators CrOF, CrOCl, CrOBr, \cite{miao20182d} and VOF \cite{liu2018screening}, AFM Mott insulators FeOY (Y = F, Cl, Br, and I) \cite{wang2020discovery}, and FM half-metals VOCl and VOBr \cite{pan2020quantum}. 

\begin{figure}[th]
\begin{center}
\includegraphics[width=0.9\textwidth]{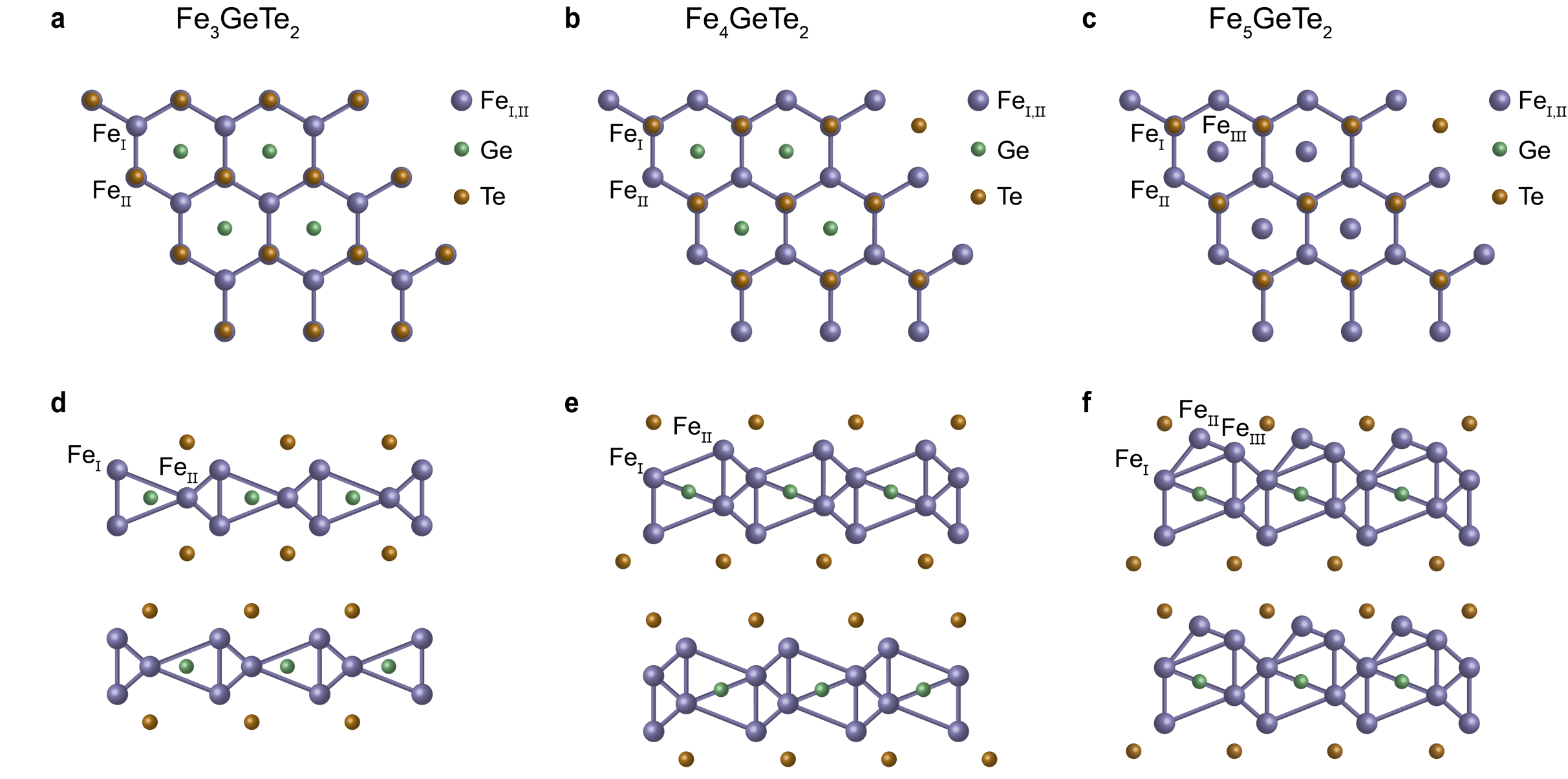}\caption{\small Summary of crystal structures of Fe$_n$GeTe$_2$. (a-c) In-plane lattice structure and (d-f) out-of-plane interlayer stacking of Fe$_n$GeTe$_2$ with $n=3, 4,$ and 5, respectively. Different types of Fe sites are labeled with Fe$_\mathrm{I}$, Fe$_\mathrm{II}$, and Fe$_\mathrm{III}$. This figure is adapted from Ref. \cite{seo2020nearly}.} 
\label{FGT}
\end{center}
\end{figure}

 \textbf{Fe$_n$GeTe$_2$ (3$\leq n \leq$5)} realizes the FM metallic ground state with strong easy-axis anisotropy \cite{seo2020nearly}, thus standing out as a unique class of ternary transition metal compounds. In each Fe$_\mathrm{n}$GeTe$_2$ layer, covalently bonded Fe$_\mathrm{n}$ heterometallic slab is sandwiched between two Te layers, where the in-plane projection of Fe sites forms a honeycomb lattice (Figures \ref{FGT}a-c). Between Fe$_\mathrm{n}$GeTe$_2$ layers, it follows the rhombohedral ABC stacking sequence (Figures \ref{FGT}d-f). For $n=3$, the structure and the valence states of Fe$_3$GeTe$_2$ can be written as (Te$^{2-}$)(Fe$_\mathrm{I}^{3+}$)[(Fe$_{\mathrm{II}}^{2+}$)(Ge$^{4-}$)](Fe$_\mathrm{I}^{3+}$)(Te$^{2-}$) with two inequivalent Fe sites, two Fe$_\mathrm{I}^{3+}$with one above and one below the Ge layer, and one Fe$_{\mathrm{II}}^{2+}$ in the same plane as Ge (Figure \ref{FGT}d) \cite{deng2018gate}. For $n=4$, similarly, Fe$_4$GeTe$_2$ also has two inequivalent Fe sites, two pairs of Fe$_\mathrm{I}^{3+}$ and Fe$_{\mathrm{II}}^{2+}$ with one pair above and one pair below the Ge layer (Figure \ref{FGT}e) \cite{zhang2020itinerant}. For $n=5$, Fe$_5$GeTe$_2$ contains three Fe sites, where Fe$_\mathrm{I}^{3+}$ is at the split sites that are either above or below the Ge layer, and two pairs of Fe$_\mathrm{II}^{3+}$ and Fe$_\mathrm{III}^{2+}$ with one pair above and one pair below the Ge plane just like in Fe$_4$GeTe$_2$ (Figure \ref{FGT}f) \cite{may2019ferromagnetism}. By gradually approaching the 3D-like network with enhanced exchange coupling while maintaining the vdW crystallographic structure as $n$ increases in Fe$_n$GeTe$_2$ (3$\leq n \leq$5), the FM critical temperature increases from 220K in Fe$_3$GeTe$_2$ to 270 K in Fe$_4$GeTe$_2$ to $~$300 K in Fe$_5$GeTe$_2$ \cite{seo2020nearly}. The metallic nature, the high Curie temperature, and the vdW layered structure make Fe$_n$GeTe$_2$ (3$\leq n \leq$5) especially attractive to the spintronics application.

\begin{figure}[th]
\begin{center}
\includegraphics[width=1.0\textwidth]{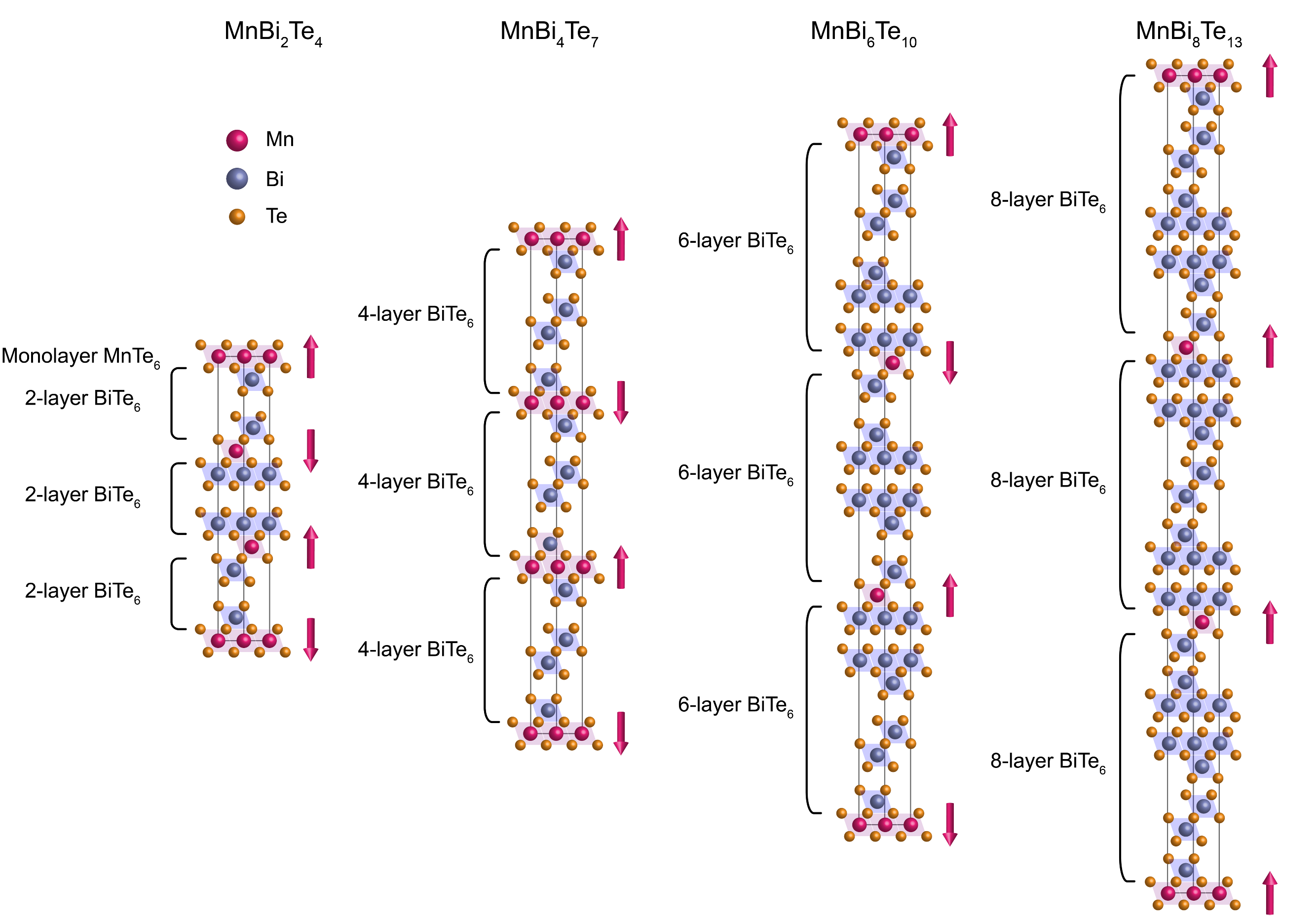}\caption{\small Summary of crystal structures of MnBi$_{2n}$Te$_{3n+1}$ with $n=1,2,3,$ and 4, respectively. The structural unit cell is labeled with ``monolayer MnTe$_6$" and ``$2n$-layer BiTe$_6$" in the sketch. The layered AFM order with out-of-plane spins is marked in the MnTe$_6$ layer. This figure is adapted from Ref. \cite{hu2020realization}.} 
\label{MBT}
\end{center}
\end{figure}

 \textbf{MnBi$_{2n}$Te$_{3n+1}$ ($n=1, 2, 3, 4$)} features a unique class of vdW magnetism because of the interplay between magnetism and topology. The crystal structure of MnBi$_{2n}$Te$_{3n+1}$ ($n=1, 2, 3, 4$) is characterized by the alternating stacking of monolayer MnTe$_6$ octahedra and $2n$-layer BiTe$_6$ octahedra, as shown in Figure \ref{MBT}g \cite{hu2020realization}. Within the MnTe$_6$ plane, it forms an FM order along the out-of-plane easy axis. As $n$ increases from 1 to 4, the separation between the Mn layers increases, and the interlayer exchange coupling switches from the interlayer AFM coupling in $n=1, 2, 3$ to the FM one in $n=4$. Because of the topological insulator nature of Bi$_2$Te$_3$, the introduction of layered magnetism in MnBi$_{2n}$Te$_{3n+1}$ gives rise to a novel modification of electronic properties \cite{hu2020realization, hu2020van}. Within this class, MnBi$_2$Te$_4$ is mostly explored in the 2D limit. Interestingly, because of the layered AFM order, the even-layer MnBi$_2$Te$_4$ has zero net magnetization whereas the odd-layer MnBi$_2$Te$_4$ gives a non-zero magnetization. It has been theoretically predicted \cite{li2019intrinsic} and experimentally shown \cite{deng2020quantum,liu2020robust,gao2021layer} that odd- and even-layer MnBi$_2$Te$_4$ are axion insulator and quantum anomalous Hall insulator, respectively. Furthermore, the theoretical prediction shows that the layered AFM topological insulator phase can be realized in a wider class of compounds MB$_2$T$_4$ with M = Ti, V, Mn, Ni, and Eu, B = Bi and Sb, and T = Te, Se, and S.  

\subsubsection{4$d$ electron-based two-dimensional vdW magnets}

The class of vdW magnets, including 4$d$ transition metals, has been much less explored than that concerning 3$d$ transition metals. Because of the increase of the orbital index and the element number, it is anticipated that the electronic correlation is weaker in 4$d$ electron-based vdW magnets as a result of the more extended 4$d$ orbitals and that the spin-orbit coupling should be stronger in 4$d$ transition metals due to their higher atomic number $Z$. The combination of these two factors can make the 4$d$ transition metal-based vdW magnets distinct from the 3$d$ ones, which is profoundly manifested in $\alpha$-RuCl$_3$, a candidate hosting the Kitaev spin model \cite{takagi2019concept}. Accumulating experimental and theoretical studies over the past years, only Nb and Ru were found to contribute to the class of 4$d$ electron-based vdW magnets. 

\textbf{4$d$ transition metal halides} vdW magnets so far include $\alpha$-RuX$_3$ (X = Cl, Br, and I) and Nb$_3$X$_8$ (X = F, Cl, Br, and I). 

For $\alpha$-RuX$_3$, it shares the same structure as 3$d$ transition metal trihalides as given in Figure \ref{material structure1}a, which is a honeycomb lattice made of RuCl$_6$ octahedra. Because of the relatively strong spin-orbit coupling, the magnetic state takes the $J_\mathrm{eff} = 1/2$ configuration: the 4$d^5$ electrons in Ru$^{3+}$ first take the low-spin configuration to partially occupy the lower $t_\mathrm{2g}$ manifold under the octahedral crystal field splitting, and then half-fill the upper energy $J_\mathrm{eff} = 1/2$ state and fully fill the lower energy $J_\mathrm{eff} = 3/2$ state after the spin-orbit coupling-induced splitting \cite{takagi2019concept} (Figure \ref{crystal field}). Such a $J_\mathrm{eff} = 1/2$ magnetic configuration, together with the honeycomb magnetic lattice, enables the bond-direction dependent Ising-type FM exchange coupling, i.e., the Kitaev spin model. However, in $\alpha$-RuX$_3$ (X = Cl, Br, and I), there are non-Kitaev terms in addition to the Kitaev term, such as the next nearest neighboring AFM coupling. $\alpha$-RuCl$_3$ and $\alpha$-RuBr$_3$ are shown to form a zigzag AFM order below the critical temperature of 6.5 K \cite{banerjee2017neutron} and 34 K \cite{imai2022zigzag}, respectively, whereas $\alpha$-RuI$_3$ remains paramagnetic down to 0.35 K \cite{ni2022honeycomb}. Despite the development of the zigzag AFM order in $\alpha$-RuCl$_3$ below 6.5 K, there is experimental evidence of Kitaev quantum spin liquid-like behavior at temperatures between 6.5 K and 60 K \cite{banerjee2017neutron,do2017majorana}, or under an in-plane magnetic field $\geq$ 7 T \cite{sears2017phase,baek2017evidence,wang2017magnetic,banerjee2018excitations,kasahara2018majorana,bruin2022robustness,czajka2021oscillations}. Further efforts have been made to exfoliate $\alpha$-RuCl$_3$ into thin flakes and examine their physical properties in the 2D limit. In 2D $\alpha$-RuCl$_3$, a new structural distortion was found, and an out-of-plane easy-axis anisotropy was reported, despite the survival of the Kitaev terms \cite{yang2023magnetic}. 

For Nb$_3$X$_8$, it realizes a trimerized kagome lattice made of Nb$_3$X$_{13}$ trimers that stack vertically following the rhombohedral ABC sequence (Figures \ref{material structure1}k and \ref{material structure1}l) \cite{haraguchi2017magnetic}. The [Nb$_3$]$^{8+}$ has seven $d$ electrons and thus yields the $S=1/2$ spin configuration. Nb$_3$Cl$_8$ and Nb$_3$Br$_8$ undergo a paramagnetic to FM phase transition at the critical temperatures of 90 K \cite{haraguchi2017magnetic} and 382 K \cite{pasco2019tunable}, and Nb$_3$I$_8$ is predicted to be FM as well \cite{conte2020layer}. In addition to the 2D vdW magnetism, Nb$_3$X$_8$ also receives much research interest in their electronic band structure, because flat and topological electronic bands are expected from the kagome lattice structure \cite{sun2022observation, regmi2022spectroscopic}. 

\subsubsection{5$d$ electron-based two-dimensional vdW magnets}

5$d$ transition metal-based vdW magnets are even less explored than 4$d$ ones, with nearly no experimental demonstrations and only a few theoretical predictions. Because of the further enhanced spin-orbit coupling in 5$d$ transition metal elements, the theoretical interest in 5$d$ electron-based vdW magnets primarily focuses on the novel electronic band topology that emerges from the interplay between magnetism and strong spin-orbit coupled electronic states. So far, only a couple of \textbf{5$d$ transition metal trihalides}, OsCl$_3$ and ReX$_3$ (X = Br, I) have been theoretically investigated. Based on DFT calculations, OsCl$_3$ is predicted to host a FM ground state with an in-plane easy axis and support the realization of quantum anomalous Hall insulating phase \cite{sheng2017monolayer}. Through DFT and self-consistent Hubbard-$U$ calculations, ReX$_3$ (X = Br, I) is expected to show robust layered AFM ground state and realizes the topologically nontrivial electronic states in even-layer films \cite{mahatara2021layer}.  

\subsubsection{$f$ electron-based two-dimensional vdW magnets}

$f$ electron-based vdW magnets form an independent class because of the distinction between $f$ and (3,4,5)$d$ electrons. Generally speaking, $f$ electrons are more localized than $d$ electrons because of the narrower $f$ orbitals, which makes the orbital overlaps between $f$ orbitals of two neighboring rare earth ions and between $f$ and $p$ orbitals of neighboring rare earth ion and the ligand anion much suppressed. Therefore, the magnetic exchange coupling between rare earth magnetic ions is typically very weak, and as a result, the magnetic onset temperatures for $f$ electron-based vdW magnets are typically expected to be low. Thus far, there have been very few $f$ electron-based vdW magnets among which Gd-based ones have been both theoretically predicted and experimentally demonstrated. GdAu$_2$ and GdAg$_2$ monolayer alloys were grown on metal substrates and verified to manifest an FM order below the critical temperatures of 19 K and 85 K, respectively \cite{ormaza2016high}. In addition, GaTe$_3$ vdW bulk was shown to be an AFM metal, and its monolayer has been mechanically isolated \cite{lei2020high}. On the theory forefront, GdI$_2$ \cite{wang2020prediction} monolayer is predicted to host high magnetic moments and a high critical temperature due to the direct exchange coupling between 5$d$ and 4$f$ orbitals of Gd$^{2+}$ ions. 

\subsubsection{Twisted two-dimensional vdW magnets}

\begin{figure}[th]
\begin{center}
\includegraphics[width=1.0\textwidth]{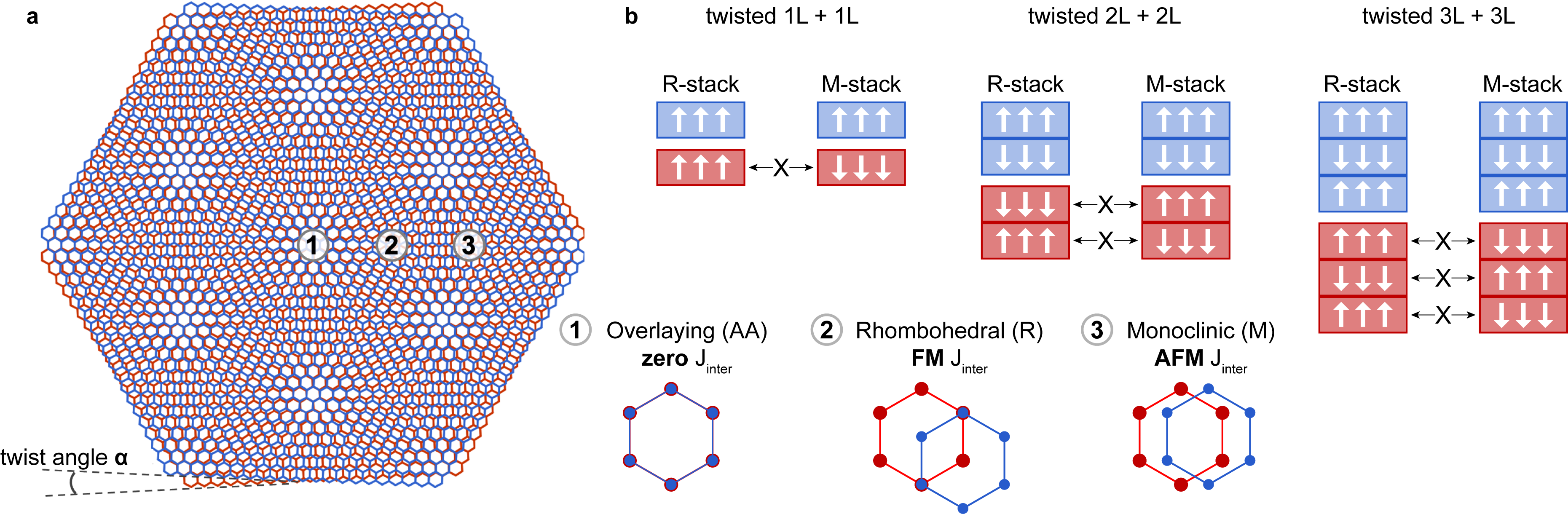}\caption{\small Moir\'e magnetism in twisted CrI$_3$ moir\'e superlattices. (a) The moir\'e superlattice formed at the interface of the two twisted CrI$_3$ flakes with twist angle $\alpha$, highlighting three stacking geometries, 1-overlaying (AA) staking with zero interlayer exchange coupling, $J_\mathrm{inter}$, 2-rhombohedral (R) stacking with FM $J_\mathrm{inter}$, and 3-monoclinic (M) stacking with AFM $J_\mathrm{inter}$; (b) The expected spin arrangements at the R and M stacking sites for twisted bilayer (1L + 1L), double bilayer (2L + 2L), and double trilayer (3L + 3L) CrI$_3$, at very small $\alpha$.} 
\label{twisted CrI3}
\end{center}
\end{figure}

Compared to the 2D magnetism in natural few-layer/monolayer vdW magnets, twisted vdW magnets form an even newer platform to realize interesting 2D spin textures that do not exist in their natural untwisted counterparts. The key theoretical concept used in twisted moir\'e magnets is the stacking-dependent interlayer exchange coupling, i.e., $\mathcal{H}_\mathrm{M}$. The total spin Hamiltonian involves those of two composing 2D magnets that are spatially homogeneous ($\mathcal{H}_1$ and $\mathcal{H}_2$) and this at the twisted interface that is spatially modulating at the moir\'e periodicity ($\mathcal{H}_\mathrm{M}$), that is $$\mathcal{H}=\mathcal{H}_1+\mathcal{H}_2+\mathcal{H}_\mathrm{M}$$. It is the competition between the homogeneous and the inhomogeneous terms to decide the magnetic ground states and the collective magnetic excitations. As the moir\'e exchange coupling can be tuned by the twist angle, this competition is tunable to realize different magnetic phases.  Fortunately, such a stacking-dependent interlayer exchange coupling has been theoretically computed \cite{sivadas2018stacking} and experimentally demonstrated \cite{li2019pressure,song2019switching} in few-layer CrI$_3$. Specifically, there are two interlayer stacking geometries, monoclinic stacking and rhombohedral stacking, which have similar elastic energy in CrI$_3$, and interestingly, distinct interlayer exchange couplings: AFM for the monoclinic and FM for the rhombohedral stacking. This observation soon triggered the theoretical investigation of twisted CrI$_3$ bilayers, whose interface includes both monoclinic and rhombohedral stackings within individual moir\'e supercells and thus features a periodically modulated interlayer exchange coupling that varies from AFM to FM at the moir\'e wavelength (Figure \ref{twisted CrI3}a).  Interesting magnetic properties are predicted to emerge due to the competition between this modulated interlayer exchange coupling between layers and the uniform Ising-type FM intralayer exchange coupling within layers. Examples include noncollinear spins \cite{hejazi2020noncollinear}, skyrmion lattices \cite{tong2018skyrmions,hejazi2021heterobilayer,akram2021skyrmions,akram2021moire,ghader2022whirling,zheng2023magnetic}, moir\'e magnons \cite{li2020moire}, topological magnons \cite{kim2022theory}, and one-dimensional magnon network \cite{wang2020stacking}.

At the experimental forefront, a few recent experiments have shown new magnetic phases in twisted CrI$_3$ moir\'e superlattices (Figure \ref{twisted CrI3}b). Twisted bilayer CrI$_3$ (twisted 1L+1L CrI$_3$) \cite{xu2022coexisting} and twisted double trilayer CrI$_3$ (twisted 3L+3L CrI$_3$) \cite{song2021direct}, where the composing monolayer and trilayer CrI$_3$ have non-zero magnetization, have been reported to show the coexisting of FM and AFM states at the rhombohedral and monoclinic stacking regions, respectively, at small twist angles. Twisted double bilayer CrI$_3$, whose composing bilayer CrI$_3$ has a zero magnetization, surprisingly shows a non-zero total magnetization over an intermediate twist angle range (i.e., 0.5$^\mathrm{o}$ -- 5$^\mathrm{o}$), whereas having zero total magnetization at very small and very large twist angles \cite{xie2022twist,cheng2023electrically,xie2023evidence}. Concurrent with the development of the nontrivial non-zero magnetization in twisted 2L + 2L CrI$_3$ is the observation of noncollinear spins that follow the same non-monotonic twist angle dependence and peak at $\sim$1.1$^\mathrm{o}$ \cite{xie2023evidence}.

In addition to CrI$_3$, CrCl$_3$ and CrBr$_3$ are known to host the similar stacking geometry dependent interlayer exchange coupling \cite{chen2019direct,si2021revealing,klein2019enhancement}, and their twisted moir\'e superlattices are predicted to show nontrivial spin textures as well \cite{xiao2021magnetization}. It is worth highlighting that, CrCl$_3$ has an easy-plane anisotropy and CrBr$_3$ hosts isotropic Heisenberg spin interactions, which are in contrast to CrI$_3$ that exhibits an easy-axis aniostropy \cite{kim2019evolution}. Due to the softness of spins in CrCl$_3$ and CrBr$_3$, it is anticipated that the moir\'e interlayer exchange coupling can make a more profound impact on the magnetic ground states and collective excitations in CrCl$_3$ and CrBr$_3$ than in CrI$_3$. 

Besides CrX$_3$ (X = Cl, Br, and I), MPS$_3$ (M = Fe, Ni, and Co) is another family of vdW magnets that have stacking-dependent magnetic orders. Here, the interlayer stacking is monoclinic \cite{joy1992magnetism}, for which there are three degenerate states with 120$^\mathrm{o}$ rotated interlayer shift vectors. Their magnetic orders are featured by the zigzag AFM states which have FM spin alignment within the zigzag chains and AFM spin coupling between the two adjacent chains \cite{joy1992magnetism}. Interestingly, the zigzag chain direction is locked to the underlying monoclinic stacking. The zigzag spin chain is aligned in the same direction as the shift vector between layers \cite{joy1992magnetism}. Therefore, in twisted MPS$_3$ (M = Fe, Ni, and Co) moir\'e superlattices, there is expected to be all three degenerate monoclinic stacking states within individual moir\'e supercells that favor 120$^\mathrm{o}$ rotated zigzag spin chains, respectively. The consequence of such a moir\'e interlayer coupling on the magnetic orders and excitations remains to be explored both theoretically and experimentally.

\subsection{Twisted moir\'e superlattices with flat bands}

Strongly correlated electron systems are a fertile playground for realizing novel magnetism, as it has been well demonstrated in many 3D correlated materials including transition metal complex oxides (e.g., high-$T_\mathrm{c}$ cuprates \cite{sachdev2003colloquium}, nickelates \cite{fowlie2022intrinsic}, maganites \cite{kimura2000layered}, iridates \cite{witczak2014correlated,rau2016spin}), heavy fermion systems \cite{coleman2006heavy}, etc. Very recently, twisted moir\'e superlattices of non-magnetic composing layers have been found capable of entering into the strongly correlated regime in the 2D limit, by the realization of ``flat bands" whose kinetic energy is quenched by the moir\'e potential \cite{bistritzer2011moire}. The significance of ``flat bands" to strongly correlated physics is two-fold: first, flat bands have a large density of states that promotes strong interactions; second, the kinetic energy of flat bands is much smaller than the Coulomb potential, which sets the systems well into the strongly correlated regime. Twisted graphene \cite{balents2020superconductivity,andrei2021marvels} and twisted TMDC moir\'e superlattices \cite{mak2022semiconductor,kennes2021moire} are two families of moir\'e electronic materials that have been extensively and successfully studied so far. Numerous novel electronic phases have been discovered in either or both material families, including Mott insulator \cite{cao2018correlated,yankowitz2019tuning}, Wigner crystal \cite{regan2020mott}, charge orders \cite{tang2020simulation,xu2020correlated,jin2021stripe}, nematic orders \cite{kerelsky2019maximized,jiang2019charge,cao2021nematicity}, unconventional superconductivity \cite{cao2018unconventional,lu2019superconductors}, Kondo physics \cite{zhao2023gate}, etc. Moreover, interesting magnetic phases have developed or been predicted to develop out of the strong correlations in both systems, for example, AFM \cite{tang2020simulation,li2021continuous}, orbital magnetism \cite{lu2019superconductors}, correlated Chern insulator \cite{sharpe2019emergent,serlin2020intrinsic,stepanov2020untying,nuckolls2020strongly,saito2021hofstadter,das2021symmetry,choi2021correlation,park2021flavour,stepanov2021competing,pierce2021unconventional,chen2020tunable,polshyn2020electrical}, fractional Chern insulator \cite{xie2021fractional,cai2023signatures,zeng2023thermodynamic}, topological magnetism \cite{wang2023topological}, spin liquid \cite{li2021continuous}, etc. It is worth highlighting that the hexagonal and honeycomb moir\'e superlattices make the correlation-driven magnetism in twisted graphene and twisted TMDCs host unique characteristics, as compared to the tetragonal and square lattices in many 3D correlated electron systems.  

\begin{figure}[th]
\begin{center}
\includegraphics[width=1.0\textwidth]{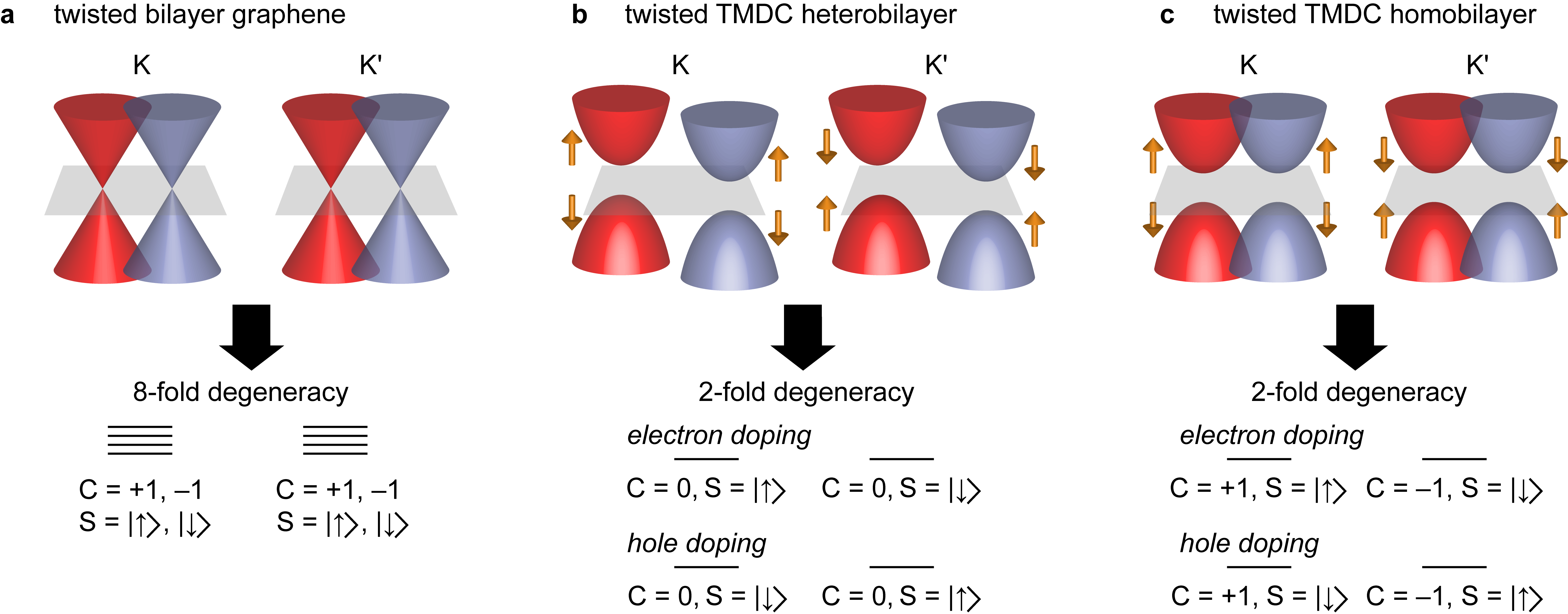}\caption{\small Sketch of electronic bands of twisted moir\'e superlattices. Electronic band alignments at K and K$'$ valleys at small twist angles, i.e., R-stacked geometry (for the TMDC cases), the resulting flat band degeneracy, Chern number (C) and spin polarization (S) for (a) twisted bilayer graphene in the chiral limit, (b) twisted TMDC heterobilayers of the type-II band alignment, and (c) twisted TMDC homobilayers. Note that the Chern numbers in (b) and (c) are labeled for the most common cases but can be different if the material parameters are fine-tuned.} 
\label{twisted moire}
\end{center}
\end{figure}

\subsubsection{Twisted graphene moir\'e superlattices}

Twisted bilayer graphene has 2-fold valley degeneracy, 2-fold valence-conduction degeneracy, and 2-fold spin degeneracy, yielding an 8-fold degeneracy of the electronic states in the chiral limit (Figure \ref{twisted moire}a)\cite{andrei2020graphene,balents2020superconductivity,andrei2021marvels}. This high level of degeneracy, together with the linear electronic band dispersion, limits the twist angle for realizing strong electron correlations over a very narrow range around the first "magic" angle, i.e., $1.10^\mathrm{o} \pm 0.05^\mathrm{o}$ \cite{andrei2020graphene,balents2020superconductivity,andrei2021marvels}, in twisted graphene moir\'e superlattices. The band-filling factor of the flat bands of twisted bilayer graphene, $\nu$, defines the number of carriers per moir\'e supercell, ranging from $\nu = -4$ when all the bands are empty, to $\nu = 0$ when half of the bands are filled (i.e., charge neutrality point that separates the valence and conduction bands), to $\nu = 4$ when all the bands are filled. It was found experimentally that twisted bilayer graphene hosts high resistance ($R_{xx}$) states at one-quarter ($\nu = \pm1$), half ($\nu = \pm2$) and three-quarters ($\nu = \pm3$) filling of the conduction and valence bands \cite{cao2018correlated,yankowitz2019tuning,lu2019superconductors,sharpe2019emergent}. 

Theoretical calculations predicted that the electronic interactions can lift up the valley and the spin degeneracies and drive the fraction-filling correlated insulating states into magnetic ones \cite{kang2019strong,xie2020nature,ochi2018possible,seo2019ferromagnetic}. Experimentally, twisted bilayer graphene aligned with hexagonal boron nitride (hBN) was shown to have a giant anomalous Hall effect in the initial experiment \cite{sharpe2019emergent} and then proven to host a precise quantization of the Hall resistivity in the later-on experiment \cite{serlin2020intrinsic}, which conclusively demonstrates the realization of a Chern insulating state with Chern number $C = 1$. The robustness of this $C = 1$ Chern insulating phase critically depends on breaking the out-of-plane two-fold rotational symmetry in twisted bilayer graphene through the interaction with the aligned hBN substrates. Further experiments revealed more Chern insulating states in ``clean" twisted bilayer graphene without requiring its registry with the hBN substrate, including $C = -1$ for $\nu = -3$ and $C = -2$ for $\nu = -2$ \cite{stepanov2020untying}, and in local probe studies of twisted bilayer graphene, including $C = \pm3$ for $\nu = \pm1$ \cite{nuckolls2020strongly}.
Moreover, Chern insulators were also found in ABC-trilayer graphene/hBN moir\'e superlattices \cite{chen2020tunable} and twisted monolayer-bilayer heterostructures \cite{polshyn2020electrical}. It is worth noting that these integer Chern insulating states ($C = \pm1, \pm2, \pm3$) happen in the flat bands and, therefore, inherent strong electron correlations from the flat bands. 

The discoveries of flat Chern bands in moir\'e graphene systems further motivated the search for fractional Chern insulating states, ideally in the zero magnetic field limit, as also suggested by earlier theoretical works \cite{sheng2011fractional,neupert2011fractional,regnault2011fractional,tang2011high,qi2011generic}. To describe the fractional Chern insulators, a pair of quantum numbers ($t,s$) satisfying the Diophantine equation $\nu = t\phi/\phi_\mathrm{0}+s$ are introduced, where $\nu$ is the filling factor at which the incompressible peaks occur in the inverse compressibility measurements, $\phi$ is the magnetic flux per moir\'e supercell, and $\phi_\mathrm{0}$ is the magnetic flux quantum. In twisted bilayer graphene, states with fractional values of ($t,s$) were observed above the magnetic field of 5 T and persist till 11 T, including $(t,s) = (2/3, 10/3), (1/3,11/3)$ within $3<\nu<4$, corresponding to $\nu_\mathrm{c} = 1/3, 2/3$ fillings, respectively, on the $C = -1$ band, $\nu_\mathrm{c}$ being the filling factor for the Chern band \cite{xie2021fractional}. More fractional Chern insulating states were found at higher magnetic fields in the same sample \cite{xie2021fractional}. We note that although the fractional Chern insulating states in twisted bilayer graphene were found by applying an external magnetic field, the role of this magnetic field is to slightly modify the Berry curvature of the intrinsic Chern bands, in contrast to the fractional Chern insulating states found in the heterostructure of AB stacked bilayer graphene aligned with hBN where the Chern bands are introduced by the external magnetic field \cite{spanton2018observation}.

\subsubsection{Twisted TMDC moir\'e superlattices}

Twisted TMDC moir\'e superlattices contain two families, the TMDC heterostructures which are made of different TMDC compounds and the TMDC homostructures which are composed of the same TMDC layers. Here, the TMDC compounds are focused on the H-phase TMDC semiconductors, mainly including WS$_2$, WSe$_2$, MoS$_2$, MoSe$_2$, and MoTe$_2$.

For \textbf{twisted TMDC heterostructures}, unlike twisted bilayer graphene with valley, spin, and valence-conduction band degeneracies, the valence-conduction band and spin degeneracies are lifted due to the broken inversion symmetry and the strong spin-orbit coupling, respectively, which leaves only the 2-fold degeneracy that comes from the valley degeneracy (Figure \ref{twisted moire}b) \cite{regan2020mott,tang2020simulation}. This low degeneracy of electronic states, together with the quadratic dispersion of electronic bands, endorses the robustness of the strongly correlated regime in twisted TMDC heterobilayers, i.e., insensitive to twist angles up to a couple of degrees. The band-filling factor, $\nu$, is defined by the number of electron/hole per moir\'e superlattice cell; positive (negative) values for electron (hole) doping. Experimental results in aligned WS$_2$/WSe$_2$ heterobilayers show strong insulating states at filling factors of $|\nu|=1, 2, 1/3, 2/3$, corresponding to Mott insulator state \cite{tang2020simulation,xu2020correlated,huang2021correlated}, moir\'e band insulator state \cite{regan2020mott,tang2020simulation,xu2020correlated,huang2021correlated}, and two generalized Wigner crystal states\cite{regan2020mott,xu2020correlated}, respectively. Moreover, weaker insulating states are observed at other fractional filling factors, for example, on the electron doping side, $\nu = 1/2, 2/5, 1/4, 1/7$ and their conjugated states at $1-\nu$ \cite{xu2020correlated}. Due to the presence of unpaired electrons within the moir\'e supercells, magnetic interactions on a triangular lattice are anticipated in these partially filled insulating states. It was shown in $\nu = 1$ of WS$_2$/WSe$_2$ heterobilayers that a negative Weiss constant of $\theta$ = -0.6 K $\pm$ 0.2 K was found, demonstrating an AFM exchange coupling for the unpaired spins of this Mott insulating state at $\nu = 1$. \cite{tang2020simulation}. Furthermore, the filling factor dependence of Weiss constant shows $\theta < 0$ for $\nu \le 1.2$ which is consistent with the AFM coupling, and $\theta \ge 0$ for $\nu > 1.2$ that suggests an FM coupling, where $\nu = 1.2$ is a critical filling factor for a quantum phase transition from AFM to weak FM \cite{tang2020simulation}. 

In addition, the evolution of magnetic properties across a continuous metal-insulator transition (MIT) has also been investigated in the R-stacked MoTe$_2$/WSe$_2$ heterobilayer at $\nu = 1$ \cite{li2021continuous}. An out-of-plane displacement field $E$ is applied to tune the bandwidth $W$ of the moir\'e electronic bands and thereby achieve the MIT when the bandwidth $W$ overcomes the onsite Coulomb repulsion $U$. A critical displacement field $E_\mathrm{c}$ was identified at which the activation energy gap vanishes and across which the scaled resistance curve collapses into two branches. A negative Weiss constant $\theta = -30 \sim - 40$ K was extracted at all displacement fields, which is consistent with an AFM superexchange coupling for the localized moment in a Mott insulator with a magnetic interaction energy scale of ~3meV on both sides of MIT. However, till the base temperature of 1.6 K (i.e., $~5\%$ of the Weiss constant), no magnetic order was discovered, suggesting that the magnetic ground state is either a spin liquid or a 120$^\mathrm{o}$ N\'eel order below 1.6 K. 

The transition into a quantum anomalous Hall insulator from a Mott insulator was realized in the H-stacked MoTe$_2$/WSe$_2$ heterobilayers at $\nu = 1$ \cite{li2021quantum}. Between the R- and H-stacked cases, important distinctions include different high-symmetry stacking sites and interlayer spin alignments \cite{li2021continuous,li2021quantum}, the latter of which is significant in determining the interlayer tunneling and the moir\'e band structure. By tuning the displacement field, a new state with weak longitudinal resistance ($R_\mathrm{xx}$) but large Hall resistance ($R_\mathrm{xy}$) emerges, suggesting its broken time-reversal symmetry. The Hall resistance was further confirmed to show the quantized value of $h/e^2$ for temperatures up to 2.5 K and remain finite (but not quantized) till 5 K -- 6 K, corresponding to the quantum anomalous Hall insulator state and the FM insulator state, respectively. 

For \textbf{twisted TMDC homostructures}, there is only the 2-fold valley degeneracy, without the valence-conduction degeneracy and the spin degeneracy (Figure \ref{twisted moire}c). Theoretically, twisted TMDC homostructures (e.g., R-stacked MoTe$_2$, R-stacked WSe$_2$) are predicted to realize topological flat bands with opposite Chern numbers in the two spin/valley sectors \cite{wu2019topological,devakul2021magic} which are key ingredients for realizing integer Chern insulators and fractional Chern insulators \cite{li2021spontaneous,crepel2023anomalous,wang2023fractional}. Very recently, experimental evidence of integer and fractional Chern insulating states have been realized in R-stacked MoTe$_2$ homobilayers \cite{cai2023signatures,zeng2023thermodynamic}. Note that the moir\'e superlattice forms a honeycomb lattice with two degenerate moir\'e potential sites, Mo-Te and Te-Mo stacking sites. An out-of-plane displacement field $E$ is applied to tune the moir\'e potential depth difference between the Mo-Te and Te-Mo stacking sites, and can turn the honeycomb moir\'e superlattice, where the carriers reside in both layers, into the triangular moir\'e superlattice where the carriers migrate into the top layer. The filling factor is defined to be the ratio between the charge carrier density over the moir\'e superlattice density, i.e., $\nu = n/n_\mathrm{M}$, and hence $\nu = -1$ corresponds to half-band filling of holes. It was consistently demonstrated in both works \cite{cai2023signatures,zeng2023thermodynamic} that $\nu = -1$ and $\nu = -2/3$ states both exhibit broken time-reversal symmetry and show linear dispersion down to zero magnetic field in the Landau fan diagram, corresponding to Chern numbers of $C = -1$ and $C = -2/3$, respectively. Furthermore, the broken time-reversal symmetry for $\nu = -1$ and $\nu = -2/3$ states happens at ~14 K and ~4.5 K, respectively \cite{cai2023signatures,zeng2023thermodynamic}, and the insulating gaps for them are ~6 meV and 0.6 meV, respectively \cite{zeng2023thermodynamic}. An additional fractional Chern insulating state $\nu = -3/5$ was observed by Cai \textit{et al} more clearly above 1T and persists down to zero magnetic field at a selected spot with a twist angle of 3.57$^\mathrm{o}$ \cite{cai2023signatures}.

\section{Fabrications of two-dimensional vdW magnets}

One of the biggest challenges in scaling up applications of 2D vdW magnets is to achieve the reliable, massive production of the samples. So far, at the research forefront, most 2D vdW magnets have been achieved by the mechanical exfoliation technique for atomically thin layers and the tear-and-stack technique for moir\'e superlattice structures. Some of the 2D vdW magnetic atomic crystals have been grown in large sizes using molecular beam epitaxy (MBE) growth and chemical vapor deposition (CVD) growth techniques. Table \ref{Fabrication Techniques} summarizes the four fabrication techniques for 2D vdW magnets.

\begin{table}[th]
    \center
    \begin{tabular}{|c||*{4}{c|}}\hline
    \backslashbox[20mm]{Prop.}{Fab.}&\makebox[3em]{Mechanical Exfoliation}&\makebox[3em]{Tear-and-Stack}&\makebox[3em]{MBE Growth}&\makebox[3em]{CVD Growth}\\\hline\hline
    Substrate Selection & flexible & flexible & in-situ preparations & flexible \\
    & (e.g., SiO$_2$/Si) & (e.g., SiO$_2$/Si) & (e.g., hBN) & (e.g., SiO$_2$/Si) \\\hline
     & & twisted CrI$_3$ & CrCl$_3$, CrBr$_3$ & TMCs\\
    Samples Fabricated & almost all vdW magnets & twisted graphene & CrTe$_2$, VSe$_2$ & CrX$_3$ \\
    &  & twisted TMDCs & Fe$_3$GeTe$_2$ & MPX$_3$ \\\hline
    Sample/Grain Size & 10 -- 20 $\mu$m & 5 -- 10 $\mu$m & sub-$\mu$m & 100 $\mu$m \\\hline
    Thickness & controllable & controllable & less controllable & less controllable \\\hline
    Crystalline Quality & low defect density & moir\'e variation & defective & defective \\\hline
    \end{tabular}
    \caption{\small Summary of the four major techniques for fabricating 2D vdW magnets and the five selected physical parameters for the fabricated 2D vdW magnetic films. }
    \label{Fabrication Techniques}
\end{table}

\subsection{Mechanical exfoliation for atomic crystals}
\begin{figure}[th]
\begin{center}
\includegraphics[width=1.0\textwidth]{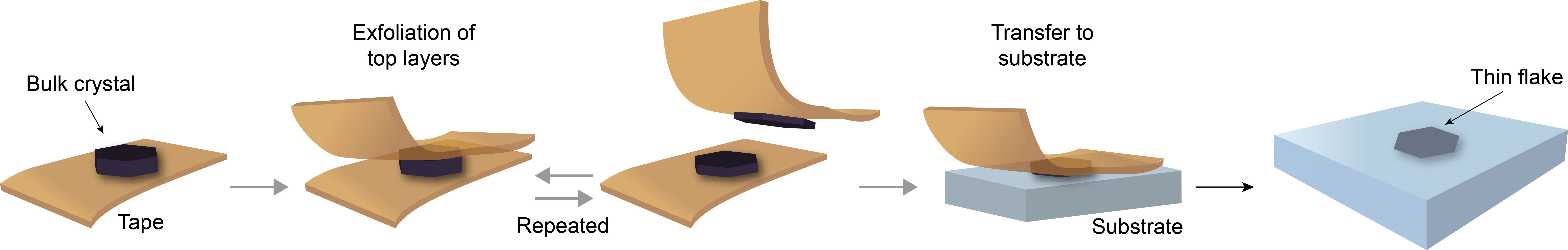}\caption{\small Schematic illustration to show the mechanical exfoliation technique for isolating $\mu$m-size 2D flakes out of their vdW bulk counterpart. This procedure starts with placing a bulk crystal on an adhesive tape, and then continues with peeling the crystal multiple times with a second adhesive tape, and finally stamping either of the adhesive tapes with a thin film of crystals onto a substrate. The substrate with thin flakes will then be examined under an optical microscope to locate the flakes and roughly estimate their thickness through the optical contrast.}
\label{exfoliation}
\end{center}
\end{figure}

Mechanical exfoliation with tapes uses the adhesion force of tapes to break the weak vdW binding between layers in vdW materials. Its success in isolating monolayer graphene in 2004 opened the era of creating nearly free-standing atomic crystals out of their vdW bulk counterpart \cite{novoselov2004electric}. A standard mechanical exfoliation process uses the adhesive tape to peel the vdW crystals into thin flakes and then stamps the flakes onto a target substrate, typically SiO$_2$/Si wafer, as illustrated in Figure \ref{exfoliation}. Over the past couple of decades, constant efforts have been made to improve both the yield and the size of 2D atomic crystals using this technique. Earlier stage efforts include oxygen plasma treatment to clean up the SiO$_2$/Si substrates, heat treatment to remove trapped gas between layers and substrates, application of viscoelastic stamps to increase the contact between stamp and layers, testing different substrates to maximize the coupling between substrates and layers, etc. \cite{yi2015review} Such efforts managed to improve the 2D atomic crystal flake size of graphene and TMDCs up to the $\sim$100 $\mu$m scale in the lateral dimension \cite{yi2015review}. More recently, to further increase flake sizes up to the mm scale, efforts have been made using evaporated Au films as exfoliation tapes \cite{desai2016gold,velicky2018mechanism}, controlling crack propagation for layer-resolved splitting \cite{shim2018controlled}, exfoliating in a layer-by-layer manner \cite{liu2020disassembling}, and engineering layers for layer number selections \cite{moon2020layer}. It is worth noting that most of these techniques have been only applied to graphene and TMDCs, but not yet to 2D vdW magnets. In addition, all these techniques for large-size flakes involve the process of first depositing metal layers, e.g., Au, Ni, and then dissolving these layers, which is possible to modify the physical properties of the isolated 2D materials. Further efforts are required to demonstrate their applicability to 2D magnets and their capability of maintaining the intrinsic properties of 2D magnets.

\subsection{Tear-and-stack technique for moir\'e superlattices}
\begin{figure}[th]
\begin{center}
\includegraphics[width=0.9\textwidth]{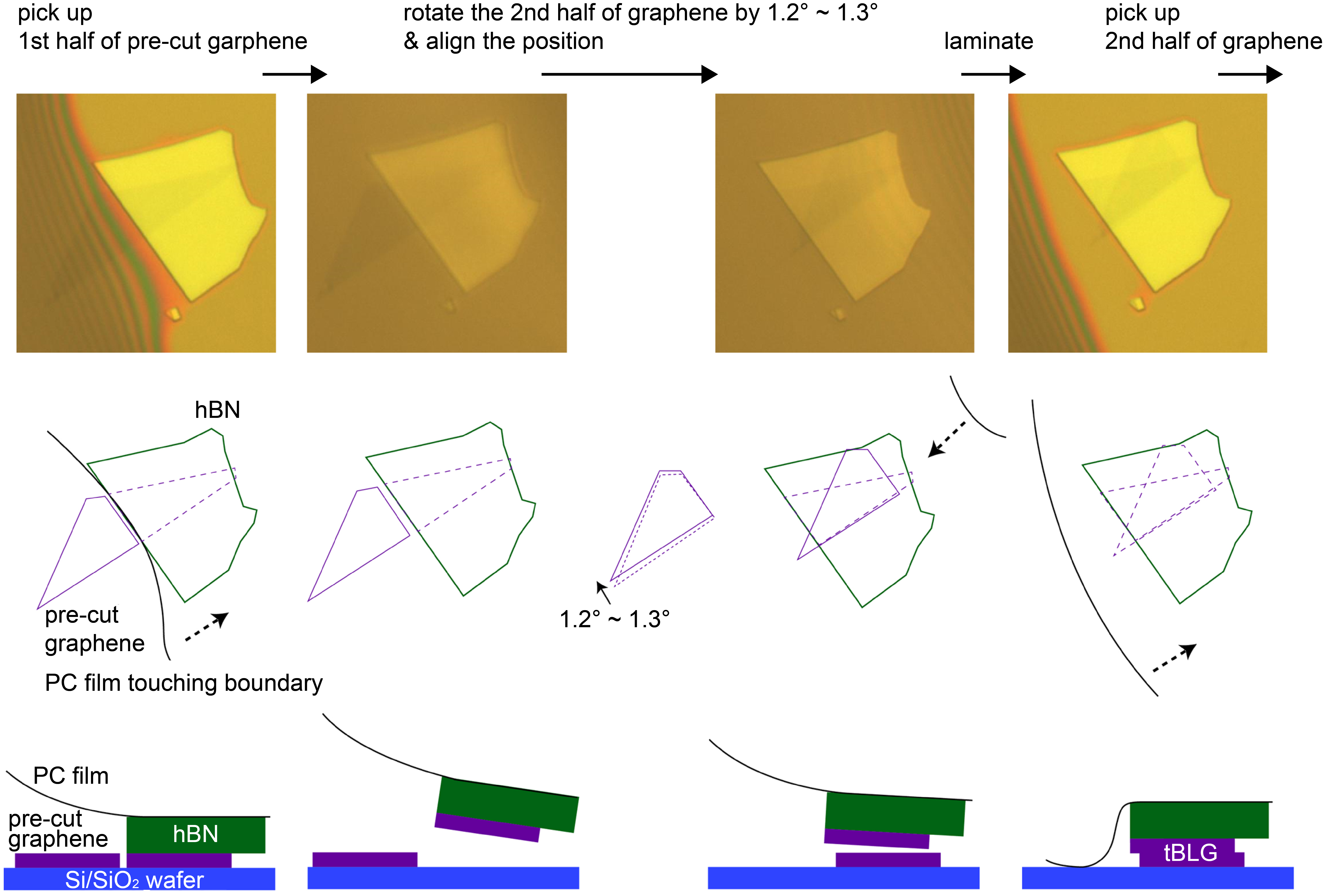}\caption{\small Optical images and schematic illustrations to show the procedure of stacking a homostructure. This procedure starts with picking up a hBN flake of proper thickness by a PC film, follows by using the PC/hBN stack to pick up the first half of a targeted 2D material (\textit{e.g.}, pre-cut graphene in this example), then rotating the second half of the 2D material by a small twist angle (\textit{e.g.}, 1.2$^\mathrm{o}$ $\sim$ 1.3$^\mathrm{o}$ here), and finally dropping the PC/hBN/2D material stack onto the rotated second half of the 2D sample on the substrate to form the twisted homobilayer structure with the hBN encapsulation (\textit{e.g.}, twisted bilayer graphene (tBLG) here). Figure is adapted from Ref. \cite{saito2020independent} }
\label{stacking}
\end{center}
\end{figure}

Moir\'e superlattices are made of vertical stacking of two mismatched atomic crystals, including lattice constant mismatch and angular alignment mismatch, with clean interfaces between the composing layers. To achieve this, a tear-and-stack technique was developed for moir\'e superlattices of homostructures \cite{kim2016van}, and a stacking process without tearing was used for those of heterostructures, both of which were built upon a deterministic and clean-interface transfer procedure that was invented for achieving ultraclean and ultraflat graphene on the hBN flakes \cite{dean2010boron}. A representative stacking procedure is shown in Figure \ref{stacking}, using the example of fabricating a twisted bilayer graphene with pre-cut graphene flakes \cite{saito2020independent}. Over the years, the deterministic placement procedure has been improved from using the PMMA as a carrier layer in the initial generation \cite{dean2010boron}, to applying Elvacite as the sacrificial layer \cite{zomer2011transfer,hunt2013massive}, to using PDMS for the dry-transfer process \cite{castellanos2014deterministic}, and then to using vdW materials (e.g. hBN) as the pick-up layer \cite{wang2013one}. The accuracy of the lateral alignment is achieved by using two sets of XYZ micrometers for the target substrate and the pick-up layer, and the precision of the rotation angle is obtained by having a rotational stage on the target substrate stage. The key challenges of the moir\'e superlattice fabrication lie in the cleanness of the interface and the homogeneity of the angular alignment. The interfacial cleanness can potentially be addressed by the assembly of heterostructure and homostructures in an inert environment or even inside the vacuum, whereas the control over twist angle homogeneity remains as an open question in the field \cite{frisenda2018recent,lau2022reproducibility}. To scale up the production of moir\'e superlattices, there have been efforts to adopt a machine learning-based algorithm to automate the exfoliation of 2D atomic crystals \cite{masubuchi2019classifying,saito2019deep,han2020deep,zichi2023physically} and implementing robotic arms to perform the stacking of heterostructures and homostructures \cite{masubuchi2018autonomous,mannix2022robotic}. We note that, up to now, the scaling-up methods discussed above have not been applied to fabricating the moir\'e superlattices with the flat bands and the twisted magnetic moir\'e superlattices.

\subsection{Molecular beam epitaxy growth}
Molecular beam epitaxy (MBE) growth is a powerful technique to achieve wafer-scale, high-quality films \cite{joyce1985molecular, henini2012molecular}, for example, complex oxides\cite{schlom2008thin,nunn2021review}, Group III nitrides \cite{zhao2017recent}, topological insulators \cite{chen2011molecular}, and many on. Recently, MBE has been applied to grow large-scale graphene films \cite{zuo2015situ}, TMDC films \cite{vishwanath2018challenges}, and some vdW magnets \cite{huang2021two}. So far, there are only a handful of 2D vdW magnets grown by MBE, such as CrCl$_3$ \cite{bedoya2021intrinsic}, CrBr$_3$ \cite{kezilebieke2021electronic,chen2019direct}, CrTe$_2$ \cite{zhang2021room}, Fe$_3$GeTe$_2$ \cite{liu2017wafer}, VSe$_2$ \cite{bonilla2018strong}, etc. However, in the current stage, the MBE grown 2D vdW materials are typically polycrystalline with an angular spread up to 1$^\mathrm{o}$, often have uncontrolled layer numbers ranging from isolated islands to multi-layers in a single sample, and host relatively large density of atomic defects with unknown characteristics. For this newly emerging topic, a few challenges have been identified. It is of top priority to select and prepare proper substrates for the MBE growth of 2D vdW materials, including and beyond 2D magnets. As of now, the popular growth substrate is hBN as it has a triangular lattice similar to graphene, TMDCs, and many honeycomb magnets, but unfortunately, 2D vdW materials grown on hBN are subject to the multigrain structures with crystalline misorientations up to 1$^\mathrm{o}$ and grain size in the order of sub-$\mu$m. Furthermore, it is important to understand the MBE growth kinetics and to achieve layer-by-layer growth for the vdW materials whose binding to the substrate is expected to be very weak. Currently, many MBE grown vdW materials have nonuniform thicknesses, i.e., the second, third, or even thicker layers start to grow before the growth of the first layer finishes. Lastly, it is also required to improve the quantification of the defect types and concentrations in the MBE grown vdW materials. By doing so, we can identify the origin and also propose solutions to reduce these defects.

\subsection{Chemical vapor deposition growth}
Chemical vapor deposition (CVD) growth is an alternative technique to MBE for getting large-size films, and has been widely applied in the growth of graphene \cite{li2009large,kim2009large,zhang2013review} and TMDCs \cite{lee2012synthesis,wong2016recent}. As compared to MBE grown graphene and TMDCs, CVD grown ones typically have much larger grain sizes (in the order of 100 $\mu$m) although they are still polycrystalline. The most accessible CVD technique is thermal CVD which mostly uses powder as the precursor and can work well for simple TMDCs, mainly transition metal sulfides. However, thermal CVD lacks suitable precursors for complex 2D materials (e.g., Fe$_3$GeTe$_2$) and further misses precision controls over precursor supply for deterministic coverage and thickness. To overcome the limitations of thermal CVD, metal-organic CVD (MOCVD) was developed for 2D material growth, in which the precursors are high vapor pressure solids, liquids, or gas and are individually controlled with precision. Up to now, MOCVD has enabled wafer-scale growth of MoS$_2$, MoSe$_2$, WS$_2$, and WSe$_2$ with sample qualities comparable to those grown by thermal CVD \cite{lin2018realizing,kang2015high}, and is considered as a promising route to produce large-size, wafer-scale, layer-controlled 2D vdW magnets \cite{och2021synthesis}. Very recently, CVD techniques have been applied to the growth of 2D vdW magnetic films, for example, mostly chalcogen-based vdW magnets, thanks to the success of CVD growth of TMDCs \cite{reale2016bulk,jiang2021synthesis}, some transition metal halides \cite{gronke2018chemical,gronke2019chromium}, and some transition metal ternary compounds \cite{cheng2020layered,shifa2018high}.

\section{Scientific topics investigated in two-dimensional magnetism}

Despite the early stage of 2D magnetism research in atomic and moir\'e crystals, an increasing number of scientific topics have developed in this emerging field. These include classic topics that have been extensively explored in 3D magnets, such as resolving the magnetic ground states,
probing the spin collective excitations, investigating the spin and other DoFs coupling, etc., and new topics that are more specific for 2D magnets, such as exploring efficient manners to control 2D magnetism, realizing new spintronic implementations, etc. In this chapter, we will provide a survey of these scientific topics that are popularly studied in 2D magnets.

\subsection{Identifying magnetic ground states of two-dimensional magnets}

\begin{wrapfigure}[18]{r}{0.33\textwidth}
\vspace{0pt} 
\includegraphics[width=0.3\textwidth]{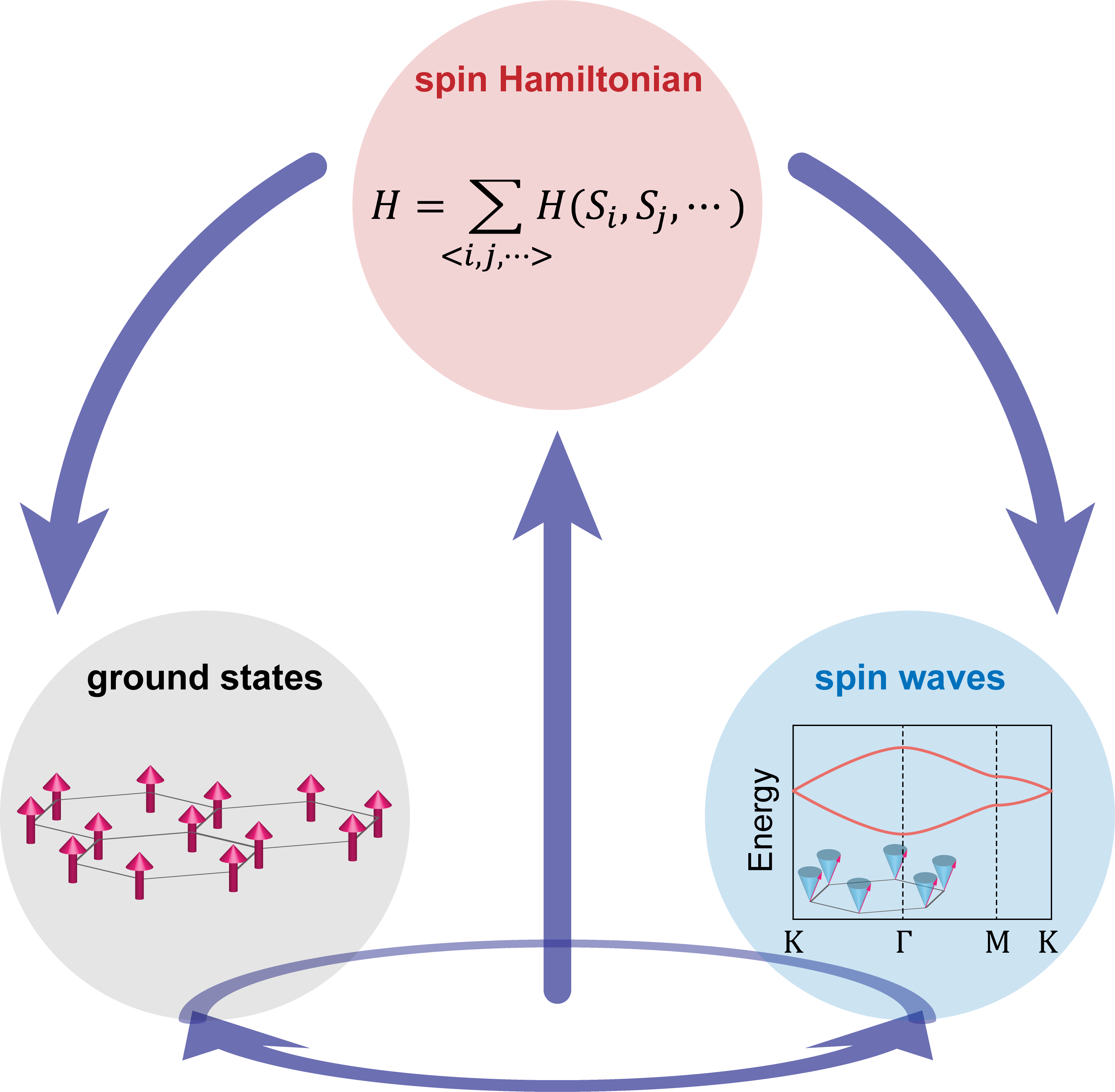}\caption{\small Sketch illustrating the relationship amongst the spin Hamiltonian, the magnetic ground state, and the spin wave excitations.}%
\label{relationship}
\end{wrapfigure}

To comprehensively describe a magnetic phase of matter, one needs to know both the magnetic ground state that depicts how the magnetic moments arrange throughout the atomic lattice of a material and the magnetic excitations that describe how the magnetic moments cooperatively
deviate from their arrangement in the ground state \cite{ashcroft2022solid}. The former is a static spin pattern with the lowest possible magnetic energy of the investigated material; the latter is a dynamic spin wave with the momentum-energy dispersion unique to the material; and the combination of both then provides the full spin Hamiltonian of the material. Conversely, with a known spin Hamiltonian, the magnetic ground state and the magnetic excitations can both be computed. Moreover, the magnetic ground state matters for the magnetic excitations, and the magnetic excitations encode
information about the magnetic ground states. The relationship among the magnetic ground state, magnetic excitations, and spin Hamiltonian is summarized in Figure \ref{relationship}.

When a material transits from a paramagnetic phase to a magnetically ordered phase, the spins at individual magnetic sites change from fluctuating randomly among all orientations to collectively pointing at selected directions. The varieties of magnetic ground states are traditionally classified by their magnetic point and space groups, 122 and 1651 in total, respectively \cite{lifshitz2004magnetic}. We note the difference in symmetry elements between point and space groups being with and without allowing for a translation after the point symmetry operation, respectively. For example, for a square lattice AFM with opposite spins between the nearest neighboring sites, it breaks the time-reversal symmetry from a space group perspective but preserves the time-reversal symmetry from a point group perspective because a time-reversal operation followed by a translation of one lattice constant along either of the two crystalline axes can recover the magnetic order before the operations. If the probe technique (e.g., X-ray diffraction, neutron scattering) is sensitive to the translational operation (typically at the length scale of nm), one should choose to use space groups. Otherwise (e.g., optical spectroscopy), one can safely use point group to describe the investigated magnetic state. Here, we first categorize the magnetism into two big families: collinear spin magnetism, where all the spins are aligned parallel or antiparallel along a single direction, and noncollinear spin magnetism, where the spins are either dynamically fluctuating or spatially modulating along more than one direction.

\subsubsection{Collinear spins in two-dimensional magnets}
\begin{figure}[th]
\begin{center}
\includegraphics[width=1.0\textwidth]{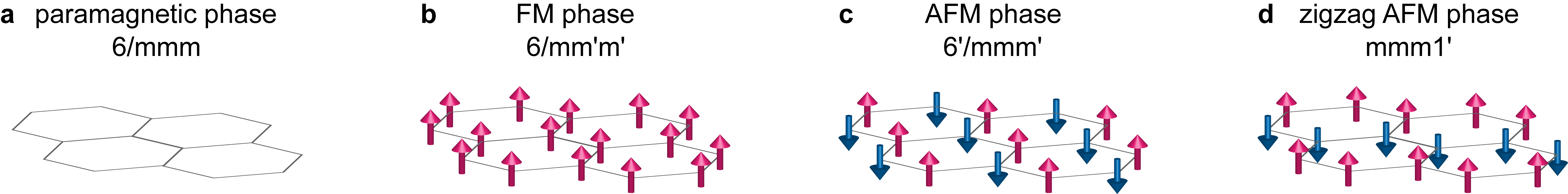}\caption{\small Examples for honeycomb magnets to illustrate broken symmetries in the magnetic ground states. (a) Paramagnetic phase with a point group 6/mmm; (b) FM phase with all spins pointing out-of-plane with a point group 6/mm$’$m$’$, e.g., magnetic sites for CrI$_3$ monolayer; (c) N\'eel-type AFM phase with spins aligning out-of-plane ferromagnetically/antiferromagnetically within/between the two sublattice(s) with a point group 6$’$/mmm$’$, e.g., magnetic sites for MnPS$_3$ monolayer; (d) zigzag-type AFM phase with spins aligning in the out-of-plane direction ferromagnetically/antiferromagnetically within/between the two sets of zigzag chains with a point group mmm1$’$, e.g., magnetic sites for FePS$_3$ monolayer.} 
\label{honeycomb mag}
\end{center}
\end{figure}

Collinear spin magnetism typically belongs to long-range magnetic orders, for which all the spins uniformly align along one single direction, being parallel for FM (or ferrimagnetism, FiM) and antiparallel for AFM. The selected spin orientation breaks the full rotational symmetry for spins in the paramagnetic states and often leads to the breaking of underlying crystalline lattice symmetries. Below, we use collinear honeycomb magnets with an out-of-plane easy-axis as examples to illustrate the variety of broken symmetries and their significance in distinguishing different magnetic phases. In Figure \ref{honeycomb mag}, we show three types of collinear magnets on a honeycomb lattice (Figure \ref{honeycomb mag}a with a structural point group 6/mmm), an FM state with all spins aligned in the same out-of-plane direction (Figure \ref{honeycomb mag}b with a magnetic point group 6/mm$’$m$’$, e.g., monolayer CrI$_3$ \cite{huang2017layer}), an AFM state with spins on the two honeycomb sublattices pointing in opposite out-of-plane directions (Figure \ref{honeycomb mag}c with a magnetic point group 6$’$/mmm$’$, e.g., monolayer MnPS$_3$ \cite{long2020persistence}), and an AFM state with spins along the same direction within the zigzag chains and along opposite out-of-plane directions between neighboring zigzag chains (Figure \ref{honeycomb mag}d with a magnetic point group mmm1$’$, e.g., monolayer FePS$_3$ \cite{lee2016ising}). Concepts illustrated in the following can be generalized to address cases of in-plane collinear spins on honeycomb magnets or collinear spins on other classes of lattices.

\textbf{Broken time-reversal symmetry} As pointed out above, from the space group perspective, all collinear magnets break time-reversal symmetry, whereas, from the point group perspective, it depends on whether a translation operation after the time-reversal operation can recover the spin pattern to the one prior to these operations. Both the FM in Figure \ref{honeycomb mag}b and the AFM in Figure \ref{honeycomb mag}c break the time-reversal symmetry, whereas the AFM in Figure \ref{honeycomb mag}d preserves the time-reversal symmetry because the spin pattern in Figure \ref{honeycomb mag}d remains the same after the time-reversal operation followed by a translation along $\Vec{b}$ by one atomic lattice constant.

Broken time-reversal symmetry often leads to non-reciprocal effects, such as magneto-optical Kerr effect (MOKE), magnetic circular dichroism (MCD), anomalous Hall effect (AHE), etc. However, we stress that broken time-reversal symmetry (even from a point group perspective) and non-reciprocal effects do not have a one-to-one correspondence. For example, the FM state in Figure \ref{honeycomb mag}b can support non-reciprocal effects, but the AFM state in Figure \ref{honeycomb mag}c cannot, despite the fact that both of them break time-reversal symmetry. More accurately and explicitly, only the magnetic phases, being FM or AFM, with magnetic point groups that are compatible with FM orders can produce non-reciprocal effects, while the magnetic order with the FM-incompatible magnetic point groups cannot. Note that within the 58 colored magnetic point groups (i.e., with explicit broken time-reversal symmetry), only 11 of them are compatible with FM orders.

\textbf{Broken spatial inversion symmetry} When the broken time-reversal symmetry cannot be directly probed for the case of the AFM state in Figure \ref{honeycomb mag}c due to the lack of non-reciprocal effects, it can be distinguished from the paramagnetic phase via other broken symmetries, one of which is the broken spatial inversion symmetry. It is worth noting that it is typically AFM, rather than FM, to break spatial inversion symmetry, because spin, or magnetic moment, is invariant under the spatial inversion operation. Examples of such cases include the A-type layered AFM in even-layer CrI$_3$ \cite{huang2017layer}, MnBi$_2$Te$_4$ \cite{gao2021layer}, CrSBr \cite{lee2021magnetic}, the N\'eel AFM order in MnPS$_3$ \cite{long2020persistence}, MnPSe$_3$ \cite{ni2021imaging}, etc. 

Broken spatial inversion symmetry manifests as the presence of second-order nonlinear optical and optoelectronic effects. One example is the magnetism-induced electric dipole second harmonic generation (ED SHG), which has been reported to be giant for the case of even-layer CrI$_3$ \cite{sun2019giant}, even-layer MnBi$_2$Te$_4$ \cite{fonseca2022anomalous} and CrSBr \cite{lee2021magnetic} and detected in MnPS$_3$ \cite{chu2020linear} and MnPSe$_3$ \cite{ni2021imaging} as well.

Magnetoelectric coupling is another effect shown up in noncentrosymmetric magnets because the electric field, which is odd under the spatial inversion operation, is allowed to couple linearly to the parity-odd (i.e., spatial inversion symmetry broken) magnetic order. Based on this effect, the electrical switch of the layered AFM was demonstrated in bilayer CrI$_3$ \cite{huang2018electrical,jiang2018controlling}. Moreover, the noncentrosymmetry of the magnetic state has also been predicted to potentially lead to a ferroelectric order and, therefore, a multiferroic state, for example, in MnPS$_3$ \cite{ressouche2010magnetoelectric}.

\textbf{Broken rotational symmetry} When collinear spins point along or pattern along a direction orthogonal to the high-symmetry rotation axis of the crystalline lattice, this rotational symmetry is likely to be broken. For example, the AFM chains in Figure \ref{honeycomb mag}d break the six-fold rotational symmetry of the underlying honeycomb lattice to the two-fold rotational symmetry, whereas the AFM sublattices in Figure \ref{honeycomb mag}c reduce the six-fold to three-fold symmetry. We note that these two examples of rotational symmetry reduction show distinct linear responses.

Broken rotational symmetry can lead to anisotropic linear responses (e.g., magnetism-induced linear dichroism or birefringence), if and only if the magnetism with broken rotational symmetry ends up into the triclinic, monoclinic, or orthorhombic crystal systems. Based on this, the AFM in Figure \ref{honeycomb mag}c with the three-fold rotational symmetry belongs to the trigonal crystal system and hence cannot host linear dichroism or linear birefringence in the linear responses, but that in Figure \ref{honeycomb mag}d with the two-fold rotational symmetry can as it falls into the orthorhombic crystal system. It has been shown in FePS$_3$ that the magnetism-induced linear dichroism can be as large as 100$\%$ \cite{zhang2022cavity}.

\textbf{Broken translational symmetry} When the magnetic primitive cell is enlarged from one of the crystal structures, certain translational symmetries of the lattice are broken in the magnetic phase. For example, the zigzag AFM state in Figure \ref{honeycomb mag}d contains four magnetic sites per magnetic primitive cell, twice the number of the two sites for the crystalline honeycomb lattice. Diffraction-based experimental tools (e.g., elastic X-ray or neutron diffraction) are ideal for exploring broken translational symmetry, as new diffraction peaks are expected to appear.

Broken translational symmetry corresponds to the expansion of the primitive cell in real space and, therefore, the reduction of the Brillouin zone in momentum space (also referred to as the folding of the Brillouin zone). This Brillouin zone folding allows for the multiplication of phonon band structures, which can be typically observed as new phonon modes in Raman spectroscopy. In FePS$_3$ \cite{lee2016ising} and CrI$_3$ \cite{li2020magnetic}, 
phonon modes from the Brillouin zone boundaries show up in the Raman spectroscopy of the magnetic phase due to the zone folding by magnetism. 

\subsubsection{Noncollinear spins in two-dimensional magnets}
Noncollinear spin magnetism can be realized in either long-range or quasi-long-range magnetic orders, where spins choose to align at angles beyond 0$^\mathrm{o}$ (parallel) and 180$^\mathrm{o}$ (antiparallel) between one and another. More interestingly, in some noncollinear magnets, the spins wind along either the longitudinal or the transverse direction and form skyrmionic \cite{fert2017magnetic}, chiral \cite{fukushima2008chiral}, or spiral \cite{kimura2007spiral} patterns. So far, there have been nearly no experimental reports on noncollinear spins in natural 2D magnetic atomic crystals, and only a couple in artificial 2D magnetic moir\'e superlattices. In this subsection, we provide the survey over noncollinear spin states proposed and/or realized in a limited number of 2D systems known thus far.

From a symmetry perspective, noncollinear spin magnetism can be differentiated by broken symmetries in a very similar way as what has been described for collinear spin magnetism above. In addition to these highlighted broken symmetries in Section 3.1.1, an additional symmetry to be considered for noncollinear spins is the broken mirror symmetry that typically develops due to the winding of spins. Broken mirror symmetry itself is often challenging to be detected by linear responses but can be reliably captured by nonlinear ones \cite{jin2020observation,luo2021ultrafast,guo2023ferrorotational}. 

\begin{figure}[th]
\begin{center}
\includegraphics[width=1.0\textwidth]{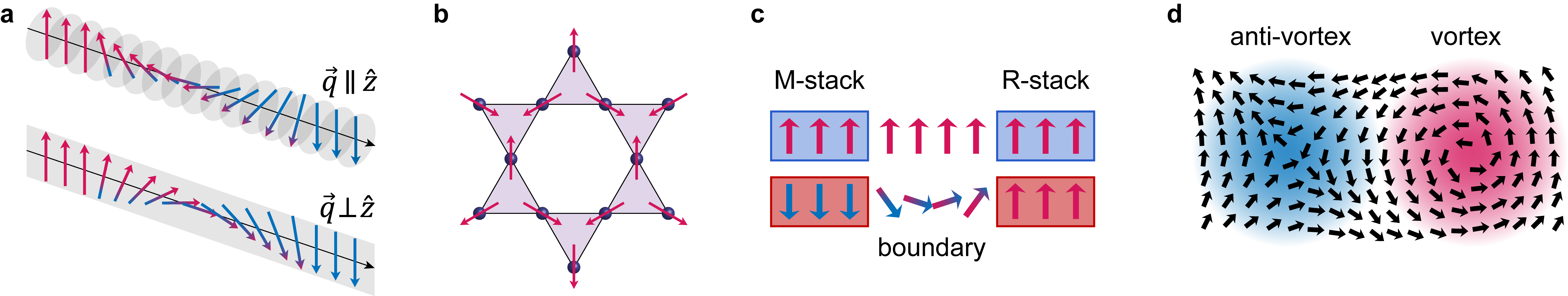}\caption{\small Examples of noncollinear spin texture in 2D magnets. (a) Spiral magnetic orders for both $\hat{q}||\hat{z}$ and $\hat{q}\perp\hat{z}$ cases; (b) All-in-all-out noncollinear AFM order in a kagome magnetic lattice, e.g., Co$_3$Sn$_2$S$_2$; (c) Moir\'e magnetic order with noncollinear spins at the boundary between two opposite interlayer exchange coupling regions within individual moir\'e supercells, e.g., between M-stacked AFM interlayer coupling and R-stacked FM interlayer coupling regions in twisted CrI$_3$ bilayers; (d) Vortex and anti-vortex pair in 2D XY-magnets, e.g., possibly in CrCl$_3$ monolayer.} 
\label{noncollinear mag}
\end{center}
\end{figure}

\textbf{Spiral magnetic order} Spiral magnetic order describes a periodic spin structure where the spins rotate about the rotation axis (typically defined as $\hat{z}$ axis) with a spin-spiral wavevector $\Vec{q}$ (equivalently, wavelength $a = 2\pi/q$). The wavevector $\Vec{q}$ can be either parallel or orthogonal to the rotation axis $\hat{z}$, both of which are illustrated in Figure \ref{noncollinear mag}a. NiI$_2$ is one and the only known vdW magnet that hosts a spiral magnetic order in its bulk form \cite{kurumaji2013magnetoelectric}. Due to the formation of spin spirals in NiI$_2$, a spontaneous polarization arises via the inverse Dzyaloshinskii–Moriya interaction, and as a result, a multiferroic state develops in NiI$_2$ \cite{kurumaji2013magnetoelectric,ju2021possible,song2022evidence}. 

\textbf{Kagome AFM order} Kagome lattice is unique in the sense that there are three magnetic sites per structural primitive cell, which allows for noncollinear spin alignments within the plane at these three magnetic sites, being 120$^\mathrm{o}$ rotated from one another, as shown in Figure \ref{noncollinear mag}b. In addition to magnetism, kagome lattice further hosts van Hove singularity points, flat bands, and Dirac band dispersion in electronic and magnetic excitation band structures \cite{yin2022topological}. So far, the physics of kagome lattice is primarily studied in quasi-two-dimensional lattices embedded in three-dimensional matrices (e.g., MnX$_3$ \cite{nakatsuji2015large}, Co$_3$Sn$_2$S$_2$ \cite{neubauer2022spin}). There have been very few layered vdW kagome lattice systems realized experimentally, with only a couple of examples such as Nb$_3$X$_8$ \cite{pasco2019tunable}.

\textbf{Moir\'e magnetic order} Moir\'e superlattice formed by twisted magnetic atomic crystals can host the interlayer exchange coupling that periodically modulates from FM to AFM. This moir\'e exchanging coupling tends to introduce modulated spin alignments where the spins are noncollinear at the boundaries where they flip into the opposite direction, as shown in Figure \ref{noncollinear mag}c. While moir\'e magnetic superlattices have been theoretically predicted to host skyrmion lattices \cite{tong2018skyrmions,hejazi2021heterobilayer,akram2021skyrmions,akram2021moire,ghader2022whirling,zheng2023magnetic} and noncollinear spin textures \cite{hejazi2020noncollinear}, there have been only early-stage experimental efforts made in twisted bilayer \cite{song2021direct,xu2022coexisting}, double bilayer \cite{xie2022twist,xie2023evidence,cheng2023electrically}, and double trilayer \cite{song2021direct} CrI$_3$ homostructures.

\textbf{Quasi-long-range order} Quasi-long-range order was predicted to develop in 2D XY-type magnetic systems. Its manifestations include, for example, the formation of vortex-and-antivortex pairs (Figure \ref{noncollinear mag}d) and the emergence of spin-induced nematicity. In the meantime, the spin fluctuations are expected to be present even down to the lowest temperature in 2D XY magnets due to the gapless nature of their spin wave excitation spectra. There have hitherto been a few 2D XY magnets that are experimentally realized, for example, NiPS$_3$ and CrCl$_3$. Yet, no direct experimental evidence shows the presence of quasi-long-range order in the 2D limit of both few-layer NiPS$_3$ and few-layer CrCl$_3$.

\subsection{Investigating spin wave excitations in two-dimensional magnets}
Spin waves describe the collective propagation of magnetic disturbance in long-range magnetic orders in solids \cite{bloch1930theorie}. It is of both fundamental importance and technological practice to study spin waves. From the fundamental science perspective, spin waves concern the dynamic properties of magnetism, encoding the time-scale and energy-scale information of the magnetism that is not accessible in static properties such as the magnetic ground states discussed above. The quanta of spin waves are magnons that are bosonic quasiparticles, similar to phonons for lattice vibrations. From the practical application point of view, magnons can be used as the carrier of information for low-power transmission, processing and computing (i.e., magnonics) \cite{chumak2015magnon}, thanks to the absence of Ohmic losses and the exploitation of wave interference. 2D vdW magnets provide a new platform for realizing ultra-compact magnonics \cite{mak2019probing}.

Theoretically, the spin waves have been computed using the linear spin-wave model via the semi-classical approach or the quantum mechanical approach. Similar to phonon band structures and vibrational modes, magnon band structures depict the magnon energy dispersion as a function of the electronic momentum within the magnetic Brillouin zone (i.e., $\bold{E}(\Vec{k})$), and the eigenvector for a given ($\Vec{k}$, $\bold{E}$) describes the relative spin alignments for this magnon mode. Experimentally, for 3D vdW bulk magnets, the spin waves are probed by inelastic neutron scattering or resonant inelastic X-ray scattering, where a complete $\bold{E}(\Vec{k})$ dispersion relationship is obtained; for 2D vdW magnet films, the magnons have only been detected by optical techniques, including Raman scattering \cite{lee2016ising,cenker2021direct,luo2023evidence} and time-resolved optical spectroscopy (e.g., time-resolved MOKE, time-resolved absorption spectroscopy) \cite{zhang2020gate}. Because the photon momentum is extremely small as compared to the electronic momentum, optical tools can only access zone center magnons (i.e., at $\Vec{k} = 0$) through the single magnon process, and sometimes in AFMs, zone boundary magnons (i.e., $\Vec{k}$ at the Brillouin zone boundaries) via two-magnon excitation processes. 

\begin{table}[th]
    \center
    \begin{tabular}{|c||*{3}{c|}}\hline
    \backslashbox[48mm]{spin dimension $n$}{lattice dimension $d$}&\makebox[3em]{$d=1$}&\makebox[3em]{$d=2$}&\makebox[3em]{$d=3$}\\\hline\hline
    $n=1:$ Ising-type $\Longleftrightarrow \alpha>1$: easy-axis & - & $\Delta \neq 0$ & $\Delta \neq 0$ \\\hline
    $n=2:$ XY-type $\Longleftrightarrow \alpha<1$: easy-plane & - & $\Delta = 0$ & $\Delta = 0$ \\\hline
    $n=3:$ Heisenberg-type $\Longleftrightarrow \alpha=1$: isotropic & - & - & $\Delta = 0$ \\\hline
    \end{tabular}
    \caption{Summary of the presence or absence of spin wave, depending on the spin dimension $n$ and the lattice dimension $d$. For the entries with \lq\lq-\rq\rq, it represents the absence of long-range or quasi-long-range orders. $\Delta$ indicates the magnon gap.}
    \label{spin wave gap}
\end{table}

The magnetic anisotropy, or equivalently, the spin dimension $n$, determines whether the spin wave excitations are gapped ($\Delta \neq 0$) or gapless ($\Delta = 0$). Table \ref{spin wave gap} summarizes the presence and absence of spin wave gap in a magnetic system depending on the spin dimension $n$ and the lattice dimension $d$.  Furthermore, the magnetic lattice symmetry and the magnetic ground state are of foundational importance to spin wave excitations. It is worth highlighting that the research on spin waves in 2D vdW magnets is at its early stage, with only a handful of theoretical and experimental results thus far. In this section, we will survey spin waves in four magnetic lattices with the simple intralayer FM order, orthorhombic lattice (e.g., CrSBr \cite{bae2022exciton}), triangular/hexagonal lattice (e.g., VSe$_2$ \cite{rudzinski2022effect}), honeycomb lattice (\textit{e.g.}, CrI$_3$ \cite{li2020magnetic}), and kagome lattice (e.g., Co$_3$Sn$_2$S$_2$, a quasi-2D example). 

\begin{figure}[th]
\begin{center}
\includegraphics[width=1.0\textwidth]{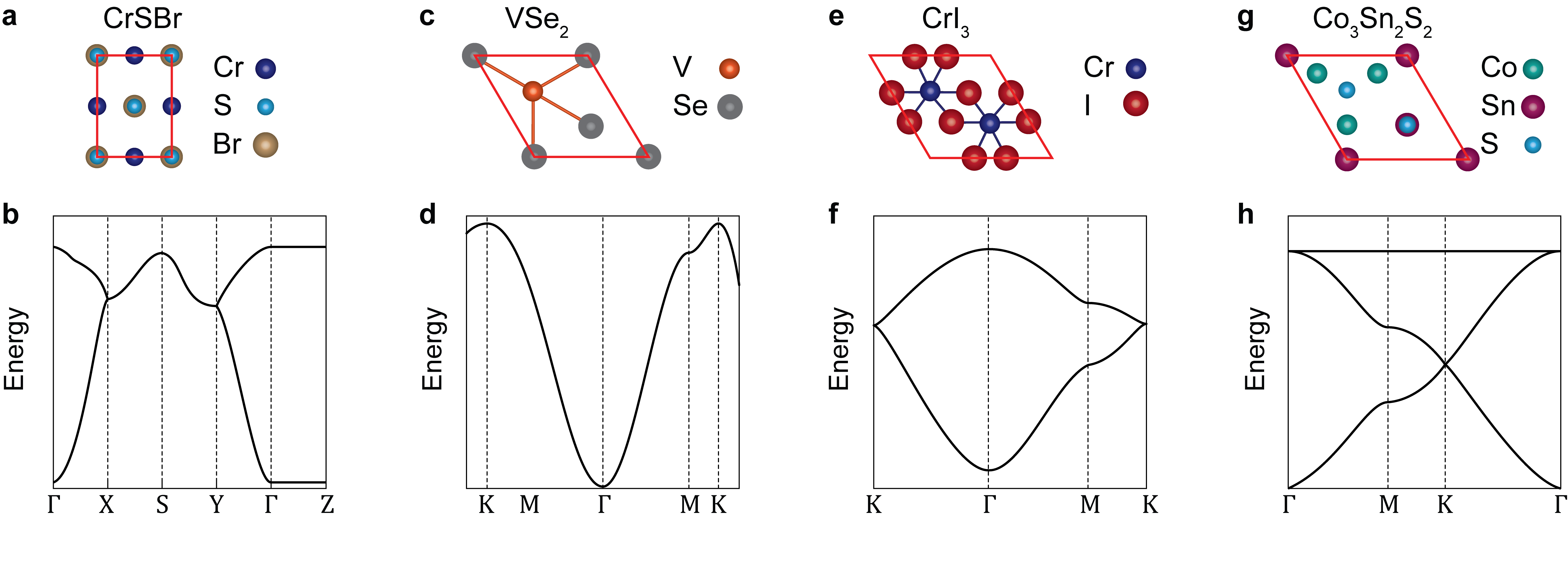}\caption{\small Spin wave dispersion $\bold{E}(\Vec{k})$ for four representative families of 2D vdW magnets. Crystalline primitive cell and computed spin wave dispersion for (a-b) orthorhombic 2D vdW magnet, e.g., CrSBr with two magnetic sites per primitive cell; (c-d) triangular 2D vdW magnet, e.g., VSe$_2$ with one magnetic site per primitive cell; (e-f) honeycomb 2D vdW magnet, e.g., CrI$_3$ with two magnetic sites per primitive cell; and (g-h) kagome 2D vdW magnet, e.g., Co$_3$Sn$_2$S$_2$ with three magnetic sites per primitive cell. Figures (b), (d), (f), and (h) are adapted from Refs. \cite{bae2022exciton}, \cite{rudzinski2022effect}, \cite{pershoguba2018dirac}, and \cite{yin2022topological}, respectively.} 
\label{spin wave}
\end{center}
\end{figure}

\subsubsection{Spin waves in 2D orthorhombic magnets}

Among the known 2D vdW magnets, only the ternary $3d$ transition metal compounds, MXY (e.g., CrSBr), have the orthorhombic crystal structure. Within each crystallographic primitive cell of MXY (Figure \ref{spin wave}a), there are two magnetic sites that naturally lead to two spin wave branches, one acoustic and one optical branches, respectively. Depending on the spin Hamiltonian with the FM or AFM intralayer exchange coupling, the dispersion of two spin wave branches can be quite different. CrXY (X = O, S, Se; Y = Cl, Br, I) monolayers are theoretically calculated \cite{jiang2018screening, hou2022multifunctional}, some experimentally verified \cite{lee2021magnetic, telford2022coupling, zhang2019magnetism}, to be a FM insulator/semiconductor, whereas FeOY (Y = F, Cl, Br, I) monolayers are predicted to be dominated by the intralayer AFM coupling \cite{wang2020discovery}. We use the FM intralayer coupling case as an example to illustrate the spin waves in this orthorhombic vdW magnet family, as the FM semiconductor CrSBr has been extensively studied \cite{bae2022exciton, diederich2023tunable, cham2022anisotropic, scheie2022spin, esteras2022magnon, cho2023microscopic, bo2023calculated}.

The spin Hamiltonian for CrSBr includes the strong FM intralayer exchange coupling ($J_{intra}<0$), the weak AFM interlayer exchange coupling ($J_{inter}$), and a biaxial anisotropy ($A_x>0$ and $A_z>0$), which is expressed as $$\mathcal{H} =\sum_{<i,j>}J_{intra}\hat{\Vec{\mathrm{S}}}_i\cdot\hat{\Vec{\mathrm{S}}}_j+\sum_k J_{inter}\hat{\Vec{\mathrm{S}}}_k\cdot\hat{\Vec{\mathrm{S}}}_{k+1}+A_x\hat{\mathrm{S}}_x^2+A_z\hat{\mathrm{S}}_z^2$$
\noindent Although the first and the third nearest neighboring DM interactions are allowed by symmetry, they are extremely small based on the inelastic neutron scattering data \cite{scheie2022spin}. With the spin Hamiltonian above, the spin wave dispersion can be calculated using the Holstein–Primakoff formalism, and the spectra along the high-symmetry lines are shown in Figure \ref{spin wave}b. The presence of the acoustic and optical branches is consistent with two magnetic sites per primitive cell per layer. The acoustic and optical branches correspond to the in-phase and out-of-phase precession of spins between the magnetic sites, respectively. The acoustic branch is gapped out at the $\Gamma$ point because of the easy-axis anisotropy, the bandwidth of the spin wave branches is determined by the intralayer exchange coupling, and the weak spin wave energy dispersion along the $\Gamma-\mathrm{Z}$ direction is from the very small interlayer exchange coupling. A zoom-in look further shows the split of the acoustic and the optical branches into two sets, in-phase and out-of-phase spin precessions between the two layers, because the interlayer AFM coupling with the easy-axis anisotropy doubles the magnetic unit cell along the $c$ axis \cite{esteras2022magnon}. 

The spin waves of CrSBr have been experimentally probed by inelastic neutron scattering for the 3D bulk case \cite{scheie2022spin} and by time-resolved absorption spectroscopy for the 2D few-layer thin film case \cite{bae2022exciton}. For the bulk CrSBr, both spin wave branches have been observed, but the spin wave gap in the acoustic branch and the AFM interlayer coupling-induced splitting were not detected because of the spectral resolution. As a result, a Heisenberg spin Hamiltonian up to the eighth nearest neighbor exchange coupling can fit the data well. For the few-layer CrSBr, only the acoustic branch at the $\Gamma$ point was detected due to the optical selection rule and the negligibly small photon momentum limitations. At the same time, thanks to the exceptionally high spectral resolution through the prolonged time-domain measurements, the acoustic branch was clearly resolved to split into the in-phase and out-of-phase spin precession modes at 24 GHz and 34 GHz, respectively. More impressively, they have exceptionally long coherent time and can propagate up to 6-7 $\mu$m \cite{bae2022exciton}. Of further interest, the spin waves are coupled strongly to the excitons in CrSBr \cite{wilson2021interlayer}.

\subsubsection{Spin waves in 2D hexagonal magnets}
The magnetic sites in transition metal dihalides (MX$_2$), transition metal dichalcogenides (H-phase MX$_2$, T-phase MX$_2$), and MnBi$_{2n}$Te$_{3n+1}$ ($n=1,2,3,4$) form the triangular lattice within the vdW layers, with one magnetic site per structural primitive cell per layer (Figure \ref{spin wave}c). Despite the fact that quite many 2D vdW magnets belong to this class, there has been very limited research on their spin waves. 

The intralayer AFM exchange coupling on a triangular lattice is subjected to geometrical frustration, which can lead to unique yet complex magnetic structures, for example, the noncollinear spins in NiBr$_2$ monolayer \cite{bikaljević2021noncollinear} and the spiral magnetic order in NiI$_2$ \cite{kurumaji2013magnetoelectric}. Such magnetic states greatly enlarge the magnetic primitive cell and typically are subjected to DM interactions, each and both of which make their spin wave research a challenging task: one would expect multiple spin wave branches with fine energy separations and subtle splittings at selected momenta. As a result, there have hardly been any experimental or theoretical investigations on spin waves for such noncollinear spin magnets in 2D.

On the other hand, the intralayer FM exchange coupling is in a much simpler situation - one magnetic site per primitive cell per layer corresponds to one spin wave branch from the intralayer exchange coupling. Exemplary 2D vdW magnets of this type include VSe$_2$ \cite{bonilla2018strong}, MnBi$_2$Te$_4$ \cite{hu2020realization}, etc., for which the spin Hamiltonian is composed of the intralayer FM coupling ($J_{intra}<0$), the interlayer AFM coupling ($J_{inter}>0$), and the out-of-plane easy-axis anisotropy ($A_z>0$), i.e., $$\mathcal{H} = \sum_{<i,j>}J_{intra}\hat{\Vec{\mathrm{S}}}_i\cdot\hat{\Vec{\mathrm{S}}}_j+\sum_k J_{inter}\hat{\Vec{\mathrm{S}}}_k\cdot\hat{\Vec{\mathrm{S}}}_{k+1}-A_z\hat{\mathrm{S}}_z^2$$
The computed spin wave dispersion along $\Gamma-\mathrm{M}-\mathrm{K}-\Gamma$ is shown in Figure \ref{spin wave}d. A single spin wave branch is observed, with a finite spin wave gap determined by the intralayer out-of-plane easy axis anisotropy. The interlayer AFM coupling doubles the magnetic primitive cell along the $c$ axis and thus the number of spin wave branches. Due to the isotropic Heisenberg type of this interlayer AFM coupling, however, the two spin wave branches remain degenerate, corresponding to spin precessions in one of the two layers but not the other. 

The spin wave of MnBi$_2$Te$_4$ has been probed by magneto-Raman spectroscopy for both the few-layer and bulk cases \cite{lujan2022magnons}. As an optical probe, Raman spectroscopy can only access zero-momentum excitations, e.g., either single magnon excitations from the Brillouin zone center $\Gamma$ or two-magnon excitations from the zone boundary. In Ref. \cite{lujan2022magnons}, only the zone-center single magnon was observed in the canted-AFM and FM phases with a finite out-of-plane magnetic field applied.

\subsubsection{Spin waves in 2D honeycomb magnets}

Transition metal trihalides (MX$_3$), ternary transition metal compounds CrXTe$_3$, MPX$_3$, and Fe$_n$GeTe$_2$ ($3 \leq n \leq 5$) belong to the 2D vdW honeycomb magnet family, where the magnetic atoms are arranged into the honeycomb lattice within the vdW layers. Honeycomb lattice is a non-Bravais lattice with two magnetic sites per structural primitive cell (Figure \ref{spin wave}e). Honeycomb magnets have been the most investigated family of vdW magnets up to the present, with a variety of magnetic ground states realized in this family, ranging from FM with easy-axis, isotropic, and easy-plane anisotropies (e.g., CrX$_3$, X = I, Br, and Cl, respectively) \cite{kim2019evolution}, N\'eel AFM with easy-axis and easy-plane anisotropies (e.g, MnPS$_3$ and MnPSe$_3$) \cite{chu2020linear,ni2021imaging}, zigzag AFM with easy-axis and easy-plane anisotropies (e.g., FePS$_3$ and NiPS$_3$) \cite{lee2016ising,kim2019suppression}, and even spin liquid candidates with the Kitaev-type spin Hamiltonian (e.g., $\alpha$-RuCl$_3$) \cite{takagi2019concept}. 

Despite the popularity of honeycomb magnets, the experimental and theoretical investigation of their spin waves has been quite limited, especially in the 2D limit. Surveying through the literature, the following examples of spin waves in honeycomb magnets were identified, including CrX$_3$ \cite{kim2019evolution, li2020magnetic,cenker2021direct}, FePS$_3$ \cite{lee2016ising,luo2023evidence}, and NiPS$_3$ \cite{belvin2021exciton,afanasiev2021controlling,jana2023magnon} at the experimental forefront, and CrX$_3$ at the theoretical forefront \cite{pershoguba2018dirac}. For CrI$_3$, the most investigated spin-wave system, its spin Hamiltonian contains the intralayer FM nearest neighbor exchange coupling ($J_{intra}<0$), the interlayer AFM/FM exchange coupling ($J_{inter}>0/J_{inter}<0$) for the monoclinic/rhombohedral stacking, and the out-of-plane easy-axis anisotropy ($A_z>0$), i.e., $$\mathcal{H} = \sum_{<i,j>}J_{intra}\hat{\Vec{\mathrm{S}}}_i\cdot\hat{\Vec{\mathrm{S}}}_j+\sum_k J_{inter}\hat{\Vec{\mathrm{S}}}_k\cdot\hat{\Vec{\mathrm{S}}}_{k+1}-A_z\hat{\mathrm{S}}_z^2$$
It turns out that the computed spin wave dispersion, $\mathbf{E}(\Vec{k})$, of monolayer CrI$_3$ shares a similar Dirac dispersion as the electronic band structure of graphene, shown in Figure \ref{spin wave}f along the $\Gamma-\mathrm{M}-\mathrm{K}-\Gamma$ momentum direction. There are, in total, two spin wave branches, acoustic and optical branches, that are consistent with two magnetic sublattices in the CrI$_3$ monolayer. At the $\Gamma$ point that corresponds to the uniform spin precession, the acoustic (optical) one shows the in-phase (out-of-phase) precession between the two spin sublattices and their energies scale with the spin anisotropy (the mean) of the intralayer exchange coupling. At the $\mathrm{M}$ point that is the van Hove singularity point, the spin wave density of states diverges. At the $\mathrm{K}$ point that is at the Brillouin zone corner, the two spin wave branches meet each other and develop a linear energy-momentum dispersion, named after the Dirac dispersion, and hence the Dirac magnons \cite{pershoguba2018dirac}. When the interlayer exchange coupling for multilayer or bulk CrI$_3$ is considered, one needs to discuss the multilayer and the bulk cases separately because of their difference in the interlayer exchange coupling type 
and the out-of-plane translational symmetry. 
First, in bulk CrI$_3$ with the FM interlayer exchange coupling and the presence of out-of-plane translational symmetry, it remains to be one layer per magnetic primitive cell, and therefore, the spin wave dispersion is nearly the same as the CrI$_3$ monolayer with minor corrections from the presence of the interlayer FM coupling. Second, in few-layer CrI$_3$ with the interlayer AFM coupling and the absence of out-of-plane translational symmetry, the magnetic unit cell expands as the layer number increases, and the net magnetization alternates between zero and finite for even and odd number of layers. As a result, the number of spin wave branches is expected to increase proportionally to the layer number, and the spin wave pairs with opposite angular momenta (i.e., $s = 1$ and $s = -1$) altering from degenerate to non-degenerate as changed from even to odd number of layers.

In addition to the simple spin Hamiltonian terms shown in $\mathcal{H}$ above for CrI$_3$, exchange coupling terms such as next nearest neighbor DM interactions \cite{chen2018topological,chen2021magnetic} and Kitaev-type interactions \cite{lee2020fundamental} have also been considered theoretically in literature. For the DM interactions, $\mathcal{H}_{\mathrm{DM}}=\sum_{<i,j>}\Vec{D}_{ij}\cdot(\hat{\Vec{\mathrm{S}}}_i\times\hat{\Vec{\mathrm{S}}}_j)$ where $<i,j>$ indicates the next nearest neighboring sites and $D_{ij}<0$, their presence maintains the spatial inversion symmetry for the FM phase in bulk and monolayer CrI$_3$, and its momentum corresponds to the $\mathrm{K}$ points in the momentum space. As a result, this next nearest neighbor DM interaction term opens up spin wave gaps at the K points, making the massless Dirac magnons massive and introducing topological magnon edge states \cite{chen2018topological,chen2021magnetic}. For the  Kitaev interactions, $\mathcal{H}_\mathrm{K}=\sum_{<i,j>\in \lambda\mu(\nu)}[KS^{\nu}_iS^{\nu}_j+\Gamma(S^{\lambda}_iS^{\mu}_j+S^{\nu}_iS^{\lambda}_j)]$ with $(\lambda,\nu,\mu)$ being permutation of $(x,y,z)$, $K$ being the Kitaev interaction strength, and $\Gamma$ being the symmetric off-diagonal anisotropy that introduces spin wave gaps at the $\Gamma$ points. This Kitaev interaction term combines with the isotropic Heisenberg interaction and gives the $\mathrm{J}-\mathrm{K}-\Gamma$ spin Hamiltonian, i.e.,$\mathcal{H}_{\mathrm{J}-\mathrm{K}-\Gamma} = \sum_{<i,j>\in \lambda\mu(\nu)}[J_{intra}\hat{\Vec{\mathrm{S}}}_i\cdot\hat{\Vec{\mathrm{S}}}_j+KS^{\nu}_iS^{\nu}_j+\Gamma(S^{\lambda}_iS^{\mu}_j+S^{\nu}_iS^{\lambda}_j)]$, which is an alternative spin Hamiltonian to describe the CrI$_3$ monolayer, and can also introduce spin wave gaps at the K points \cite{chen2021magnetic,lee2020fundamental}.

Experimentally, the spin waves in CrI$_3$ have been probed by multiple experimental techniques. For bulk CrI$_3$, inelastic neutron scattering spectroscopy provides the momentum-resolved spin wave spectra \cite{chen2018topological,chen2021magnetic}, Raman spectroscopy \cite{li2020magnetic}, and ferromagnetic resonance spectroscopy focuses on the high energy resolution of the acoustic spin wave branch at the $\Gamma$ point \cite{lee2020fundamental}. For the monolayer and bilayer CrI$_3$, Raman spectroscopy and time-resolved MOKE mainly resolve the spin wave gap in the acoustic branch(es) at the $\Gamma$ point \cite{cenker2021direct,zhang2020gate}, and possibly the optical branch frequency in Raman spectroscopy \cite{cenker2021direct}. For the few-layer CrI$_3$, inelastic magneto-tunneling junction probes the spin waves at $\Gamma$ point for the spin wave gaps and spin waves at $\mathrm{M}$ points with divergent densities of states \cite{kim2019evolution,klein2018probing}.

\subsubsection{Spin waves in 2D kagome magnets}

Only a couple of vdW magnets, i.e., Nb$_3$X$_8$ (X = Cl, Br, and I) \cite{pasco2019tunable}, have been identified for realizing the 2D vdW kagome magnetism whose magnetic sites form a kagome lattice. Kagome lattice is another non-Bravais lattice with three magnetic sites per structural primitive cell, as shown in Figure \ref{spin wave}g. Among all the 2D vdW magnets, kagome magnets are the least investigated family, despite the recent popularity of 3D kagome magnets and superconductors \cite{yin2022topological}. 

While there have been experimental realizations of the 2D form of Nb$_3$X$_8$ \cite{pasco2019tunable} via mechanical exfoliation, few characterizations of their magnetic properties have been performed as yet, leaving their magnetic ground states and spin waves elusive. Theoretical calculations have suggested that monolayer Nb$_3$X$_8$ are semiconducting FMs. Due to the specialty of the kagome lattice, even the simple kagome FMs can host nontrivial spin wave excitations \cite{yin2022topological,chisnell2015topological}, whose spin Hamiltonian includes the nearest neighboring FM exchange coupling ($J_{intra}<0$) and the out-of-plane easy-axis anisotropy ($A_z>0$), i.e., $$\mathcal{H}_{\mathrm{monolayer}}=\sum_{<i,j>}J_{intra}\hat{\Vec{\mathrm{S}}}_i\cdot\hat{\Vec{\mathrm{S}}}_j-A_z\hat{\mathrm{S}}_z^2$$
Consistent with three magnetic sites per primitive cell, there are in total three spin wave branches for this kagome FM, whose energy-momentum dispersion is shown in Figure \ref{spin wave}h along the $\Gamma-\mathrm{M}-\mathrm{K}-\Gamma$ momentum direction. The two lower-energy spin wave branches show similar properties as the spin wave dispersion spectra of a honeycomb FM as shown in Figure \ref{spin wave}f, including the Dirac dispersion at the $\mathrm{K}$ points, the van Hove singularity at the $\mathrm{M}$ points, and the spin wave gap at the $\Gamma$ point. The third spin wave branch forms the so-called magnon ``flat band" where the magnon energy is constant across the entire Brillouin zone. Furthermore, this flat band touches the second spin wave band at the $\Gamma$ point. Importantly, the spin wave band touchings at the $\Gamma$ and $\mathrm{K}$ points are symmetry-protected by the nearest neighboring FM spin Hamiltonian $\mathcal{H}$ above.

When a DM interaction between the nearest neighboring sites is included, i.e., $\mathcal{H}_{\mathrm{DM}}=\sum_{<i,j>}\Vec{D}_{ij}\cdot(\hat{\Vec{\mathrm{S}}}_i\times\hat{\Vec{\mathrm{S}}}_j)$ with $\Vec{D}_{ij}$ being the out-of-plane DM vector $D_z$, it opens up gaps between the three spin wave bands at the original band touching $\Gamma$ and $\mathrm{K}$ points. Of more interest are the topological spin wave bands and the chiral magnon edge states, after the gap opening via this DM interaction \cite{chisnell2015topological,zhang2013topological}. The 2D vdW kagome FM, Nb$_3$X$_8$, can be another family of potential candidates for exploring topological spin waves.  

\subsection{Exploring spin-lattice-charge coupling in two-dimensional magnets}
Three primary degrees of freedom, lattice, charge, and spin, are intimately related in 2D vdW magnets for a couple of reasons. First, 2D vdW magnets involve transition metals of $d$ orbital or rare-earth elements of $f$ orbital and, therefore, naturally result in the strongly correlated interaction regime involving multiple degrees of freedom. Second, the reduction of the lattice dimensionality often leads to modifications of the electronic bands and, furthermore, the magnetic exchange interactions, allowing these three degrees of freedom to closely couple with each other. 

The examination of the coupling among these three degrees of freedom has been carried out by inspecting the relationship between them, in both theoretical and experimental studies. For example, one can vary the crystalline structure and check if the magnetic and electronic properties respond to this structural change, or one can tune the magnetism and examine if the structural and electronic structures vary in response. The 2D-ness of vdW magnets provides additional possibilities to selectively tune one degree of freedom, such as thickness dependence, electric field effect, etc., to impact other degrees of freedom. 

The microscopic origins for the couplings among different degrees of freedom are often material-specific. So far, the spin-lattice-charge coupling has been investigated in a handful 2D vdW magnets, including CrX$_3$ \cite{mcguire2015coupling, mcguire2017magnetic, mcguire2017crystal}, MPS$_3$ (M = Ni, Fe) \cite{brec1986review}, CrSBr \cite{wilson2021interlayer,cenker2022reversible}, and MnBi$_2$Te$_4$ \cite{deng2020quantum,gao2021layer}. Below, we focus our effort on surveying the two types of identified couplings in these compounds: the spin-lattice coupling and the spin-charge coupling. 

\subsubsection{The spin-lattice coupling in two-dimensional magnets}
The crystalline lattice is of fundamental importance to the physical properties of solids. In the case of 2D vdW magnets, two aspects of crystalline lattice in determining the magnetic properties have been noted.  First, for nearly all vdW magnets, the magnetic cations (M) are surrounded by the ligand anions (X) and the exchange couplings between magnetic sites are mediated through the ligand sites. As a result, changes in the distances between the magnetic sites ($d_\mathrm{{M-M}}$) or between the magnetic and ligand sites ($d_\mathrm{{M-X}}$) and variations in the angles between magnetic sites through bridging ligands ($\alpha_\mathrm{{M-X-M}}$) are expected to modify the magnetic exchange coupling strength and sign. Second, due to the vdW nature, the structural coupling between adjacent layers is weak and often allows for different interlayer stacking geometries with similar elastic energies. Consequently, different interlayer stacking geometries can lead to distinct interlayer, and sometimes intralayer, magnetic exchange coupling, in terms of both the strength and the sign. 

\begin{figure}[th]
\begin{center}
\includegraphics[width=1.0\textwidth]{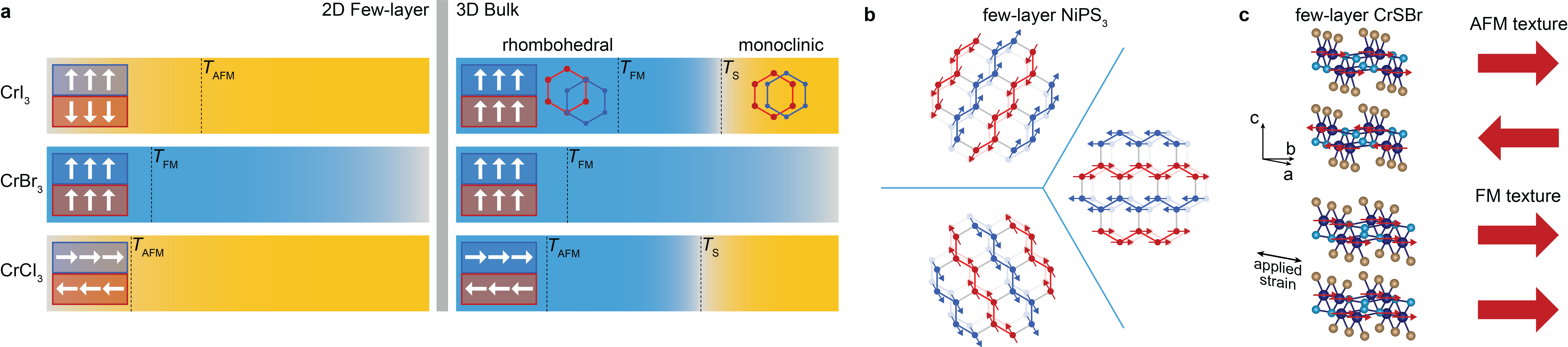}\caption{\small Examples to illustrate the spin-lattice coupling in 2D vdW magnets. (a) In CrX$_3$ (X=Cl, Br, I), the comparison between bulk and few-layer cases showing the dependence of interlayer magnetism on the interlayer stacking geometry; (b) In MPS$_3$ (M=Ni, Fe), the in-plane zig-zag chain direction being locked to the shift vector between adjacent layers; (c) In CrSBr, the in-plane lattice parameters determining the interlayer exchange coupling. Figure (b) is adapted from Ref. \cite{zhou2022dynamical}.} 
\label{spin lattice coupling}
\end{center}
\end{figure}

\textbf{Interlayer magnetism vs interlayer stacking geometry} The coupling between layered magnetism and interlayer stacking was discovered and extensively explored in CrX$_3$ (X = Cl, Br, and I) \cite{huang2017layer,chen2019direct, klein2019enhancement}. For individual layers of CrX$_3$ (X = Cl, Br, and I), the Cr$^{3+}$ cation is located at the center of the octahedral cage of six X$^{-}$ anions, and this CrX$_6$ building block arranges in an edge-sharing way to form the honeycomb lattice. Both bulk CrCl$_3$ and CrI$_3$ single crystals undergo a first-order structural phase transition from the monoclinic (space group $C2/m$) to the rhombohedral (space group $R\overline{3}$) interlayer stacking geometry when cooling across the critical temperature of $\sim$240 K \cite{mcguire2017magnetic} and $\sim$220 K \cite{mcguire2015coupling}, respectively, whereas bulk CrBr$_3$ single crystal remains in the rhombohedral (space group $R\overline{3}$) stacking phase down to low temperatures \cite{hong2023pressure}. In the 2D limit, few-layer CrCl$_3$ \cite{klein2019enhancement} and CrI$_3$ \cite{huang2017layer} films no longer experience any structural phase transitions and remain in the monoclinic space group $C2/m$ while few-layer CrBr$_3$ stays in the rhombohedral space group $R\overline{3}$. The comparison between bulk and few-layer CrX$_3$ is summarized in Figure \ref{spin lattice coupling}a.

Closely coupled to the interlayer stacking geometry, the layered magnetism in CrX$_3$ varies between their bulk and film forms, but in different ways, as detailed below. For CrI$_3$, its bulk form hosts a FM ground state with spins aligned in the same out-of-plane direction within and across layers below $T_C =$ 61 K \cite{mcguire2015coupling}, and its few-layer form realizes a layered AFM order with spins pointing in the same out-of-plane direction within layers and in the opposite directions between adjacent layers below $T_N =$ 45 K \cite{huang2017layer}. CrI$_3$ illustrates the fact that rhombohedral/monoclinic stacking leads to FM/AFM interlayer exchange coupling, which was further confirmed via theoretical calculations \cite{sivadas2018stacking} and other experiments \cite{li2019pressure, song2019switching}. For CrBr$_3$, the bulk and the exfoliated few-layer form a FM order with all spins aligned along the same out-of-plane direction below $T_C =$ 37 K and $\sim$32 K \cite{sun2021magnetic, kim2019tailored}, respectively. In scanning tunneling microscopy (STM) study of MBE-grown CrBr$_3$ \cite{chen2019direct}, bilayer CrBr$_3$ could have both R-stacking, same as exfoliated bilayers, and H-stacking, 180$^\mathrm{o}$ rotated between the two layers, which realize an FM and a layered AFM state, respectively. Different from CrI$_3$, CrBr$_3$ demonstrates that the R/H-stacking results in an FM/AFM interlayer exchange coupling. For CrCl$_3$, both the bulk and the few-layer cases host the layered AFM state where the spins are aligned along the in-plane direction ferromagnetically within the layers and antiferromagnetically between the layers \cite{klein2019enhancement}. Despite the same magnetic ground state, the few-layer CrCl$_3$ has shown a substantial enhancement of interlayer exchange coupling as compared to the bulk counterpart \cite{klein2019enhancement}. The relationship between interlayer magnetism and lattice structure is shown in Figure \ref{spin lattice coupling}a.

\textbf{Intralayer magnetism vs interlayer stacking geometry} NiPS$_3$ and FePS$_3$ are the platforms for investigating the coupling between zigzag AFM and interlayer stacking. From the structural perspective, the transition metal ions, Ni$^{3+}$ and Fe$^{3+}$, arrange into the honeycomb lattice within the layer, and further stack in the monoclinic geometry to form few-layers and bulk NiPS$_3$ and FePS$_3$\cite{lee2016ising, kim2019suppression}. Due to the triangular lattice of NiPS$_3$ and FePS$_3$ monolayers, three degenerate monoclinic stacking domain states are expected in their thicker forms, as shown in Figure \ref{spin lattice coupling}b. 

From the magnetic point of view, NiPS$_3$ and FePS$_3$ both feature a zigzag AFM order where the spins align ferromagnetically within the zigzag chains and antiferromagnetically between adjacent chains, with a distinction that the spins orient in-plane for NiPS$_3$ and out-of-plane for FePS$_3$ \cite{brec1986review}. Because of the honeycomb lattice of the Ni$^{3+}$ and Fe$^{3+}$ ion sites, one would naturally expect three zigzag orientations that are 120$^\mathrm{o}$ rotated from one another. It is until very recently that the zigzag chain direction is directly shown to be locked to the monoclinic interlayer stacking geometry in few-layer cases as shown in Figure \ref{spin lattice coupling}b \cite{zhou2022dynamical, kim2023anisotropic}. 

Furthermore, the zigzag AFM in FePS$_3$ has been shown to couple with the interlayer shear motion, where the shear mode amplitude increases by more than 30 times upon cooling below the N\'eel temperature \cite{zong2023spin}. The visualization of this motion in the real space corresponds to a macroscopic mechanical motion where micro-patches of FePS$_3$ films oscillate coherently with a locked-phase and along a common axis. The temperature dependence of this shear mode correlates with that of the demagnetization of the zigzag AFM order after photoexcitation, confirming the conversion between the magnetic and elastic energy in FePS$_3$. Additional investigations of the harmonics of the acoustic shear mode reveals the distinction between even and odd harmonics and shows the transition from out-of-plane shear to the in-plane traveling wave across the N\'eel temperature \cite{zhou2023ultrafast}.

\textbf{Interlayer magnetism vs intralayer lattice structure} CrSBr is the first, and so far, the only vdW magnet for investigating the coupling between interlayer magnetism and the in-plane lattice structure \cite{cenker2022reversible}. CrSBr has an orthorhombic lattice and realizes the layered AFM order \cite{goser1990magnetic, telford2020layered}. For even-/odd-layer CrSBr flakes, its crystalline structure is centrosymmetric whereas its magnetic order breaks/preserves the inversion symmetry \cite{lee2021magnetic}. It has been realized very recently in a 20nm-thick CrSBr flake that an in-plane uniaxial strain along the $a$ axis can introduce a layered AFM to FM phase transition without the application of an external magnetic field, as illustrated in Figure \ref{spin lattice coupling}c \cite{cenker2022reversible}. From symmetry analysis, it is intuitively unexpected that the strain can couple directly with the magnetism. The density functional theory (DFT) calculations reveal the fact that this uniaxial strain tunes the in-plane lattice constant of CrSBr layers and affects the exchange pathways between Cr ions in adjacent layers, without altering the interlayer stacking geometry. 

In addition to the modification of interlayer exchange coupling, tuning the in-plane lattice structure via uniaxial strain was also found to affect the magnetic anisotropy. The experimental observation includes the significant reduction of the saturation magnetic field and the switch into the linear magnetic field dependence for the spin flop process under an out-of-plane magnetic field \cite{cenker2022reversible}. The DFT calculations suggest the suppression of the magnetic anisotropy and the change from the higher-order to quadratic anisotropy under the tensile strain along the $a$ axis \cite{cenker2022reversible}. Furthermore, theory calculations also predict the control of critical temperatures in CrSBr by applying uniaxial strains \cite{esteras2022magnon}. 

\subsubsection{The spin-charge coupling in two-dimensional magnets}
In 2D vdW magnets, two forms of the charge degree of freedom have been considered when discussing the coupling between spin and charge degrees of freedom. First, it refers to the electronic band structure, the charged single-particle excitation spectrum. The emergence of magnetic orders selectively breaks symmetries in vdW magnets, propagating the different impacts on the electronic properties. In light of the recent interest in altermagnetism \cite{vsmejkal2022emerging}, magnetic topological materials \cite{bernevig2022progress}, and kagome magnets \cite{yin2022topological}, this type of spin-charge coupling is of particular importance. Second, it discusses the exciton, the electron-hole two-particle excitation that is present in many 2D vdW semiconductors. The magnetic order and the magnon excitation can modify the electronic band structure statically and dynamically, which further impacts the binding energy of electron-hole pairs, excitons. Considering the prospects of excitons in 2D semiconductors \cite{mueller2018exciton}, this second type of spin-charge coupling is of unique interest. 

\begin{figure}[th]
\begin{center}
\includegraphics[width=1.0\textwidth]{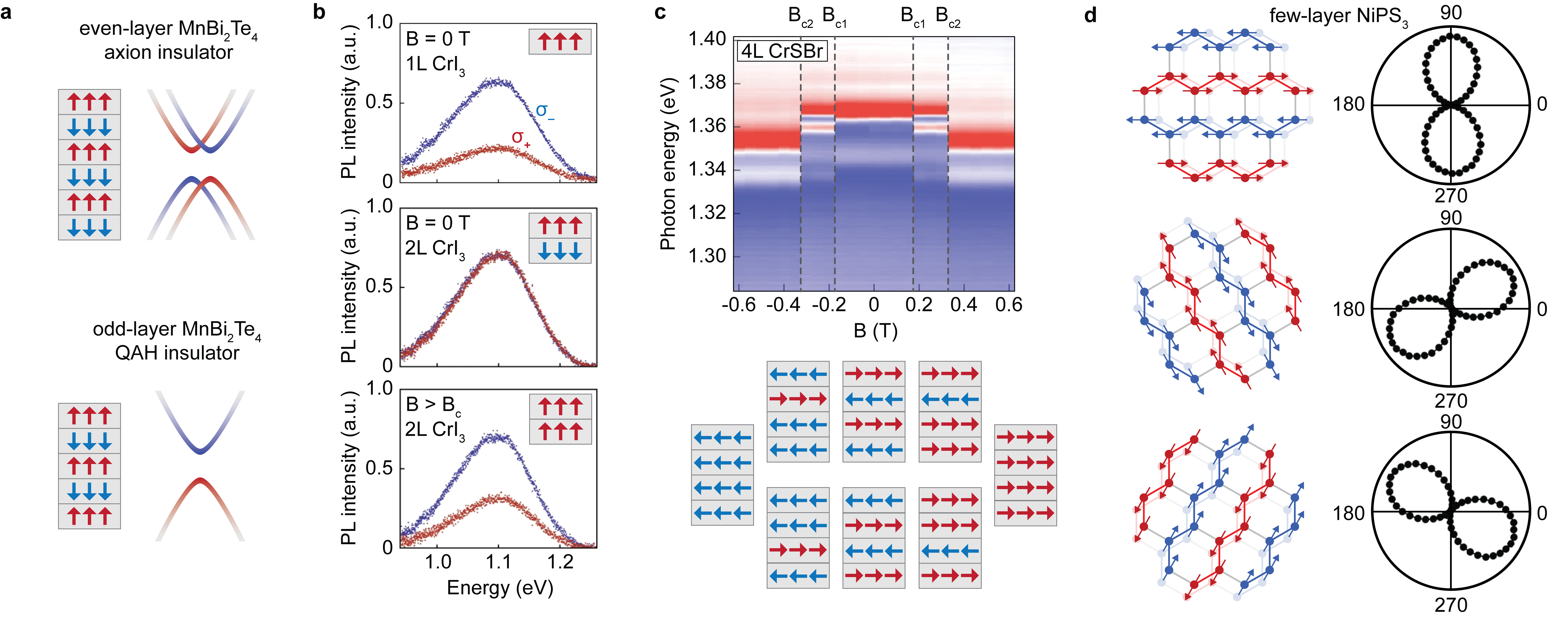}\caption{\small Examples to show the spin-charge coupling in 2D vdW magnets. (a) The layered AFM order determining the electronic bands, e.g., even-layer and odd-layer MnBi$_2$Te$_4$ corresponding to the axion insulator and quantum hall insulator phase, respectively; (b) The layered magnetism influencing the photoluminescence (PL) polarization, e.g., monolayer CrI$_3$ and bilayer CrI$_3$ above B$_\mathrm{c}$ showing circular dichroism in PL whereas bilayer CrI$_3$ below B$_\mathrm{c}$ does not; (c) The layered magnetism affecting the PL frequency, e.g., four-layer CrSBr showing PL frequency shifts across spin-flip transitions; (d) The intralayer zigzag magnetism deciding the PL linear polarization direction, e.g., few-layer NiPS$_3$ showing PL linear polarization locked to the zigzag direction. Figures (a), (b), (c), and (d) are adapted from Refs. \cite{gao2021layer}, \cite{seyler2018ligand}, \cite{wilson2021interlayer}, and \cite{hwangbo2021highly}, respectively.} 
\label{spin charge coupling}
\end{center}
\end{figure}

\textbf{Magnetism vs electronic bands} The coupling between magnetism and electronic band structure is of timely interest in several subfields of quantum materials research, including altermagnetism \cite{vsmejkal2022emerging}, magnetic topological materials \cite{bernevig2022progress}, and kagome magnets \cite{yin2022topological}. In the research of 2D vdW magnets, MnBi$_2$Te$_4$ is one outstanding example to illustrate the coupling between magnetism and electronic bands. MnBi$_2$Te$_4$ hosts the layered AFM order with spins aligned along the out-of-plane direction. Even-layer MnBi$_2$Te$_4$ is $\mathcal{PT}$-symmetric, which breaks both time-reversal and spatial-inversion symmetries but preserves the product of them with zero net magnetization. Odd-layer MnBi$_2$Te$_4$ breaks T symmetry but preserves P symmetry with a non-zero net magnetization. Based on their distinct magnetic properties and symmetries, odd-layer MnBi$_2$Te$_4$ is confirmed to realize the quantum anomalous Hall insulator phase \cite{deng2020quantum} whereas even-layer MnBi$_2$Te$_4$ is known to host the axion insulator phase \cite{gao2021layer,qiu2023axion,gao2023quantum} (summarized in Figure \ref{spin charge coupling}a). 

In addition to even/odd-layer MnBi$_2$Te$_4$,  Nb$_3$X$_8$ (X=Cl, Br, and I) are another family of 2D vdW platforms for investigating the interplay between magnetism and electronic bands. Nb$_3$X$_8$ forms a kagome lattice whose electronic band structure features Dirac dispersion $\textbf{E}(\Vec{k})$ at the Brillouin zone corner K points, van Hove singularities at the M points, and a flat band across the entire Brillouin zone \cite{sun2022observation}. The predicted FM long-range order in Nb$_3$X$_8$ is expected to lift the Kramers' degeneracy and, therefore, cause the energy band splitting in a similar manner as the recently reported case in FeGe \cite{teng2023magnetism}, giving the possibility of tuning the energy separation between the Fermi level and the three characteristic electronic features listed above. 

\textbf{Magnetism and magnons vs excitons} Exciton consists of a pair of bounded electron and hole, whose binding energy is sufficiently increased in 2D semiconductors due to the lack of electronic screening in atomic layers, such as in monolayer TMDCs \cite{wilson2021excitons}. 2D magnetic semiconductors are another class of materials that realized excitons with strong binding energies, such as CrI$_3$ \cite{seyler2018ligand}, CrSBr \cite{wilson2021interlayer}, and NiPS$_3$ \cite{kang2020coherent}. The presence of the spin degree of freedom adds further tunability to the excitons in 2D magnets. 

In CrI$_3$, the differential reflectance spectroscopy shows three peak features, a weak peak around 1.5 eV and two stronger ones at 2.0 eV and 2.7 eV. The photoluminescence spectroscopy exhibits one weak peak around 1.1 eV \cite{seyler2018ligand}. The two stronger peaks at 2.0 eV and 2.7 eV in the differential reflectance spectra are attributed to the dipole-allowed transition metal Cr$^{3+}$ $d$ orbital to ligand I$^{-}$ $p$ orbital transitions. The weak features at 1.5 eV in the differential reflectance spectra and at 1.1 eV in the photoluminescence spectra are from the excitons of the same $d-d$ transition in the ligand crystal fields, and the 400 meV Stoke shift between the reflectance and photoluminescence peaks results from Franck–Condon principle and strong electron–lattice coupling. The photoluminescence spectra show a strong dependence on the magnetic ground states, showing strong circular dichroism in the layered AFM state of odd-layer CrI$_3$ with a net magnetization but none in the even-layer case (Figure \ref{spin charge coupling}b). Moreover, the strong electron-lattice coupling further manifests as phonon-dressed exciton replicas, up to the eighth order, in the electronic Raman spectroscopy \cite{jin2020observation}, and this electron-phonon coupling is found to be enhanced by more than 50$\%$ with the layered AFM order in few-layer CrI$_3$.

In CrSBr, the differential reflectance and the photoluminescence spectra show an absorption peak at $\sim$1.36 eV and a photoluminescence peak at $\sim$1.28 eV, corresponding to the lowest excitonic excitation across the electronic band edges (i.e., between valence band maximum at the $\Gamma$ point and conductance band minimum degenerate at the $\Gamma$ and $\mathrm{X}$ points) \cite{lee2021magnetic,wilson2021interlayer}. This exciton absorption exhibits a strong anisotropy, which is only observable when the linear light polarization is aligned along the $b$ axis, and depends on the layered AFM order, where the exciton absorption energy clearly redshifts across the spin-flip transition and the spin-flop process (Figure \ref{spin charge coupling}c) \cite{wilson2021interlayer}. The linear polarization of the exciton mode is due to the nearly one-dimensional electronic band in the conduction band, i.e., with dispersion along $\Gamma-\mathrm{Y}$ and a nearly flat band along $\Gamma-\mathrm{X}$. The frequency redshift of the exciton mode from the layered AFM to the FM states is due to the electronic band reduction from the layered AFM phase to the FM order, the former of which forbids the electronic tunneling between adjacent layers with opposite spin alignments whereas the latter allows. Moreover, this exciton frequency is further periodically modulated when coherent magnons are launched in the layered AFM phase, showing an intimate exciton-magnon coupling despite their drastically different energy scales \cite{bae2022exciton}.

In NiPS$_3$, an exceptionally sharp photoluminescence peak at 1.4756 eV, i.e., with a full width at half maximum being $\sim$350 $\mu$eV, has been reported below the N\'eel temperature of 155 K (Figure \ref{spin charge coupling}d) \cite{kang2020coherent,hwangbo2021highly,wang2021spin,kim2023anisotropic}. The origin of this extremely narrow exciton photoluminescence spectrum is currently debated between Zhang-Rice singlet-related \cite{kang2020coherent} and magnetic defect-induced \cite{kim2023anisotropic}. Despite the debate, this feature vanishes either when the temperature is above the N\'eel temperature or when the NiPS$_3$ film is thinned down to the 2D limit without the long-range AFM order, suggesting its close relationship with the magnetic properties of NiPS$_3$. Moreover, this exciton peak is highly anisotropic with a high linear polarization that is aligned normal to the spin zig-zag chain direction (Figure \ref{spin charge coupling}d), which further establishes the relationship between spins and excitons. Furthermore, there is strong electron-lattice coupling in NiPS$_3$ as manifested by the phonon-dressed exciton replica states in frequency-resolved linear dichroism spectra \cite{hwangbo2021highly}, time-resolved reflectance spectra \cite{ergeccen2022magnetically}, and photoluminescence spectra \cite{kim2023anisotropic}.

\subsection{Developing controls over two-dimensional magnets}
Controlling magnetic orders and spin waves lies at the heart of spintronics and magnonics research \cite{mak2019probing}. Traditionally, FM orders are primarily controlled by the magnetic field due to the symmetry-allowed direct coupling between the net magnetization $\Vec{M}$ and the $\Vec{B}$ field. In contrast, due to the absence of net magnetization in AFM orders, it has always been challenging to control AFM with the external magnetic field. Indirect controls, which rely on magneto-electric effect \cite{fiebig2005revival}, magneto-optical effect \cite{haider2017review}, etc., have been explored in materials such as multiferroics, magnetoelectrics, and others.  

Thanks to the thinness of 2D vdW magnets, diverse efficient and flexible controls over 2D magnetism have been proposed and applied. Examples so far include electric field, carrier doping, uniaxial strain, and electromagnetic wave irradiation that are specifically powerful in the 2D limit. In addition to magnetic fields, stoichiometry and pressure with accumulated experiences in the 3D case can also be extended into the 2D regime. 

As of now, the practice of these controls over the static magnetic orders has been carried out in a few 2D vdW magnets, including CrI$_3$, CrSBr, NiPS$_3$, MnBi$_2$Te$_4$, etc. In contrast, the manipulation of the dynamic spin waves has been rarely explored, primarily because the study of 2D spin waves is quite limited as discussed in Section 3.2 above.

\begin{figure}[th]
\begin{center}
\includegraphics[width=1.0\textwidth]{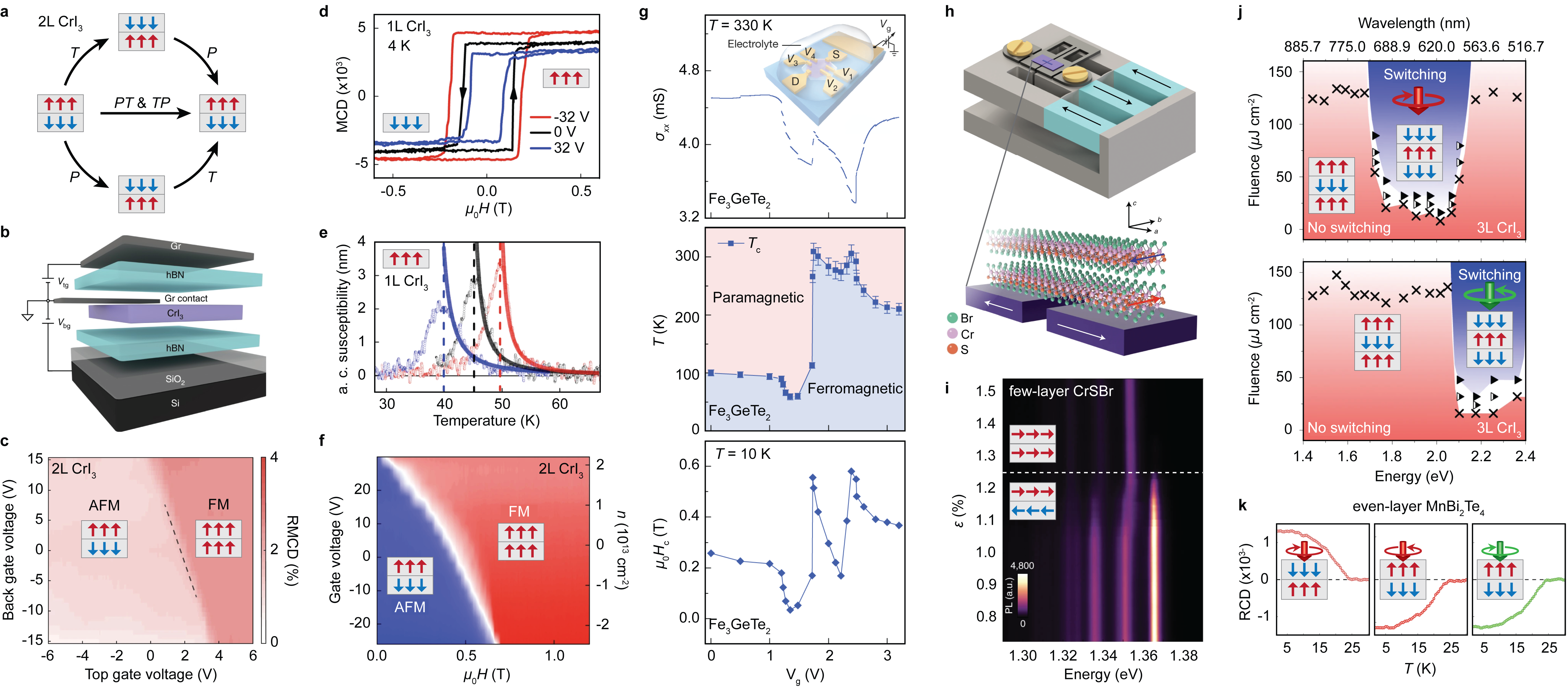}\caption{\small Examples of diverse controls over 2D vdW magnetism. (a) Illustration of $\mathcal{PT}$-symmetric layered AFM order that breaks time reversal ($\mathcal{T}$) and spatial inversion ($\mathcal{T}$) symmetries but preserves the product of them. Examples of such include even-layer CrI$_3$ and MnBi$_2$Te$_4$; (b) Sketch of a dual gate device of CrI$_3$ that tunes the displacive electric fields and charge carrier density independently; (c) Out-of-plane electric field switches the layered AFM to FM for bilayer CrI$_3$; (d-e) Carrier doping tunes the coercive fields and magnetic onset temperature for monolayer CrI$_3$; (f) Carrier doping also introduces spin-flip transitions in bilayer CrI$_3$; (g) Ionic gating dramatically changes itinerant FM in Fe$_3$GeTe$_2$; (h-i) Strain engineering introduces the spin-flip transition in few-layer CrSBr; (j) Circularly polarized light switches the layered AFM order between two time-reversal-related degenerated ground states in odd-layer CrI$_3$; (k) Circularly polarized light induces the layered AFM order in even-layer MnBi$_2$Te$_4$. Figures (a), (b-c), (d-f), (g), (h-i), (j), and (k) are adapted from Refs. \cite{sun2019giant}, \cite{huang2018electrical}, \cite{jiang2018controlling}, \cite{deng2018gate}, \cite{cenker2022reversible}, \cite{zhang2022all}, and \cite{qiu2023axion}, respectively.}
\label{control 2d magnetism}
\end{center}
\end{figure}

\subsubsection{Electric field control over parity-odd two-dimensional magnetism}

Electric field, $\Vec{E}$, is a polar vector field with an odd parity under the spatial inversion symmetry operation. It is popularly used to control order parameters with the same symmetry properties, being polar and parity-odd, for example, ferroelectricity. In fact, there are AFM orders with broken spatial inversion symmetry in both 3D bulk and 2D film forms. Such parity-odd AFM states are expected to show magnetoelectric effects by symmetry and allow for the control of the magnetic states with external electric fields. For the same applied voltage ($V$), the effective electric field inside the materials is much strong in 2D than in 3D (i.e., $\Vec{E} = \nabla V$) and, therefore, is expected to be much more efficient in driving the electric field-induced magnetic phase transitions. 

Even-layer CrI$_3$ and MnBi$_2$Te$_4$ are two examples of layered AFMs that break both spatial inversion and time-reversal symmetries, but preserve the product of the two (Figure \ref{control 2d magnetism}a). Dual gate electric devices, with both top and back gatings, were designed to apply an out-of-plane electric field without introducing charge carrier doping (Figure \ref{control 2d magnetism}b). Through experiments over bilayer CrI$_3$ \cite{jiang2018electric,huang2018electrical}, it was found that the application of this electric field introduces an interlayer potential difference between the two layers, leads to a change in the interlayer exchange coupling, and eventually switches the layered AFM state into the FM one (Figure \ref{control 2d magnetism}c). On the other hand, the experiment over even-layer MnBi$_2$Te$_4$ \cite{gao2021layer} shows that a combination of electric and magnetic fields, i.e., $\Vec{E}\cdot\Vec{B}$, produces an axion field that couples linearly to the axion insulating state. By sweeping either the electric or the magnetic field, a phase transition from the layered AFM state to the FM state can be achieved.

In addition to the layered AFM state mentioned above, the N\'eel orders in the honeycomb lattice, such as MnPS$_3$ \cite{long2020persistence} and MnPSe$_3$ \cite{ni2021imaging}, are another type of AFM with broken spatial inversion and time-reversal symmetry. Yet, little about the electric control over these N\'eel orders has been reported thus far. 

\subsubsection{Carrier doping control over two-dimensional magnetism}

The application of gating for introducing charge carriers to 2D vdW magnetic materials provides another knob for tuning 2D magnetism. Different from the electric field control that is limited to parity-odd magnets, carrier doping can, in principle, be applied to all 2D magnets. So far, electrostatic gating and ionic gating have been used to study 2D semiconducting (e.g., monolayer and bilayer CrI$_3$ \cite{jiang2018controlling}) and metallic (e.g., Fe$_3$GeTe$_2$ \cite{deng2018gate}) magnets, respectively, by introducing sufficient carrier density to modify their magnetic properties. Usually, electrostatic gating with dielectric layers such as hBN can introduce a carrier concentration on the order of $10^{13}$ cm$^{-2}$, whereas ionic gating with electrolyte can induce even higher carrier densities. 

In monolayer CrI$_3$, the electrostatic gating (without a net electric field via the dual-gate geometry) was shown to enhance/suppress the coercive field, increase/decrease the saturation moment, and raise/reduce the Curie temperature through hole/electron doping (Figures \ref{control 2d magnetism}d and e). In bilayer CrI$_3$, carrier doping significantly changes the spin-flip transition field $H_c$ from that of the intrinsic bilayer, 0.5 T, enhanced/suppressed on the hole/electron doping side. Remarkably, a transition from the layered AFM to FM state was established across a critical electron carrier density, $n_c \sim 2.5 \times 10^{13}$ $\mathrm{cm}^{-2}$ (Figure \ref{control 2d magnetism}f), above which the critical field $H_c$ approaches to a small value for the coercive field of the FM state. The results in both monolayer and bilayer CrI$_3$ demonstrate the capability of using electrostatic doping to modify intralayer and interlayer magnetic exchange coupling.

In few-layer Fe$_3$GeTe$_2$, 2D Mermin-Wagner fluctuations suppress the Curie temperature monotonically from $\sim$180 K in fifty-layer to $\sim$100 K in three-layer to $\sim$20 K in monolayer \cite{deng2018gate}. The ionic gating ($V_g$) of a trilayer Fe$_3$GeTe$_2$ flake shows a non-monotonic change of its Curie temperature, decreasing from $\sim$100 K to $\sim$50 K at $V_g=1.5$ $\mathrm{V}$ first, then increasing abruptly to $\sim$300K at $V_g=1.9$ $\mathrm{V}$, exhibiting minor non-monotonic variations around $\sim$280 K till $V_g=2.5$ $\mathrm{V}$, and eventually dropping down to $\sim$200 K at $V_g=3.2$ $\mathrm{V}$ (Figure \ref{control 2d magnetism}g). This ionic gating to drastically tune the Curie temperature can be explained by the Stoner model whose criteria suggest the formation of FM depending on the density of states at the Fermi level. The extreme electron doping through the ionic gating shifts the Fermi level inside the electronic bands of few-layer Fe$_3$GeTe$_3$, and therefore leads to the large variation of density of states at the Fermi level that is responsible for the dramatic changes in FM properties (e.g., magnetic onset temperature, magnetic coercive field, etc.).

\subsubsection{Uniaxial strain control over two-dimensional magnetism}

Magnetic exchange couplings and single-ion anisotropy closely depend on the crystalline lattice. The application of a uniaxial strain modifies the lattice constant, induces ligand field distortions, shifts the relative positions between metal and ligand atoms, etc., all of which can contribute to the change of both magnetic exchange coupling and single-ion anisotropy. In 3D bulk, the strain-induced lattice changes are typically on the order of 0.1$\%$ before the mechanical breakdown of the crystals. In 2D films, it has been demonstrated in various systems, including 2D magnets, that large strain, on the order of 1$\%$, can be achieved while retaining the crystalline integrity. It has been shown in few-layer CrSBr \cite{cenker2022reversible} and few-layer MnPSe$_3$ \cite{ni2021imaging} that magnetic orders can be switched by external strain.

Few-layer CrSBr hosts the layered AFM order with a biaxial anisotropy, where the spins orient along the $b$ axis ferromagnetically within the layers and antiferromagnetically between the adjacent layers. Using a strain cell made of three parallel piezo stacks glued to a titanium framework, a tensile uniaxial strain up to $\sim$2$\%$ can be applied to the CrSBr flake that is placed at the gap of the strain cell (Figure \ref{control 2d magnetism}h). A magnetic phase transition from a layered AFM to FM order happens at a critical strain of $\sim$1.2$\%$ applied along the $a$ axis (Figure \ref{control 2d magnetism}i). This magnetic phase transition is not intuitive through the symmetry-based analysis as the applied uniaxial strain does not share the same symmetry as the layered AFM/FM order. Theoretical calculations demonstrated that the applied strain changes the CrSBr lattice constant and further impacts the interlayer exchange pathways, introducing this observed magnetic phase transition. 

Few-layer MnPSe$_3$ has an Ising-type N\'eel order parameter with a strong XY-type spin anisotropy, for which the spins align along the same direction for each sublattice and in the opposite directions between the two sublattices of the Mn honeycomb lattice. By stretching the PDMS stamp where the few-layer MnPSe$_3$ film was attached, an uniaxial tensile strain as large as 2$\%$ was applied to the MnPSe$_3$ sample. The N\'eel vector was shown to reorient along the strain direction whereas the N\'eel temperature remained nearly the same under various strains. The ability to rotate the N\'eel vector by the uniaxial strain is due to the in-plane spin isotropy within few-layer MnPSe$_3$. The insensitivity of the N\'eel temperature to the applied strain is because the XY-type anisotropy is much stronger than the strain-induced in-plane anisotropy.

\subsubsection{Electromagnetic wave irradiation control over two-dimensional magnetism}

Electromagnetic wave from a laser can act as the symmetry-breaking field to switch magnetic orders, because circularly polarized light effectively breaks the time-reversal symmetry that couples with magnetism. While the approach has been extensively used in 3D bulk materials with substantial success, its application to 2D films has unique advantages. First, as the 2D flake is thinner than the penetration depth of the control and the probe light, there are no concerns arising from the depth mismatch that commonly happens in similar studies of 3D bulk materials. Second, because the entire thickness of the sample is illuminated by light, it eliminates the challenges created by the pinning field from the surrounding unilluminated region that is ubiquitously present in 3D bulk cases. The application of circularly polarized light to switching 2D magnetism has been demonstrated in odd-layer CrI$_3$ \cite{zhang2022all} with an uncompensated nonzero magnetization and even-layer MnBi$_2$Te$_4$ \cite{qiu2023axion} with fully compensated zero magnetization. 

Odd-layer CrI$_3$, such as three-layer CrI$_3$, has an uncompensated net magnetization in the layered AFM state. Using circularly polarized light pulses, it has been shown recently that this net magnetization, as well as the layered AFM ground state, can be switched between the two time reversal-related domain states \cite{zhang2022all}. Moreover, the switch of the magnetization is further shown to depend on the excitation photon energy and polarization. For a prepared layered AFM state with a net magnetization aligned along the out-of-plane direction for a three-layer CrI$_3$ flake, using $\sigma_{-}$ (right-handed circular) polarization, the optical switch to the down magnetization state happens at photon frequencies between 1.7 -- 2.1 eV, whereas using $\sigma_{+}$ (left-handed circular)polarization, the active frequency range shifts to 2.1 -- 2.4 eV (Figure \ref{control 2d magnetism}j). The photon frequency and helicity dependence of this observed optical switch rules out the inverse Faraday effect mechanism that should maintain the same light helicity across all frequencies. The absence of thermal demagnetization at fluences much higher than the critical fluence for the switching excludes the magnetic circular dichroism mechanism for which thermal demagnetization is expected. The proposed mechanism for the optical switch in odd-layer CrI$_3$ is spin angular momentum transfer from the photoexcited carriers to local magnetic moments.

Even-layer MnBi$_2$Te$_4$ has fully compensated zero magnetization and realizes the axion insulating ground state. Despite the absence of magnetization, it has been shown lately that this compensated layered AFM state can be induced and switched using continuous wave (CW) and pulsed circularly polarized light, respectively \cite{qiu2023axion}. It was observed that both the optical induction and switch depend on the photon frequencies. For the optical induction of a six-layer MnBi$_2$Te$_4$, $\sigma_{+}$ light above and below $\sim$600nm leads to two opposite layered AFM states that are related by the time-reversal operation (Figures \ref{control 2d magnetism}k). This helicity-dependent optical induction and switch are explained by the axion field ($\Vec{E}\cdot\Vec{B}$) created by the circularly polarized light ($(\Vec{E^{*}} \times \Vec{E})\cdot \hat{z}$), in which the rotating electric field in the circularly polarized light (i.e., $\Vec{E^{*}} \times \Vec{E}$) serves as an effective magnetic field $B_z$ and the surface normal (i.e., $\hat{z}$) acts as an effective electric field $E_z$.  

\subsubsection{Other controls over two-dimensional magnetism}
External magnetic field, stoichiometry, and pressure are control knobs routinely used in 3D bulk magnets and have recently been applied to control 2D magnetism. The external magnetic field has been used to induce spin-flip transitions in layered AFMs such as CrI$_3$ \cite{huang2017layer}, CrCl$_3$ \cite{wang2019determining}, MnBi$_2$Te$_4$ \cite{deng2020quantum}, and CrSBr \cite{telford2020layered, telford2022coupling} when applied along the N\'eel vector direction, spin-flop processes in these aforementioned layered AFMs when applied along directions orthogonal to the N\'eel vector, and N\'eel vector reorientation in XY AFMs, such as NiPS$_3$ \cite{wang2021spin} when applied in the easy plane but in a different direction from the N\'eel vector. Chemical composition has been tuned to adjust the magnetic anisotropy in CrCl$_{3-x-y}$Br$_x$I$_y$ \cite{tartaglia2020accessing}, modify the interlayer exchange coupling in MnBi$_{2n}$Te$_{3n+1}$ ($n=$1,2,3,4) \cite{hu2020realization}, and enhance the magnetic onset temperature in Fe$_n$GeTe$_2$ ($3 \le n \le 5$) \cite{seo2020nearly}. Hydrostatic pressure has been applied to vary the interlayer stacking geometry and hence the interlayer magnetic exchange coupling in CrI$_3$ \cite{li2019pressure, song2019switching}. 

\section{Experimental methods to study two-dimensional magnetism}

Up to now, the most popular way of achieving 2D vdW magnets is via mechanical exfoliation which limits the lateral dimensions to be in the order of $~10$ $\mu$m. Due to the small volume sizes of such peeled 2D vdW magnets ($\mu$m in length and nm in thickness), powerful techniques for studying 3D magnetism, such as X-ray and neutron scattering, bulk magneto-transport, etc., face critical challenges in 2D magnetism. 
In this chapter, we survey the experimental tools to probe 2D magnetism, including sensitive optical probes, magneto-transport of nanodevices, and magnetic microscopy, and highlight their main achievements in the research of 2D magnetism.

\subsection{Magneto-optical spectroscopy to investigate two-dimensional magnetism}
Optical spectroscopy techniques are non-invasive and contactless tools for studying 2D materials, including 2D magnets. First, the optical light can be focused down to the optical diffraction limit, in the order of $\mu$m, which is sufficiently smaller than mechanically exfoliated 2D magnet flakes. Second, for 2D magnetic semiconductors, there are optical resonances that can significantly enhance the signal-to-noise ratio and, therefore, achieve sufficient sensitivity for studying small volume-sized magnets. Third, the recent development in the detection scheme of many magneto-optical techniques further pushes the experimental sensitivity that is instrumental in the research of 2D magnetism. 

\subsubsection{Photoluminescence and reflectance/absorbance for studying two-dimensional magnetism}
2D semiconducting magnets are a large family of 2D vdW magnets, in which the electronic band gap is typically in the order of 1 eV, and the excitonic binding energy is at the scale of 100 meV. The presence of such excitons in the visible -- near-infrared (nIR) frequency range leads to peak features at visible -- nIR frequencies in the photoluminescence and absorbance, or equivalently, differential reflectance spectra for thin films. Examples include but are not limited to CrX$_3$ \cite{seyler2018ligand,jin2020observation}, Ni/FePS$_3$ \cite{kang2020coherent, wang2021spin}, and CrSBr \cite{wilson2021interlayer}.  

\begin{figure}[th]
\begin{center}
\includegraphics[width=1.0\textwidth]{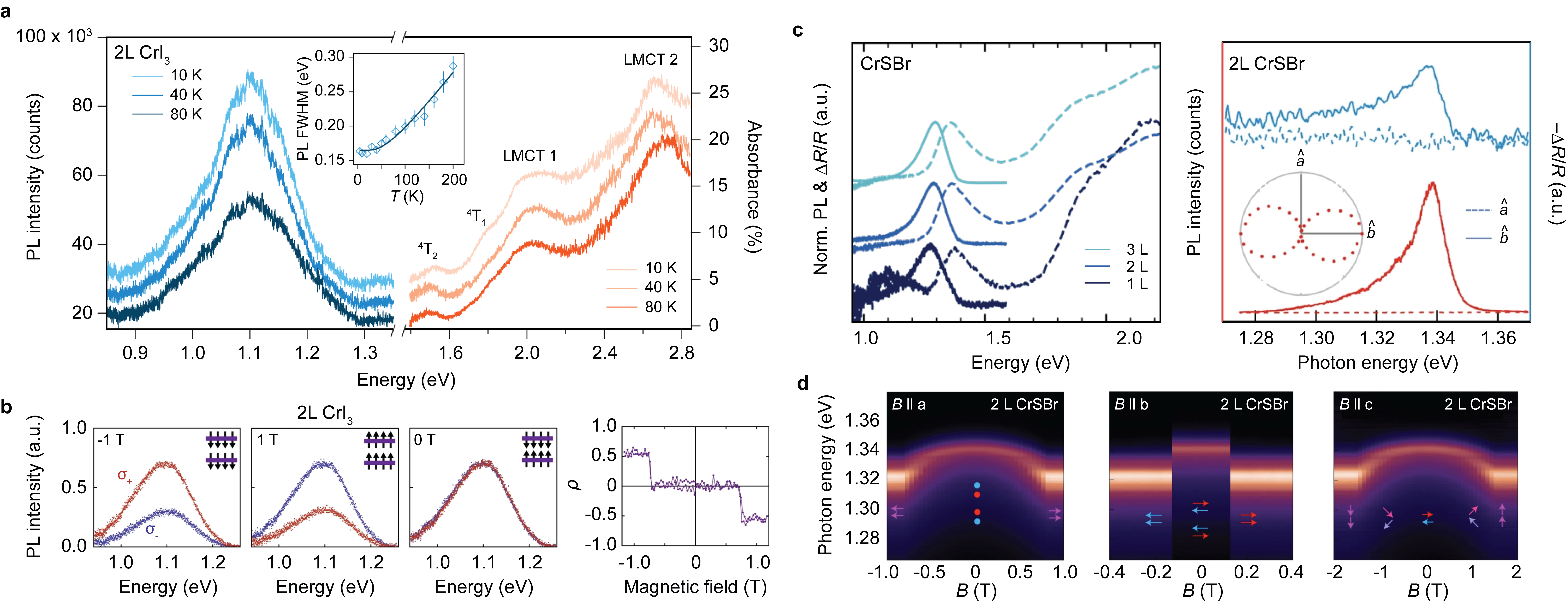}\caption{\small Examples of photoluminescence (PL) and absorption (or differential reflection) spectroscopy used in the research of 2D vdW magnetism. (a) PL (blue) and absorption (orange) spectra for bilayer CrI$_3$ at three selected temperatures; Inset showing the temperature dependence of PL full width at half maximum (FWHM); (b) PL spectra of bilayer CrI$_3$ in the two circularly polarized channels, $\sigma_+$ and $\sigma_-$, at three magnetic states, FM with spin up, FM with spin down, and layered AFM; The last figure showing the magnetic field dependence of the PL degree of circular polarization; (c) PL and absorption spectra for monolayer, bilayer, and trilayer CrSBr (left), and linear polarization dependent PL spectra for bilayer CrSBr (right); (d) Magnetic field dependence of PL spectra for four-layer CrSBr with magnetic field applied along a, b, and c crystalline directions. Figures (a), (b), (c), and (d) are adapted from Refs. \cite{jin2020observation}, \cite{seyler2018ligand}, \cite{lee2021magnetic}, and \cite{wilson2021interlayer}, respectively.}
\label{PL}
\end{center}
\end{figure}

Using few-layer CrI$_3$ as an example, its photoluminescence spectra feature a ligand field exciton peak at $\sim$1.1 eV. Its absorption/differential reflectance spectra resolve four peaks at $\sim$ 1.5 eV and $\sim$1.9 eV that result from the ligand field transitions from $^{4}\mathrm{A}_2$ to $^{4}\mathrm{T}_2$ and $^{4}\mathrm{T}_1$, and at $\sim$2.2 eV and $\sim$2.7 eV that originate from the ligand-metal charge transfer transitions (Figure \ref{PL}a) \cite{seyler2018ligand, jin2020observation}. The photoluminescence spectra show strong helicity dependence on the magnetic ground state; in the paramagnetic phase, there is no intensity contrast between $\sigma_+$ and $\sigma_-$ circularly polarized photoluminescence; in the FM state of monolayer CrI$_3$, clear dichroism is observed between $\sigma_+$ and $\sigma_-$ circularly polarized photoluminescence whose relative strength depends on the FM magnetization orientation; in the layered AFM state of bilayer CrI$_3$, equivalent spectra are detected in $\sigma_+$ and $\sigma_-$ circularly polarized photoluminescence channels; and in the layered FM state of bilayer CrI$_3$ above the spin-flip transition field, distinct spectra are shown between $\sigma_+$ and $\sigma_-$ circularly polarized photoluminescence (Figure \ref{PL}b). 

For the cases of NiPS$_3$ and CrSBr where the spins are aligned within the vdW layers, the photoluminescence from their band-edge excitons exhibits a high degree of linear polarization, rather than the circular polarization in the FM state of CrI$_3$. NiPS$_3$ shows an exceptionally narrow exciton peak in the photoluminescence spectra below the N\'eel temperature, whose intensity is the strongest and nearly zero when the polarization is normal and parallel to the zigzag chain direction, respectively (Figure \ref{spin charge coupling}d), due to the strong coupling between the exciton and the zigzag AFM order \cite{zhang2021spin,hwangbo2021highly}. CrSBr also hosts a linearly polarized exciton peak in the photoluminescence spectra, with its polarization aligned along the $b$ axis that is the same as the spin orientation in the magnetic phase (Figure \ref{PL}c). The linear polarization of excitons in CrSBr results from the dispersionless conduction band, instead of the magnetism, but the frequency of the exciton peak shifts between layered AFM and FM states (Figure \ref{PL}d).    

On the other hand, TMDC monolayers are used as a sensor or an enhancer for weak magneto-optical responses because they host strong exciton photoluminescence and absorption. In order to enhance the optical pumping of spin waves in bilayer CrI$_3$ \cite{zhang2020gate}, a WSe$_2$ monolayer was stacked onto bilayer CrI$_3$ in the device, and the optical excitation is chosen to be at the resonance of the WSe$_2$ exciton frequency. In order to optically detect the quantum oscillations in twisted MoTe$_2$ moir\'e superlattices, a high-quality WSe$_2$ monolayer was placed in close proximity to the twisted MoTe$_2$ with a thin hBN spacer in between.

\subsubsection{Magnetic circular dichroism and magneto-optical Kerr effect to study two-dimensional magnetism}
\begin{figure}[th]
\begin{center}
\includegraphics[width=1.0\textwidth]{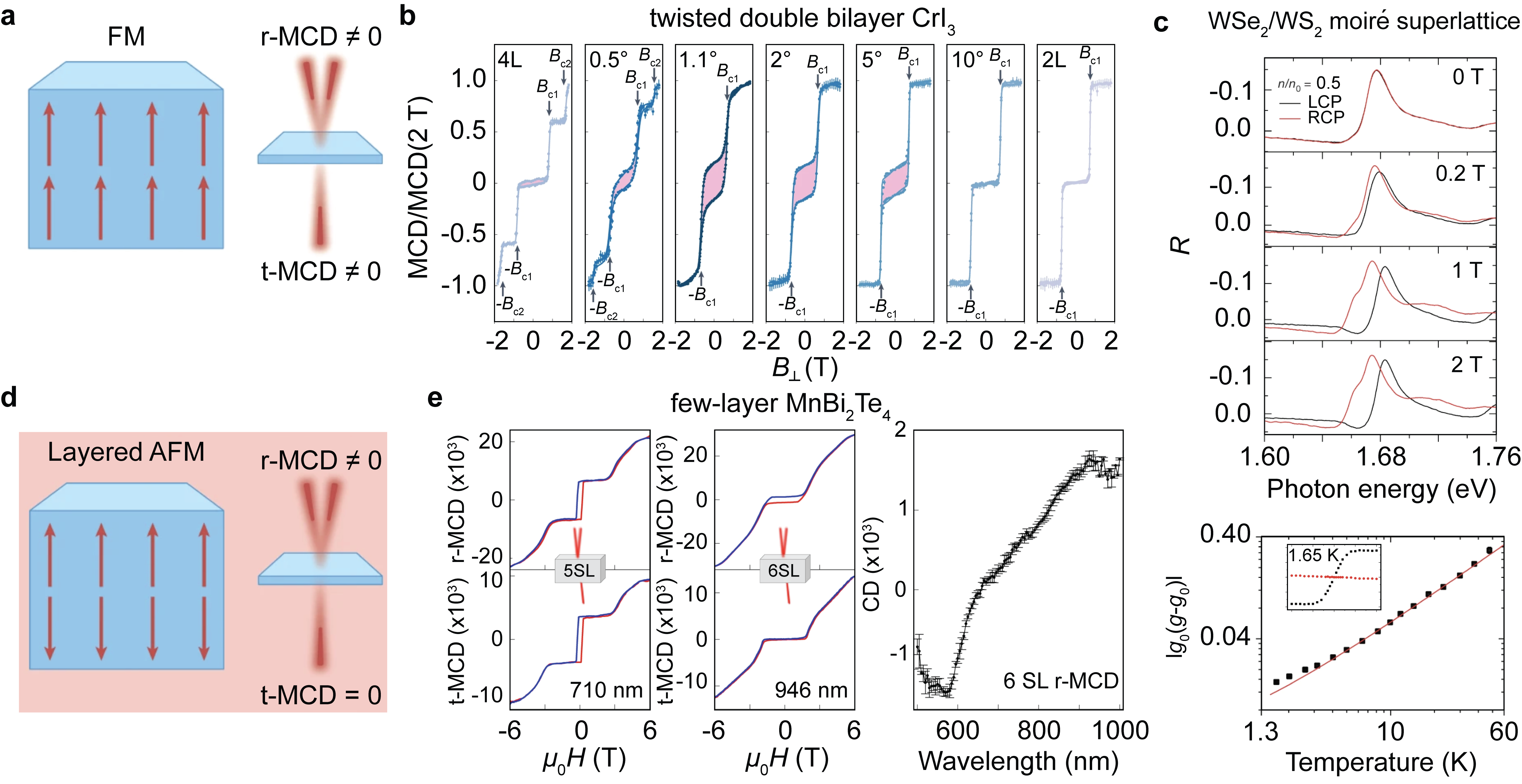}\caption{\small Examples of MCD employled in studying 2D vdW magnetism. (a) A diagram to show the presence of both t-MCD and r-MCD for an FM with a non-zero net magnetization; (b) MCD spectra for four-layer, bilayer, and twisted double bilayer CrI$_3$ at selected twist angles of 0.5$^\mathrm{o}$, 1.1$^\mathrm{o}$, 2.0$^\mathrm{o}$, 5.0$^\mathrm{o}$, and 10.0$^\mathrm{o}$; (c) Helicity-resolved differential reflection spectra for the half-filled ($\nu$ = 1) WSe$_2$/WS$_2$ moir\'e superlattice at selected magnetic fields (top) and the temperature dependence of the exciton valley g-factor (bottom); (d) Diagram to show the presence of r-MCD and the absence of t-MCD for an even-layer layered AFM with a zero magnetization; (e) r-MCD and t-MCD spectra for five-layer and six-layer MnBi$_2$Te$_4$ and the wavelength dependence of r-MCD for six-layer MnBi$_2$Te$_4$. Figures (a) and (d), (b), (c), and (e) are adapted from Refs. \cite{ahn2023flipping}, \cite{xie2023evidence},\cite{tang2020simulation}, and \cite{qiu2023axion}, respectively.}
\label{MCD}
\end{center}
\end{figure}

Magnetic circular dichroism (MCD) and magneto-optical Kerr effect (MOKE) probe the absorption and phase difference, respectively, between $\sigma_{+}$ and $\sigma_{-}$ circularly polarized light. Traditionally, they are thought to probe the presence of a net magnetization ($\Vec{M}=\sum_{\Vec{r}}\Vec{m}(\Vec{r})$, with $\Vec{m}(\Vec{r})$ being the local magnetic moment at site $\Vec{r}$), which shows up as antisymmetric terms in the optical susceptibility tensors, $$\chi_{ij} = \chi_{ij}^{0}|_{\Vec{m}=0}+\sum_{\Vec{r}}\frac{\partial{\chi_{ij}}}{\partial{m_k}}\Big|_{\Vec{m}=0}m_k(\Vec{r})$$ for which the Onsager reciprocal relationship applies, $\chi_{ij}(\Vec{M})=\chi_{ji}(-\Vec{M})$ \cite{rivlin1975electro}. In studying 2D magnets with $\mu$m lateral dimensions, MCD and MOKE are typically carried out in the normal incidence geometry and, therefore, only probe the magnetic moment along the out-of-plane direction. In magnets with net magnetization, MCD and MOKE are present in both the reflection and transmission geometries (i.e., $r$-MCD $\neq$ 0, $t$-MCD $\neq$ 0) (Figure \ref{MCD}a).

MCD and MOKE have been widely used in studying 2D magnetism, including both 2D magnetic atomic crystals and moir\'e superlattices. For 2D magnetic atomic crystals, MCD and MOKE have been applied to capture the spin-flip transitions and spin-flop processes in few-layer CrI$_3$ (bilayer and four-layer CrI$_3$ in Figure \ref{MCD}b) \cite{huang2017layer} and Mn$_2$Te$_4$ \cite{qiu2023axion}, and detect the Curie onset of FM magnetization in Fe$_3$GeTe$_2$ \cite{fei2018two}. For the 2D magnetic moir\'e superlattices, MCD and MOKE have been used to verify the coexistence of FM and AFM states in twisted bilayer and double-trilayer CrI$_3$ \cite{xu2022coexisting,song2021direct,cheng2023electrically}, and reveal the presence of collinear and noncollinear spins in twisted double bilayer CrI$_3$ (Figure \ref{MCD}b) \cite{xie2023evidence}. For the flat band-driven 2D magnetism in twisted TMDC moir\'e superlattices, MCD and MOKE have been used to evaluate the magnetic susceptibility via $\chi_M \propto \frac{\partial{\mathrm{MCD}}}{\partial{B_z}}\Big|_{B_z=0}$ for the Mott insulating state (e.g., $\nu$ = 1) and frustrated magnetic state at fractional fillings (e.g., $\nu$ = 2/3) \cite{tang2023evidence} in WSe$_2$/WS$_2$ (Figure \ref{MCD}c) \cite{tang2020simulation}, and more recently, to confirm the broken time-reversal symmetry in the integer and fractional anomalous Hall states in twisted MoTe$_2$ \cite{cai2023signatures, anderson2023programming}. The MCD and MOKE signal out of 2D magnetic systems such as CrI$_3$ flakes and TMDC moir\'e superlattices are especially strong at wavelengths that match their exciton energies, and on the contrary, could be negligibly small at wavelengths far away from optical resonance. For example, for monolayer and bilayer CrI$_3$, the MOKE signal is giant at 633 nm which hits the charge transfer (\textit{i.e.}, Wannier) exciton resonance, whereas it is undetectable at 718 nm where no optical resonance is present. \cite{huang2017layer,wu2019physical,gibertini2019magnetic}

In addition to this traditional perspective of MCD and MOKE above, a new understanding of them has emerged very recently that they can be present in $\mathcal{PT}$-symmetric AFMs which have zero magnetization, breaks both time-reversal and spatial inversion symmetries, but preserves the product of them. Taking MCD as an example, it is anticipated to show up in the reflection geometry, but not in the transmission geometry (i.e., r-MCD $\neq$ 0, t-MCD $=$ 0), as shown in Figure \ref{MCD}d. 

Examples of $\mathcal{PT}$-symmetric AFMs in 2D vdW magnets include even-layer CrCl$_3$ \cite{wang2019determining}, even-layer CrI$_3$ \cite{sun2019giant}, even-layer CrSBr \cite{lee2021magnetic}, even-layer MnBi$_2$Te$_4$ \cite{gao2021layer,qiu2023axion}, MnPS$_3$ \cite{long2020persistence}, MnPSe$_3$ \cite{ni2021imaging}, etc. Because MCD and MOKE in the normal incidence geometry are only sensitive to the out-of-plane spin moment, only even-layer CrI$_3$, even-layer MnBi$_2$Te$_4$, and MnPS$_3$ are considered possible to show nonzero r-MCD but zero t-MCD. In a recent experiment by Qiu \textit{et al.}, even-layer MnBi$_2$Te$_4$ has been shown to clearly exhibit nonzero/zero r/t-MCD despite its fully compensated zero magnetization (Figure \ref{MCD}e) \cite{qiu2023axion}. It is further noted that the magnitude and the sign of the nonzero r-MCD are strongly dependent on the optical photon energies. For even-layer MnBi$_2$Te$_4$, r-MCD shows a nearly monotonic trend as a function of photon wavelength over the range of 500 nm -- 1000 nm, with a sign flip across the wavelength of $\sim$600 nm (Figure \ref{MCD}e). It is further worth noting that even-layer CrI$_3$, of the same layered AFM order as even-layer MnBi$_2$Te$_4$, shows nearly zero r-MCD (Figure \ref{MCD}b) at a charge transfer resonance wavelength of 633 nm \cite{xie2023evidence}. It remains an open question whether even-layer CrI$_3$ should show r-MCD, and thus further investigations are required.  

Complementary to MCD, polarization-resolved photocurrent measurements have also been applied to study 2D magnets, for example, few-layer CrI$_3$ \cite{song2021spin}. Graphene electrodes are placed on the top and at the bottom of CrI$_3$ flakes to collect the photocurrent upon the light illumination. Both the magnitude and the sign of the photocurrent exhibit the polarization and the wavelength dependence in a consistent way, as the absorption spectroscopy does. This consistency is rooted in the fact that the photocurrent scales with the light absorption through which the electron-hole pairs are created and then dissociated to generate photocurrent. A dichroism of photocurrent between left- and right-handed polarized light can be observed when a net magnetization is present in CrI$_3$ flakes, which also corroborates with the corresponding MCD signal.

\subsubsection{Magnetic linear dichroism and magnetic linear birefringence to study two-dimensional magnetism}
Magnetic linear dichroism (MLD) and magnetic linear birefringence (MLB) measure the absorption and phase difference, respectively, between two orthogonal linearly polarized lights. In contrast to MCD and MOKE that are proportional to the linear term of order parameters, MLD and MLB probe the quadratic term of order parameters that show up in the symmetric terms in the optical susceptibility tensors, i.e., $$\chi_{ij} = \chi_{ij}^{0}|_{\Vec{m}=0}+\sum_{\Vec{r}}\frac{\partial^2{\chi_{ij}}}{\partial{m_k}\partial{m_l}}\Big|_{\Vec{m}=0}m_km_l(\Vec{r})$$ with the Onsager reciprocal relationship considered. Therefore, they are not sensitive to the broken time-reversal symmetry, but are best for the broken rotational symmetries. Moreover, MLD and MLB are capable of detecting both FMs and AFMs as long as their long-range orders break the rotational symmetry of the underlying crystal lattices. 

\begin{figure}[th]
\begin{center}
\includegraphics[width=1.0\textwidth]{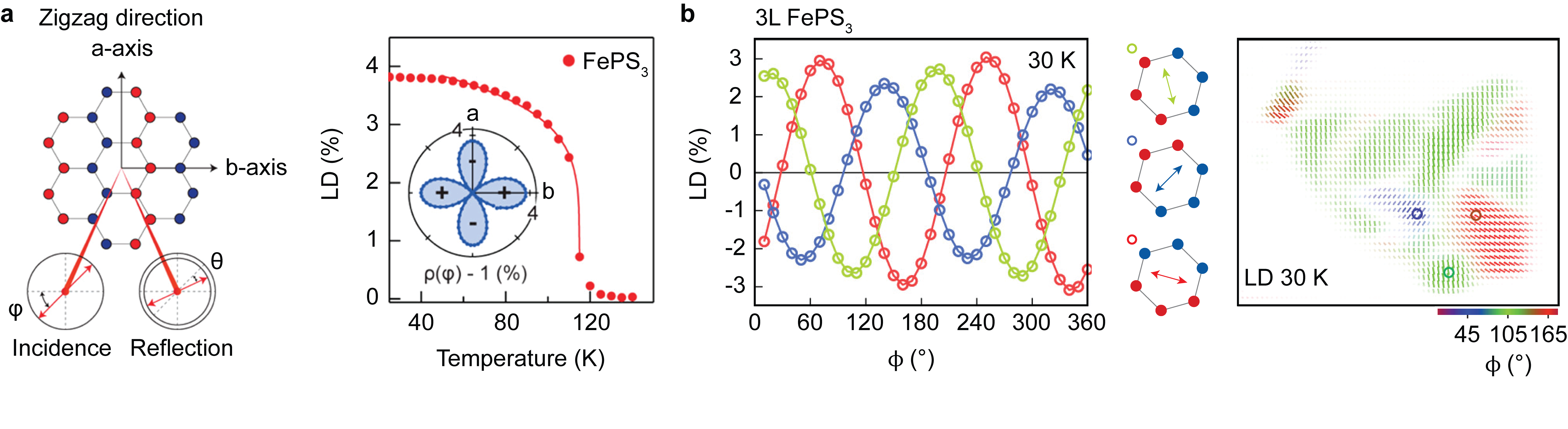}\caption{\small Examples of MLD used in studying 2D vdW magnetism. (a) Temperature dependence of MLD on few-layer FePS$_3$; Inset showing the polarization dependence of MLD; (b) Three possibilities of polarization-dependent MLD for trilayer FePS$_3$ that corresponds to three domain states of the zigzag AFM in FePS$_3$. Figures (a) and (b) are adapted from Refs. \cite{zhang2021observation} and \cite{ni2022observation}, respectively.}
\label{MLD}
\end{center}
\end{figure}

MLD and MLB have been used to investigate the zigzag AFM of NiPS$_3$ \cite{hwangbo2021highly} and FePS$_3$ \cite{zhang2021observation, zhang2022cavity} where the zigzag AFM chains break the three-fold rotational symmetry of the triangular lattice of individual layers, despite the spin orientations being along in-plane and out-of-plane directions for NiPS$_3$ and FePS$_3$. Three magnetic domain states are expected, which are related by the 120$^\mathrm{o}$ rotation operations. Taking FePS$_3$ as an example, MLD shows a steep increase when cooling across the critical temperature of $\sim$120 K (Figure \ref{MLD}a). The angular-dependent MLD exhibits a four-lobe feature with two pairs of opposite signs, whose peaks align along the zigzag and its orthogonal directions (inset of Figure \ref{MLD}a). Three 120$^\mathrm{o}$-rotated MLD patterns are expected for the three degenerate zigzag AFM domain states (Figure \ref{MLD}b). It is worth commenting that since MLD and MLB couple with the quadratic term, rather than the linear term, of order parameters, they can be finite when the short-range correlation is present. In other words, the presence of MLD or MLB does not necessarily correspond to the presence of a long-range order. For example, it remains an open question whether a long-range zigzag AFM order is present in a few-layer NiPS$_3$, an XY-type magnet with strong 2D fluctuations. 

\subsubsection{Magnetism-induced second harmonic generation to study two-dimensional magnetism}
Second harmonic generation (SHG) is a frequency double process through nonlinear light-matter interaction. Conventionally, it is believed to be only present in noncentrosymmetric materials or phases, which only considers the leading order electric dipole (ED) contribution to SHG, i.e., $$P_{i} (2\omega) \propto \chi^{\mathrm{ED}}_{ijk} (2\omega; \omega, \omega) E_j(\omega) E_k(\omega)$$ where $P_i(2\omega)$ is the induced SH ED polarization, $\chi^{\mathrm{ED}}_{ijk} (2\omega; \omega, \omega)$ is the ED SHG susceptibility tensor, and $E_{j/k} (\omega)$ is the electric field of incident fundamental light. To the leading order, magnetism-induced SHG appears in noncentrosymmetric magnetic orders hosted in centrosymmetric crystals. For such cases, the structural ED SHG is fully suppressed, i.e., $\chi^{\mathrm{ED,s}}_{ijk} = 0$, whereas the magnetism-induced ED SHG is proportional to the order parameter ($\Vec{L}$), i.e., $\chi^{\mathrm{ED,m}}_{ijk} \propto \Vec{L}$. As a result, $\chi^{\mathrm{ED,m}}_{ijk}$ is odd under the time reversal operation (i.e., $c$-tensor) and exhibits an order parameter-like onset when cooling across the magnetic critical temperature. 

\begin{figure}[th]
\begin{center}
\includegraphics[width=1.0\textwidth]{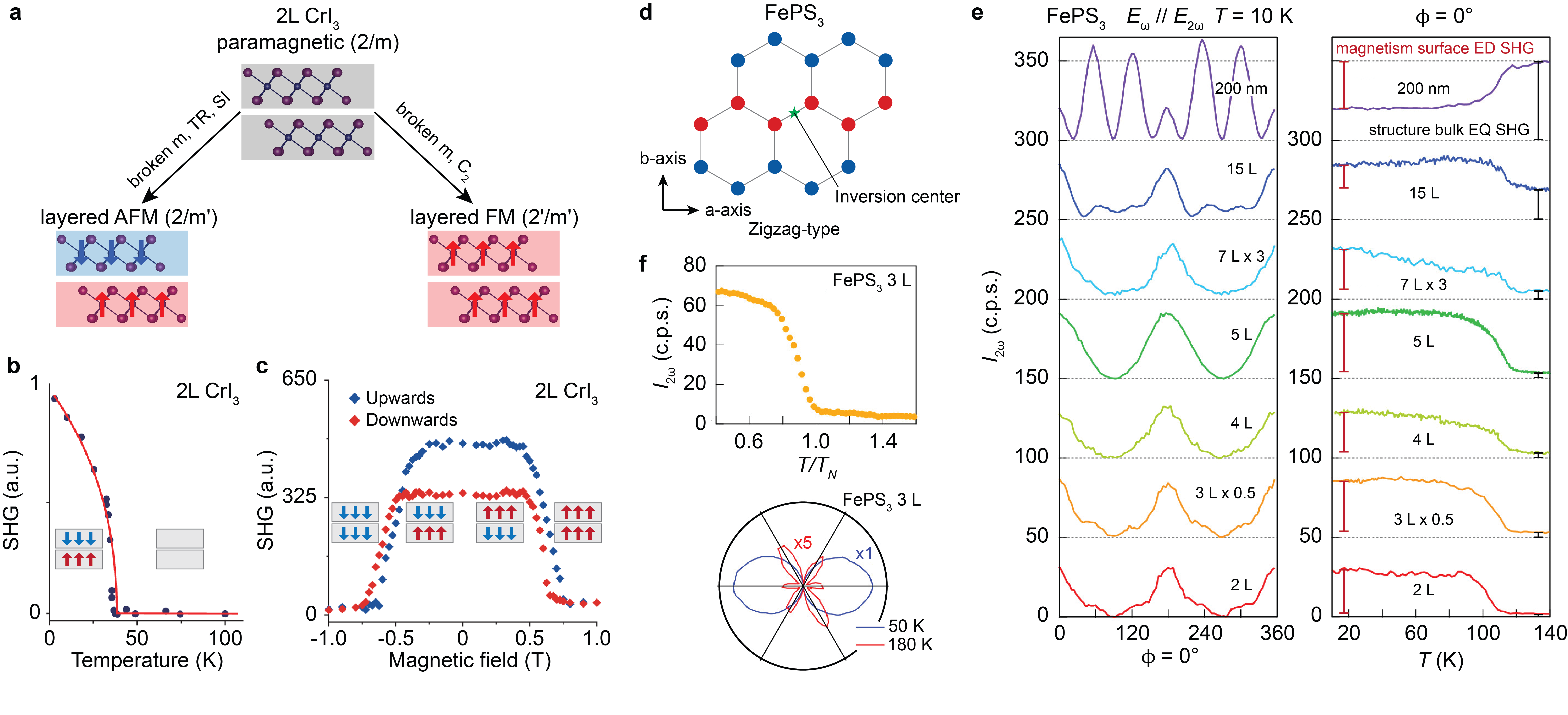}\caption{\small Examples of using SHG to investigate 2D vdW magnetism. (a) Sketch to illustrate the symmetry evolution for paramagnetic, layered AFM, and FM states for bilayer CrI$_3$; (b-c) Temperature and magnetic field dependencies of SHG for bilayer CrI$_3$; (d) Sketch to show the presence of inversion center in the zigzag AFM of FePS$_3$; (e) Layer number dependence of SHG for FePS$_3$; (f) Temperature and polarization dependencies of SHG for trilayer FePS$_3$. Figures (b) and (c) are adapted from Ref. \cite{sun2019giant}, and Figures (d-f) are adapted from Ref. \cite{ni2022observation}.}
\label{SHG}
\end{center}
\end{figure}

Examples of 2D vdW magnets that host noncentrosymmetric magnetism on top of centrosymmetric crystal lattices mainly include $\mathcal{PT}$-symmetric AFMs, such as even-layer CrCl$_3$ \cite{wang2019determining}, even-layer CrI$_3$ \cite{sun2019giant}, even-layer CrSBr \cite{lee2021magnetic}, even-layer MnBi$_2$Te$_4$ \cite{gao2021layer,qiu2023axion}, MnPS$_3$ \cite{long2020persistence}, and MnPSe$_3$ \cite{ni2021imaging}. So far, magnetism-induced ED SHG has been shown in even-layer CrI$_3$ \cite{sun2019giant}, even-layer CrSBr \cite{lee2021magnetic}, even-layer MnBi$_2$Te$_4$ \cite{fonseca2022anomalous}, MnPS$_3$ \cite{chu2020linear}, and MnPSe$_3$ \cite{ni2021direct}. Taking bilayer CrI$_3$ as an example \cite{sun2019giant}, its crystallographic structure has a centrosymmetric point group $2/m$, and its layered AFM ordered state corresponds to a magnetic point group $2/m'$. Time reversal, spatial inversion, and vertical mirror symmetries are broken across the second-order magnetic phase transition from paramagnetic to layered AFM state (Figure \ref{SHG}a). In addition, the layered FM state has a centrosymmetric magnetic point group $2'/m'$. Two-fold rotational symmetry is further broken, and spatial inversion symmetry is restored across the first-order magnetic phase transition from layered AFM to FM state (Figure \ref{SHG}a). The temperature dependence of SHG shows a clear onset of the SHG signal below $T_N = 45$ K (Figure \ref{SHG}b), whereas the magnetic field dependence of SHG demonstrates the presence/absence of SHG in the layered AFM/FM state that breaks/preserves the spatial inversion symmetry (Figure \ref{SHG}c). 

ED SHG can also originate from surfaces of a centrosymmetric bulk state. This property has been used to study the centrosymmetric zigzag magnetism in few-layer FePS$_3$ \cite{ni2022observation}. The crystal structure of few-layer FePS$_3$ obeys a structural point group $2/m$, and the zigzag AFM ordered phase has a magnetic point group of $2/m1'$, without obvious symmetry breaking across the magnetic phase transition from a point group perspective, despite broken translational symmetry and time reversal symmetry from a space group point of view (Figure \ref{SHG}d). Through the thickness dependence of SHG in few-layer FePS$_3$, it is found that the structural contribution to SHG decreases whereas the magnetism contribution remains nearly the same as the thickness reduces (Figure \ref{SHG}e). This observation suggests that the structural contribution originates from the bulk electric quadrupole (EQ) SHG process (i.e., $\chi^{\mathrm{bulk \: EQ, s}}_{ijkl}$) whereas the magnetism-induced contribution results from the surface ED SHG process (i.e., $\chi^{\mathrm{surf \: ED, m}}_{ijk}$). Due to the lack of point symmetry and time-reversal symmetry breaking across the magnetic phase transition, the surface ED SHG is expected to scale with the quadratic term of the order parameter, $\chi^{\mathrm{surf \: ED, m}}_{ijk} \propto \Vec{L}\cdot\Vec{L}$ and is even under the time-reversal operation (i.e., $i$-tensor). The temperature dependence of SHG shows a slow increase at the temperatures above $T_N$ and then a steep increase right across $T_N$ (Figure \ref{SHG}f), consistent with the coupling to the $\Vec{L}\cdot\Vec{L}$ that probes both the spin fluctuations and the square of order parameters. 

We further comment that there can also be magnetism-induced SHG from centrosymmetric magnetic phases, via EQ or magnetic dipole (MD) SHG process, i.e., $$Q_{ij}(2\omega) \propto \chi^{\mathrm{EQ}}_{ijkl}(2\omega; \omega, \omega)E_k(\omega)E_l(\omega)$$
$$M_{i}(2\omega) \propto \chi^{\mathrm{MD}}_{ijk}(2\omega; \omega, \omega)E_j(\omega)E_k(\omega)$$
Depending on the broken symmetries across the magnetic phase transition, the leading order magnetism-induced SHG can be either $c$-type with broken time-reversal symmetry (i.e., $\chi^{\mathrm{EQ,m}}_{ijkl} \propto \Vec{L}$, $\chi^{\mathrm{MD,m}}_{ijk} \propto \Vec{L}$) or $i$-type with preserved time-reversal symmetry (i.e., $\chi^{\mathrm{EQ,m}}_{ijkl} \propto \Vec{L}\cdot\Vec{L}$, $\chi^{\mathrm{MD,m}}_{ijk} \propto \Vec{L}\cdot\Vec{L}$). Since these have been achieved in centrosymmetric bulk magnets, their applications to 2D vdW magnets require substantial improvement of SHG sensitivity due to the extremely small signal level.

\subsubsection{Magneto-Raman spectroscopy to investigate two-dimensional magnetism}
Raman spectroscopy is a dynamic tool to detect incoherent excitations inside materials. When applied to magnetic materials, there can be several types of excitations that contribute to the understanding of magnetism. First, phonons are the most frequently observed excitations inside solids. Zone-center phonons (i.e., $P_{\Vec{k}=0}$) help to understand the symmetry of the crystalline lattice. Sometimes, zone-boundary phonons, (i.e., $P_{\Vec{k}=-\Vec{k}_m}$) can be accessed when the translational symmetry breaking magnetic order parameter with a wave vector $\Vec{k}_m$, $\Vec{L}_{\Vec{k}=\Vec{k}_m}$, participates in the Raman scattering process to ensure the total momentum being zero. In this magnetism-assisted Raman process, one can also gain insights into the magnetic order. Second, single magnon excitation with magnon from the zone center (i.e., $M_{\Vec{k}=0}$) provides key information about magnetic anisotropy. Third, two-magnon excitation with pairs of magnon from the opposite sides of the zone boundaries (i.e., $M_{\Vec{k}=-\Vec{k}_b}M_{\Vec{k}=\Vec{k}_b}$ with $\Vec{k}_b$ being the zone-boundary momentum) encodes main messages about the magnetic exchange coupling and spin fluctuations.  

\begin{figure}[th]
\begin{center}
\includegraphics[width=1.0\textwidth]{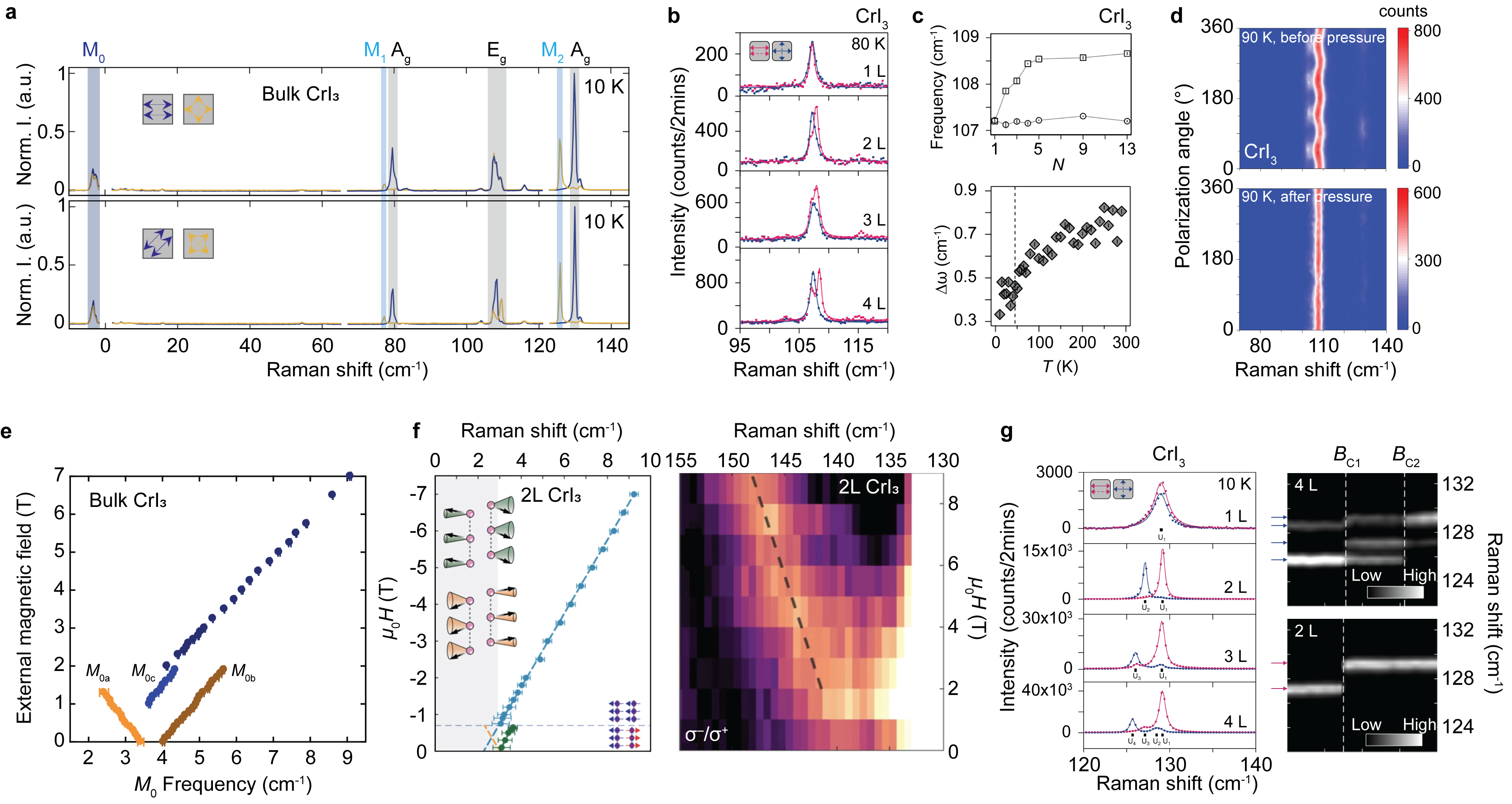}\caption{\small Examples of using magneto-Raman spectroscopy to study 2D vdW magnetism. (a) Linear polarization-resolved Raman spectra for bulk CrI$_3$, with four types of excitations marked as M$_0$, M$_{1,2}$, A$_\mathrm{g}$, and E$_\mathrm{g}$; (b) Layer number dependence of an E$_g$ phonon mode at $\sim$ 107 cm$^{-1}$; (c) Layer number and temperature dependence of the frequency shift of the E$_g$ phonon mode at $\sim$ 107 cm$^{-1}$; (d) Pressure dependence of the E$_g$ phonon mode at $\sim$ 107 cm$^{-1}$; (e) Magnetic field dependence of magnon modes (M$_0$) probed in bulk CrI$_3$; (f) Acoustic and optical magnon modes probed in bilayer CrI$_3$; (g) Layer number and magnetic field dependence of M$_2$ in few-layer CrI$_3$. Figures (a) and (e), (b) and (c), (d), and (f) are adapted from Refs. \cite{li2020magnetic}, \cite{guo2021structural}, \cite{li2019pressure}, and \cite{cenker2021direct}, respectively. Figure (g) is adapted from Refs. \cite{xie2022twist} and \cite{jin2020tunable}.}
\label{Raman}
\end{center}
\end{figure}

Raman spectroscopy has been extensively used to study 2D vdW magnetism, including phonon, magnetic order, and single- and two-magnons. Exemplary 2D vdW magnets studied by Raman spectroscopy include CrI$_3$ \cite{li2020magnetic,jin2020observation,jin2020tunable,xie2022twist,cenker2021direct,zhang2019magnetic,mccreary2020distinct,huang2020tuning}, MnBi$_2$Te$_4$ \cite{lujan2022magnons}, NiPS$_3$ \cite{kim2019suppression}, FePS$_3$ \cite{lee2016ising,luo2023evidence}, etc. Taking magneto-Raman spectra of bulk CrI$_3$ of $R\overline{3}$ space group as an example, there are four types of Raman modes \cite{li2020magnetic}: $A_g$ phonon modes with identity Raman tensors, $E_g$ phonon modes with anisotropic but symmetric Raman tensors, $M_{0}$ single-magnon modes with anisotropic but symmetric Raman tensors, and $M_{1,2}$ with isotropic but antisymmetric Raman tensors, which are listed below considering only in-plane $x,y$ components and marked in Figure \ref{Raman}a, $$R_{A_g}=\begin{pmatrix} a & 0\\ 0 & a \end{pmatrix},\quad R_{E_g}=\begin{pmatrix} b & c\\ c & -b \end{pmatrix}, \quad R_{M_0}=\begin{pmatrix} d & f\\ f & e \end{pmatrix}, \quad R_{M_{1,2}}=\begin{pmatrix} 0 & g\\ -g & 0 \end{pmatrix}$$  

$E_g$ phonons at $\sim$107 cm$^{-1}$ are excellent signatures for differentiating interlayer stacking geometries \cite{li2019pressure, song2019switching}. Bulk CrI$_3$ ($R\overline{3}$) has the rhombohedral stacking at low temperatures, and therefore, doubly degenerated $E_g$ phonons are expected. Similarly, monolayer CrI$_3$ ($R\overline{3}m$) has three-fold rotational symmetry and hence host doubly degenerated $E_g$ phonons. In contrast, few-layer CrI$_3$ ($C2/m$) features the monoclinic stacking at all temperatures, and therefore, the doubly degenerated $E_g$ phonons in the monolayer split into two nondegenerate $A_g$ and $B_g$ phonons as shown in Figure \ref{Raman}b \cite{guo2021structural}. The frequency split has been shown to decrease when lowering temperatures and increase when adding layer numbers (Figure \ref{Raman}c). This stacking monoclinicity-induced mode splitting can be eliminated when the lattice of few-layer CrI$_3$ undergoes a monoclinic to rhombohedral transition under pressure (Figure \ref{Raman}d) \cite{li2019pressure, song2019switching}. 

$M_0$ single magnons at $\sim$2--3 cm$^{-1}$ in bulk CrI$_3$ reveal further details through their Zeeman splitting under an out-of-plane magnetic field. Summarized in Figure \ref{Raman}e are all single magnon branches from Stokes and anti-Stokes in all polarization channels. Three magnon branches are observed below a critical magnetic field of $B_c=2$T, and only one branch remains above $B_c$, where $B_c$ corresponds to the spin-flip transition field between layered AFM and FM states in few-layer CrI$_3$. The three magnon branches below $B_c$ include a pair with opposite Zeeman splitting slopes and a third one with a positive slope, which together with $B_c$ confirm the coexistence of surface layered AFM and bulk FM in 3D bulk CrI$_3$. Single magnons in monolayer and bilayer CrI$_3$ have also been probed by magneto-Raman spectroscopy, expecting a single branch in monolayer and two branches (one detected) in bilayer (Figure \ref{Raman}f) \cite{cenker2021direct}. For bilayer CrI$_3$ where the spatial inversion symmetry is broken in the layered AFM phase, the optical magnon branch at the zone center is further observed at $\sim$140 cm$^{-1}$ (Figure \ref{Raman}f) \cite{cenker2021direct}. 

$A_g$ and $M_{1,2}$ around $\sim$75 and $\sim$125 cm$^{-1}$ are from the same optical phonon branches for bulk CrI$_3$, with the $A_g$ phonon from the zone center and those involved in $M_{1,2}$ from the $c$-axis zone boundaries folded by the layered AFM \cite{li2020magnetic}. Down to the few-layer regime, the $c$-axis translational symmetry is broken, and the Davydov splitting of phonons from the interlayer coupling arises to replace the $k_c$ dispersion of phonons. Taking the group of phonons around $\sim$125 cm$^{-1}$ as an example, the number of phonons is the same as the layer number, consistent with the Davydov splitting of the $A_g$ phonon into $N$ phonons in $N$-layer CrI$_3$ (labeled as $\mathrm{U}_{1,2,...,N}$ in Figure \ref{Raman}g) \cite{jin2020tunable}. Using a linear chain model, the eigenvectors, $\Vec{U}_{1,2,...,N}$, and eigenenergies, $\omega_{1,2,...,N}$, of $\mathrm{U}_{1,2,...,N}$ phonons can be computed. Further analysis also shows that the modes shown in the parallel channel are simple phonons with identity Raman tensors ($R^{S}_i \propto \Vec{U}_i \cdot \Vec{S}$ with $\Vec{S}=(1,1,...,1)$) and that the modes appearing in the crossed channel are from the layered AFM assisted Raman scattering of the phonons with antisymmetric Raman tensors ($R^{AS}_i \propto \Vec{U}_i \cdot \Vec{M}$ with $\Vec{M}=(1,-1,...,(-1)^{N-1})$). The magnetic field dependence of Raman modes in the crossed channel well captures the spin-flip transitions for both bilayer and four-layer CrI$_3$ (Figure \ref{Raman}g).

\subsubsection{Time-resolved optical spectroscopy to investigate two-dimensional magnetism}
Time-resolved optical spectroscopy in the pump-probe scheme has been used to detect the carrier dynamics and coherent excitations. The pump light pulse at a selected frequency with a chosen polarization excites the material of interest at time zero, and the probe light pulse of a particular kind investigates the excited material at a time delay $t$ with respect to the pump pulse. The pump can be designed to produce photocarriers across the band gap, provide impulsive excitations of phonons, generate coherent excitations of zone-center magnons, or else. The probe can be chosen to be differential reflectance/absorption spectroscopy for $A_g$ phonons, MCD/MOKE for magnons, MLD/MLB for $E_g$ phonons or magnons, or even SHG for more types of quasi-particles. 

\begin{figure}[th]
\begin{center}
\includegraphics[width=1.0\textwidth]{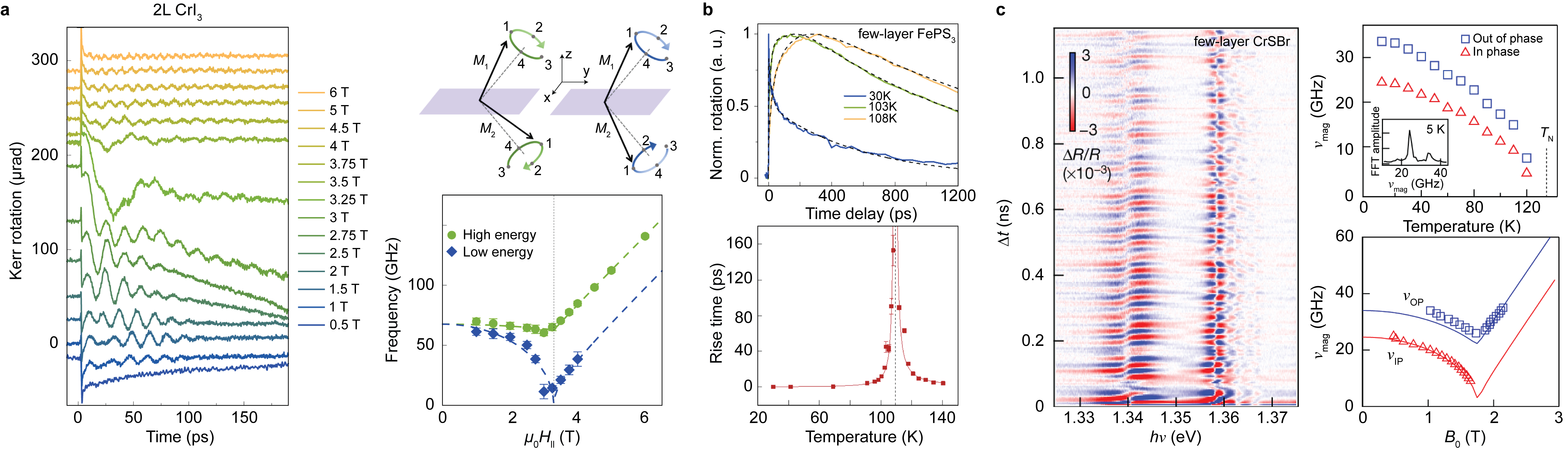}\caption{\small Examples of time-resolved ultrafast optical spectroscopy used in the research of 2D vdW magnetism. (a)Time-resolved MOKE spectra of bilayer CrI$_3$ taken with an in-plane magnetic field B$_{||}$; (b) Time-resolved MLD spectra of few-layer FePS$_3$ taken as a function of temperature; (c) Time-resolved absorption spectra of few-layer CrSBr as functions of temperature and magnetic field. Figures (a),(b), and (c) are adapted from Refs. \cite{zhang2020gate},\cite{zhang2021spin},and \cite{bae2022exciton}, respectively.}
\label{ultrafast}
\end{center}
\end{figure}

Time-resolved optics have recently been introduced to investigate 2D magnetism, with a particular focus on detecting coherent magnon excitations in 2D vdW magnets including bilayer CrI$_3$ \cite{zhang2020gate}, few-layer FePS$_3$ \cite{zhang2021spin}, and few-layer CrSBr \cite{bae2022exciton}. In the study of bilayer CrI$_3$ \cite{zhang2020gate}, time-resolved MOKE was applied to track the coherent acoustic magnons. Because nearly normal incidence MOKE only probes the out-of-plane component of spin waves, an in-plane magnetic field ($H_{||}$) is applied to tilt the static out-of-plane Ising spins into the in-plane direction whose spin waves, therefore, contain an out-of-plane component. Below the spin saturation field, $H_{||,s}=\sim3.3$ T that fully align the spins along the in-plane direction, two initially degenerate modes at 0T gradually split and soften till the lower frequency one drops to zero frequency at $H_{||,s}$ (Figure \ref{ultrafast}a). These two modes correspond to the in-phase and out-of-phase precession of spins between the two bilayers (inset of Figure \ref{ultrafast}a). Above $H_{||,s}$, both modes exhibit the linear FM-like Zeeman shifts (Figure \ref{ultrafast}a). In the study of few-layer FePS$_3$ \cite{zhang2021spin}, time-resolved MLB was used to track the demagnetization and recovery processes. By tracking the temperature dependence of the demagnetization rate and recovery rate, both rates were found to peak at the N\'eel temperature of $T_N=$110 K (Figure \ref{ultrafast}b). In the study of few-layer CrSBr \cite{bae2022exciton}, time-resolved differential reflection was used to track the exciton frequency shift due to the excitation of coherent magnons. Two magnon modes, the in-phase and out-of-phase precession of anti-aligned spins, soften towards zero frequency when heating up to the N\'eel temperature of $T_N=$132 K and softens under an out-of-plane external magnetic field with a critical field of $H_{\perp,s}=\sim$1.7 T (Figure \ref{ultrafast}c).

Time-resolved optics are more widely used in studying vdW magnets in the 3D form. Examples include Floquet engineering of the electronic band gap in MnPS$_3$ using time-resolved SHG \cite{shan2021giant}, magnetic brightening of the dark electron-phonon bounded states in NiPS$_3$ using time-resolved reflectivity \cite{ergeccen2022magnetically}, coherent detection of hidden spin-lattice coupling in NiPS$_3$ using time-resolved coherent phonon spectroscopy \cite{ergeccen2023coherent}, dynamically introducing an exciton-driven AFM metal in NiPS$_3$ using time-resolved terahertz conductivity \cite{belvin2021exciton}, optical control of magnetic anisotropy in NiPS$_3$ using time-resolved polarization rotation \cite{afanasiev2021controlling}, and spin-lattice coupling in few-layer CrI$_3$ using helicity dependent time-resolved reflectivity \cite{padmanabhan2022coherent}.

\subsection{Magneto-transport measurements to study two-dimensional magnetism}
Spin-dependent electron transport measurements are another venue for investigating 2D magnetism, including both static magnetic orders and dynamic spin waves. A unique advantage of 2D vdW magnets is that they can be flexibly integrated into various geometries of devices, such as tunneling junctions and Hall bars. 

\subsubsection{Tunneling magnetoresistance to study two-dimensional magnetism}
\begin{figure}[th]
\begin{center}
\includegraphics[width=1.0\textwidth]{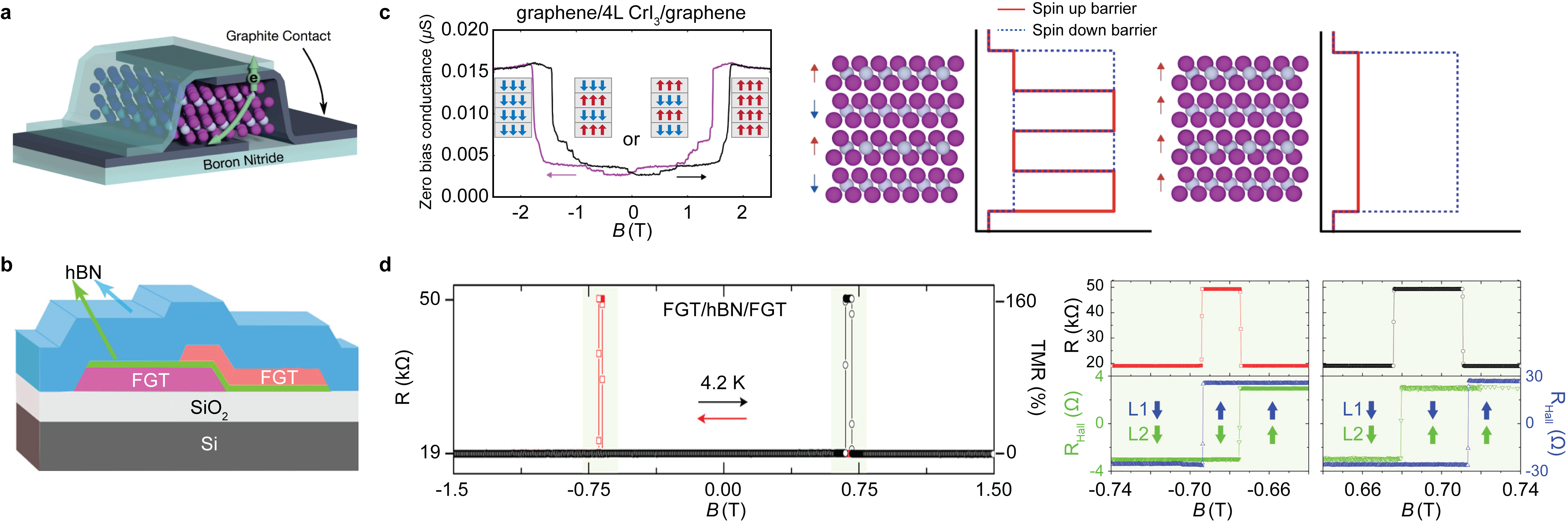}\caption{\small Examples of TMR used in researching 2D vdW magnetism. (a-b) Two types of MTJ structures used in the study of 2D vdW magnets; (c) Magneto-tunneling-conductance for a graphite/four-layer CrI$_3$/graphite MTJ as a function of the out-of-plane magnetic field; Tunneling barriers are shown for spin up and spin down cases in both layered AFM and FM states; (d) Magneto-tunneling-resistance for a Fe$_3$GeTe$_2$/hBN/Fe$_3$GeTe$_2$ MTJ as a function of the out-of-plane magnetic field; The magnetic states of the two Fe$_3$GeTe$_2$ flakes are shown to understand the tunneling resistance. Figures (a) and (c), and (b) and (d) are adapted from Refs. \cite{klein2018probing},\cite{wang2018tunneling}, respectively.}
\label{TMR}
\end{center}
\end{figure}

Tunneling magnetoresistance (TMR) probes the magnetism-dependent electron tunneling in a magnetic tunneling junction (MTJ) which is composed of a thin insulating layer sandwiched by two conducting layers. There are two types of MTJ that have been explored in the study of 2D vdW magnets. First, the thin insulating layer is a 2D vdW semiconducting magnet such as layered AFMs while the two conducting layers are graphene flakes (Figure \ref{TMR}a). Second, the two conducting layers are 2D vdW FM metals while the thin insulating layer is a non-magnetic insulating 2D material such as hBN (Figure \ref{TMR}b). 

Few-layer CrI$_3$ has been extensively explored for the first type of MTJ, a graphene/few-layer CrI$_3$/graphene heterostructure as shown in Figure \ref{TMR}a \cite{klein2018probing,song2018giant,wang2018very,kim2018one}. Below the spin-flip transition field $H_s$, few-layer CrI$_3$ host the layered AFM order where the spins align ferromagnetically within the layer and antiferromagnetically between the adjacent layers, i.e., alternating between spin-up and spin-down across the layers. This spin texture creates high tunneling barriers for both spin-up and spin-down tunneling electrons whereas the layered AFM state results in a low tunneling conductance (Figure \ref{TMR}c). Above $H_s$, few-layer CrI$_3$ transitions into the FM order with all spins aligned in the same direction within and across layers. Such a spin texture leads to a low tunneling barrier for tunneling electrons with the same spin direction as the FM state and results in a high tunneling conductance for this FM state (Figure \ref{TMR}c). The magnetoresistance, defined as $$MR = 100\%\times\frac{G_{FM}-G_{AFM}}{G_{AFM}}$$ is reported to be 95$\%$ for bilayer, 300$\%$ for trilayer, and 550$\%$ for four-layer CrI$_3$ in Ref.\cite{klein2018probing}; 310$\%$ for bilayer, 2,000$\%$ for trilayer, and 19,000$\%$ for four-layer CrI$_3$ in Ref.\cite{song2018giant}; 10,000$\%$ for 10-nm thick CrI$_3$ in Ref.\cite{wang2018very}; and as high as 10$^6\%$ in 14-layer CrI$_3$ in Ref.\cite{kim2018one}. In addition to few-layer CrI$_3$, few-layer CrBr$_3$ \cite{kim2019evolution, kim2019tailored} and CrCl$_3$ \cite{kim2019evolution, kim2019tailored,klein2019enhancement,wang2019determining} have also been used for MTJs. 

Few-layer Fe$_3$GeTe$_2$ has been applied for the second type of MTJ, a Fe$_3$GeTe$_2$/few-layer hBN/Fe$_3$GeTe$_2$ multilayer structure as shown in Figure \ref{TMR}b \cite{wang2018tunneling}. The switching field for the FM state of Fe$_3$GeTe$_2$ depends slightly on the flake geometry. This results in the top and bottom Fe$_3$GeTe$_2$ having slightly different switching fields, between which the two Fe$_3$GeTe$_2$ flakes have opposite spin alignments and produce a low tunneling conductance. Otherwise, the two Fe$_3$GeTe$_2$ have the same spin alignment that leads to a high tunneling conductance (Figure \ref{TMR}d). 

\subsubsection{Anomalous Hall effect to study two-dimensional magnetism}
Anomalous Hall effect (AHE) occurs when time-reversal symmetry is broken, typically in a magnetic phase with a nonzero out-of-plane magnetization. In the research of 2D vdW magnets, AHE has been used as a diagnosis tool for broken time-reversal symmetry whereas quantum AHE (QAHE) and fractional QAHE (FQAHE) have been further used to discern topological phases.  

\begin{figure}[th]
\begin{center}
\includegraphics[width=1.0\textwidth]{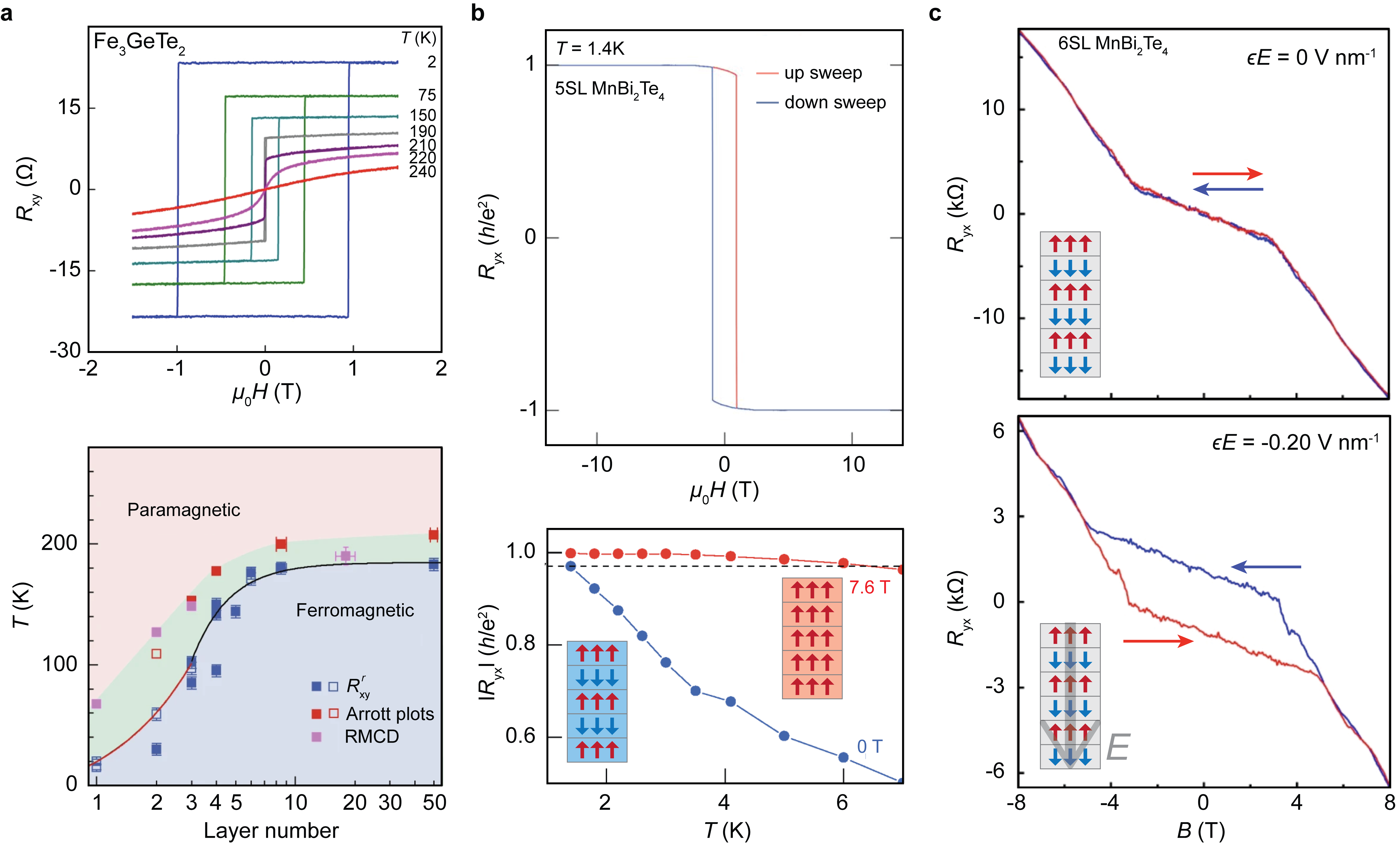}\caption{\small Examples of AHE applied in the research of 2D vdW magnetism. (a) Magnetic field, temperature, and layer number dependence of Hall resistance of few-layer Fe$_3$GeTe$_2$; (b) Quantized AHE observed in a five-layer MnBi$_2$Te$_4$, and the temperature dependence of Hall resistance at 0T and 7T magnetic fields; (c) Magnetic field dependence of the Hall resistance in a six-layer MnBi$_2$Te$_4$ with zero and 0.2V nm$^{-1}$ out-of-plane electric fields. Figures (a), (b), and (c) are adapted from Refs. \cite{deng2018gate,fei2018two}, \cite{deng2020quantum}, and \cite{gao2021layer}, respectively.}
\label{AHE}
\end{center}
\end{figure}

AHE has been observed in 2D metallic FMs such as Fe$_3$GeTe$_2$ as a probe for the FM ground state and the Curie temperature \cite{deng2018gate,fei2018two}. For thin Fe$_3$GeTe$_2$ flake (i.e., 12 nm) \cite{fei2018two}, the Hall resistance, $R_\mathrm{xy}$ at low temperatures shows a rectangular hysteresis loop with steep jumps at the coercive field, $\mu_0H_c$, under an out-of-plane external field, $\mu_0H$ (Figure \ref{AHE}a). The remanent $R_\mathrm{xy}$ at 0 T signifies a spontaneous magnetization for the itinerant FM phase. Furthermore, the 1T value of $\mu_0H_c$ at 2 K and the steep jumps at $\mu_0H_c$ suggest the single FM domain over the region of the AHE device. The remanent $R_\mathrm{xy}$ and the coercive field $\mu_0H_c$ vanish above around 240 K, suggesting the Curie temperature for the FM to paramagnetic phase transition. 

QAHE has been realized in 2D magnetic topological insulators, odd-layer MnBi$_2$Te$_4$ as direct evidence for its topological nature \cite{deng2020quantum}. MnBi$_2$Te$_4$ can be viewed as a layered topological insulator Bi$_2$Te$_3$ with its Te-Bi-Te-Bi-Te quintuple layers intercalated by an additional Mn-Te bilayer. MnBi$_2$Te$_4$ remains as a topological insulator and hosts a layered AFM order with out-of-plane spins. For odd-layer MnBi$_2$Te$_4$, the layer magnetization does not cancel completely, and the system can be considered as an FM topological insulator. A zero-field QAHE has been observed in a five-septuple-layer MnBi$_2$Te$_4$ at 1.4 K (Figure \ref{AHE}b), and this quantized resistance can survive up to 6.5 K under an external magnetic field of 7.6 T that aligns all layers ferromagnetically. 

Layer AHE has been detected in a 2D axion insulator, even-layer MnBi$_2$Te$_4$ as a probe for its axion ground state \cite{gao2021layer}. Unlike odd-layer MnBi$_2$Te$_4$ with a net magnetization, even-layer MnBi$_2$Te$_4$ has zero total magnetization due to the perfect cancellation of the layer magnetization. For a bilayer MnBi$_2$Te$_4$, the Berry curvature in the top and bottom layers are opposite, which leads to electrons deflecting in opposite directions in the two layers. By applying an out-of-plane external electric field, $\epsilon E$, the balance between opposite Berry curvature breaks down, and as a result, a large AHE signal is introduced (Figure \ref{AHE}c).

In addition to 2D vdW magnetic atomic crystals, AHE has also been applied to investigate Chern insulating phases in moir\'e superlattices of twisted graphene and twisted TMDCs. Giant AHE \cite{sharpe2019emergent} and QAHE \cite{serlin2020intrinsic} have been reported in twisted bilayer graphene aligned with the hBN substrate, and QAHE has been further realized in trilayer graphene aligned with the hBN substrate \cite{chen2020tunable} and twisted monolayer-bilayer graphene \cite{polshyn2020electrical}. Furthermore, QAHE has been observed in H-stacked MoTe$_2$/WSe$_2$ upon applying a displacement field \cite{li2021quantum}. Not only the FQAH states have been reported in R-staked twisted MoTe$_2$ homostructure \cite{cai2023signatures,zeng2023thermodynamic}, but also very recently, there have been direct experimental measurements of quantized $R_\mathrm{xy}$ at fractional factors ($\nu$ = 3/2 and 5/3) \cite{park2023observation, xu2023observation}. 

\begin{figure}[th]
\begin{center}
\includegraphics[width=1.0\textwidth]{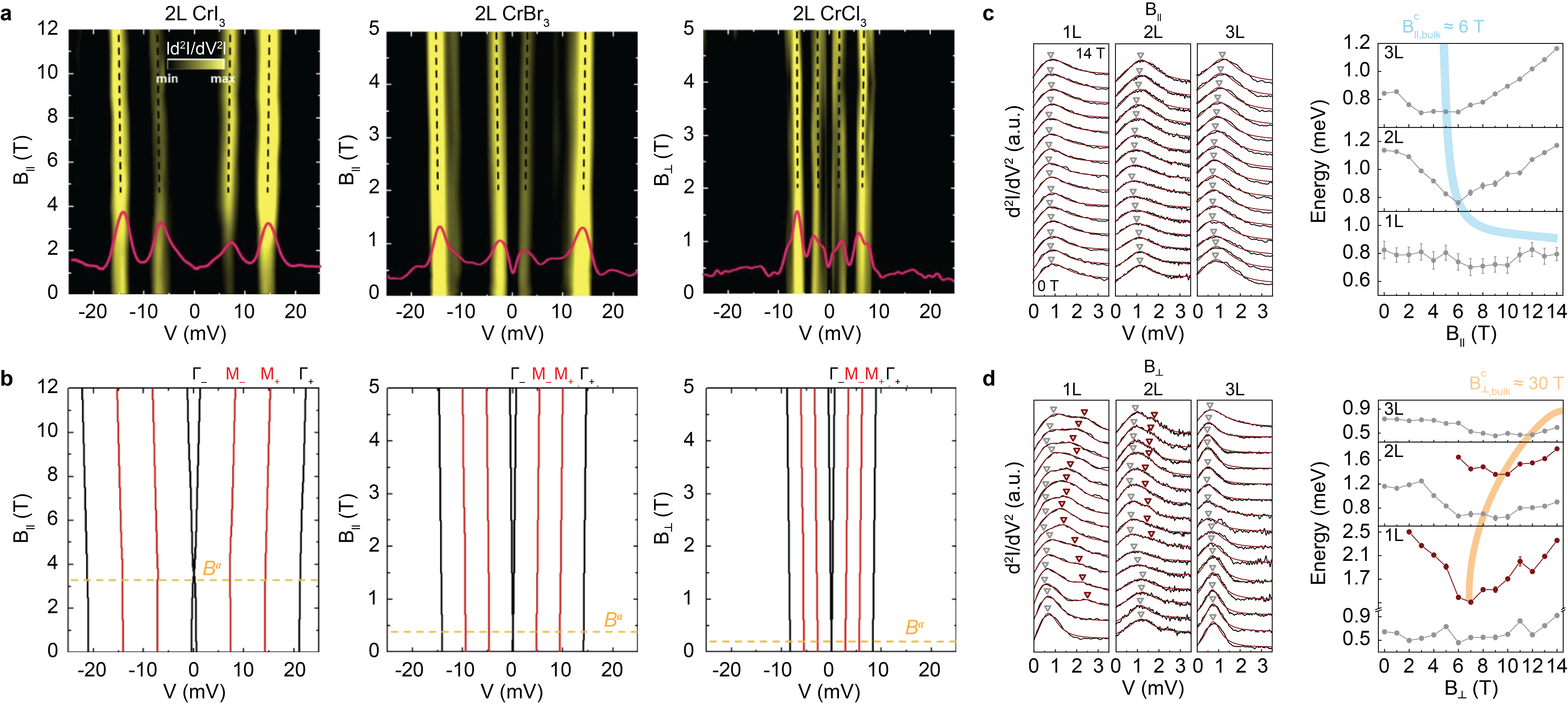}\caption{\small One example of using inelastic TMR in studying 2D vdW magnetism. (a) Magnetic field dependence of inelastic TMR spectroscopy ($d^2I/dV^2$) for bilayer CrCl$_3$, CrBr$_3$, and CrI$_3$; (b) Computed magnon Zeeman shift under the applied magnetic field for magnons at the $\Gamma$ and $M$ momentum points. (c) In-plane (B$_{||}$) and (d) out-of-plane (B$_\perp$) magnetic field dependence of the $d^2I/dV^2$ for monolayer, bilayer, and trilayer $\alpha$-RuCl$_3$ and their corresponding critical field for the magnon energy minimum. Figures are adapted from Refs. \cite{kim2019evolution} and \cite{yang2023magnetic}.}
\label{inelastic TMR}
\end{center}
\end{figure}

\subsubsection{Inelastic magneto-tunneling spectroscopy to study two-dimensional magnetism}

Magneto-tunneling spectroscopy probes the differential conductance ($dI/dV$) as a function of an applied DC voltage ($V_\mathrm{DC}$) under a given magnetic field ($B$) for an MTJ with a magnetic semiconductor as the tunneling barrier. The $dI/dV$ versus $V_\mathrm{DC}$ spectrum scales with the electronic DOS of the magnetic semiconductor. When the tunneling process involves energy transfer between electron and other quasiparticles (\textit{e.g.}, phonon, magnon, \textit{etc}.), subtle step features are expected in the $dI/dV (V_\mathrm{DC})$ spectra at the energies of the involved quasiparticles. To better highlight these quasiparticle features, the derivative of $dI/dV$, $d^2I/dV^2$, is typically used and referred to as the inelastic magneto-tunneling spectroscopy \cite{lambe1968molecular}.

Inelastic magneto-tunneling spectroscopy has been applied to probe magnons inside 2D magnetic systems. Examples include CrX$_3$ (X = Cl, Br, I) using the graphene/CrI$_3$/graphene MTJ structure \cite{klein2018probing,kim2019evolution} and $\alpha$-RuCl$_3$  using the T$_d$ MoTe$_2$/$\alpha$-RuCl$_3$/T$_d$ MoTe$_2$ MTJ structure \cite{yang2023magnetic}. For all three CrX$_3$, two pairs of peaks have been observed in the $d^2I/dV^2 (V_\mathrm{DC})$ spectra at $(V_\mathrm{DC})$ of opposite signs but with the same magnitude, and exhibit clear dispersions with the applied in-plane magnetic field $B_{||}$ (Figure \ref{inelastic TMR}a). Analysis show that the pair of peaks originates from the two M-point magnons in the two spin wave branches of honeycomb CrX$_3$, where M point is the van Hove singularity point with a divergent magnon density of states. The frequency separation between the two M-point magnons decreases as X changes from I to Br to Cl, suggesting the decrease of the spin wave bandwidth. Using a linear spin wave model with the nearest neighboring exchange coupling and an additional Zeeman term, $\mathcal{H}=-J\sum_{<i,j>}(\hat{S}^x_i\hat{S}^x_j+\hat{S}^y_i\hat{S}^y_j+\alpha \hat{S}^z_i\hat{S}^z_j)-g\mu_B\sum_i\Vec{B}\cdot\hat{\Vec{S}}_i$, and fitting the M-point magnon frequencies to the ones from the $d^2I/dV^2 (V_\mathrm{DC})$ spectra, the FM intralayer exchange coupling $J$ and the anisotropy factor $\alpha$ can be extracted for CrX$_3$. The computed results for two magnons at the M point (i.e., $\mathrm{M}_-$ and $\mathrm{M}_+$) and two magnons at the $\Gamma$ point (i.e., $\Gamma_-$ and $\Gamma_+$) are shown in Figure \ref{inelastic TMR}b. For $\alpha$-RuCl$_3$, the in-plane (B$_{||}$) and out-of-plane (B$_\perp$) magnetic field dependencies of magnons have been tracked for monolayer, bilayer, and trilayer samples (Figures \ref{inelastic TMR}c and \ref{inelastic TMR}d). The field at which the magnon energy reaches the minimum indicates the point across which the magnetic ground state changes. As the thickness reduces for $\alpha$-RuCl$_3$, the B$_{||}$ critical field increases whereas the B$_\perp$ critical field decreases, suggesting the evolution from the easy-plane to the easy-axis magnetic anisotropy (Figures \ref{inelastic TMR}c and \ref{inelastic TMR}d).

An advantage of using inelastic magneto-tunneling spectroscopy to study magnons in 2D vdW magnets is its capability of accessing magnons at finite momentum points, such as M point in CrX$_3$ above. This is complementary to magneto-optics, including magneto-Raman spectroscopy and time-resolved MOKE/MCD, as optical tools can only probe single magnons with zero momentum, i.e., $\Gamma$ point single magnons. 

\subsection{Microscopy and nanoscopy to image two-dimensional magnetism}
Long-range magnetic orders are spontaneous symmetry-breaking phases, for which multiple degenerate magnetic ground states related by broken symmetries are anticipated. Therefore, in a macroscopic sample, magnetic domains and domain walls naturally develop. Magneto-optical microscopy can detect domains down to the $\mu$m scale, whereas magnetic nanoscopy, such as scanning magnetic force microscopy, scanning NV microscopy, etc., can resolve magnetic inhomogeneity down to a few nm. 

\begin{figure}[th]
\begin{center}
\includegraphics[width=1.0\textwidth]{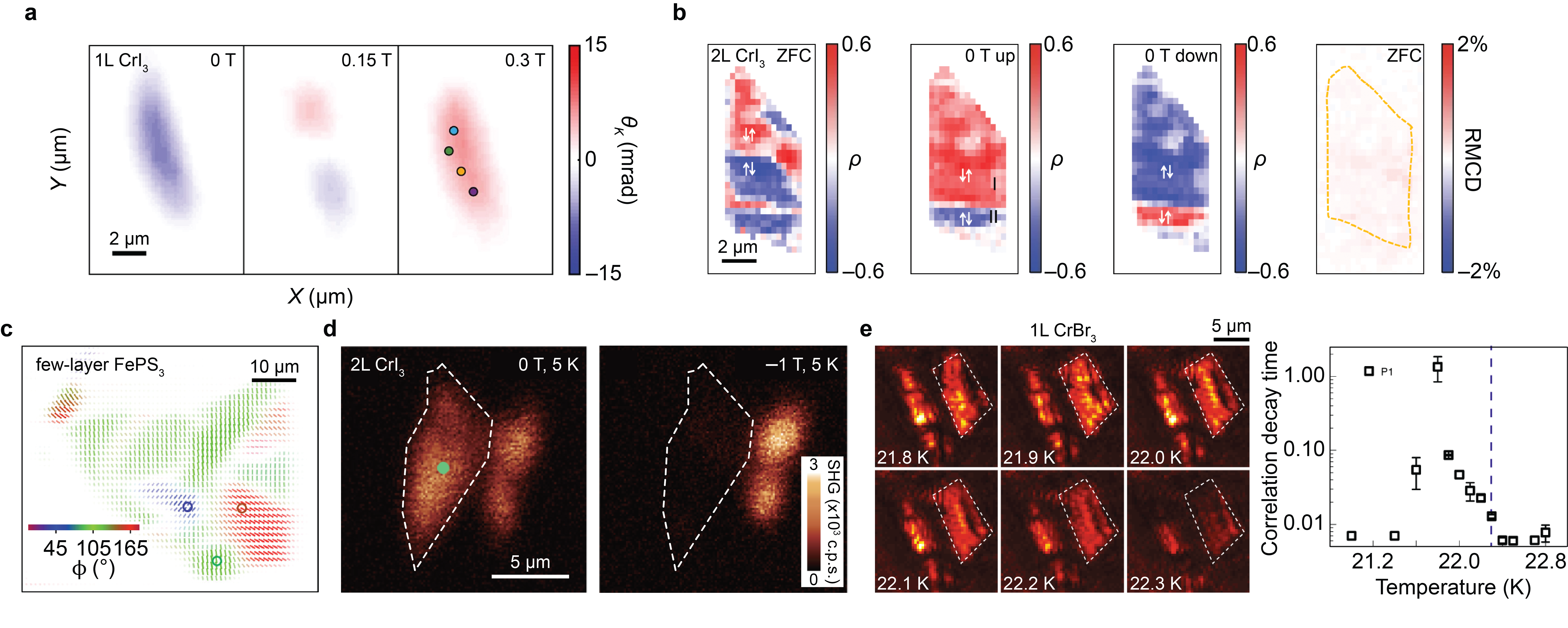}\caption{\small Examples of magneto-optical microscopy used in imaging 2D vdW magnetism. (a) MOKE images for monolayer CrI$_3$ taken at 0 T, 0.15 T, and 0.3 T out-of-plane magnetic field; (b) Helicity resolved PL microscopy for bilayer CrI$_3$ with a WSe$_2$ sensor layer; (c) MLD microscopy image for a few-layer FePS$_3$; (d) SHG image for a bilayer CrI$_3$ in the layered AFM and FM states; (e) Polarization-enhanced MCD microscopy for monolayer CrBr$_3$. Figures (a),(b),(c),(d), and (e) are adapted from Refs. \cite{huang2017layer},\cite{zhong2020layer},\cite{ni2022observation},\cite{sun2019giant}, and \cite{jin2020imaging}, respectively.}
\label{MO microscopy}
\end{center}
\end{figure}

\subsubsection{Magneto-optical microscopy to image two-dimensional magnetism}
Magneto-optics is the most popularly used experimental tool for investigating 2D magnetism, due to its compatibility with 2D materials and flexibility in experimental implementation, as discussed in Section 4.1 above. The spatial resolution of far-field magneto-optics is limited by the optical diffraction limit, that is in the order of $\mu$m, similar to the lateral dimension of mechanically exfoliated 2D magnets, $\sim$1 -- 10 $\mu$m. Nonetheless, magneto-optical microscopy has been applied to investigate magnetic domains in 2D magnets.

For probing static magnetic orders, photoluminescence, MOKE, MLB, and SHG microscopy have been used so far. The experimental implementation discussed in Section 4.1 is straightforward, either by rastering the sample with piezoelectric stages or performing wide-field imaging with a CCD camera. Examples include MOKE microscopy of domains of the FM order in monolayer CrI$_3$ (Figure \ref{MO microscopy}a) \cite{huang2017layer}, photoluminescence microscopy of layered AFM domains in bilayer CrI$_3$ proximated to WSe$_2$ (Figure \ref{MO microscopy}b) \cite{zhong2020layer}, MLB microscopy of zigzag AFM domains in FePS$_3$ (Figure \ref{MO microscopy}c) \cite{ni2022observation}, and SHG microscopy of layered AFM order in bilayer CrI$_3$ (Figure \ref{MO microscopy}d) \cite{sun2019giant}.

For detecting dynamic fluctuations, a polarization-enhanced MCD microscopy was developed that achieves high temporal (up to 100 frames per second (f.p.s.)) and high spatial ($\sim$600 nm) resolutions with high extinction ratio ($>3\times10^4$), via separating the effective NA for illumination and imaging with a long focal length lens paired with a high NA objective. Using this technique to image monolayer CrBr$_3$ \cite{jin2020imaging}, a Heisenberg FM with in-plane spins, the critical fluctuations were visualized to emerge within a narrow temperature window of $\sim$0.5 K around the critical temperature of 22.3 K, shown in Figure \ref{MO microscopy}e for the temperature-dependent fluctuation amplitude map of the magnetization, $\delta M_z(\Vec{r})=\sqrt{<M_z(\Vec{r},t)^2>-<M_z(\Vec{r},t)>^2}$, where $M_z(\Vec{r},t)$ is the temporal MCD image.

\begin{figure}[th]
\begin{center}
\includegraphics[width=1.0\textwidth]{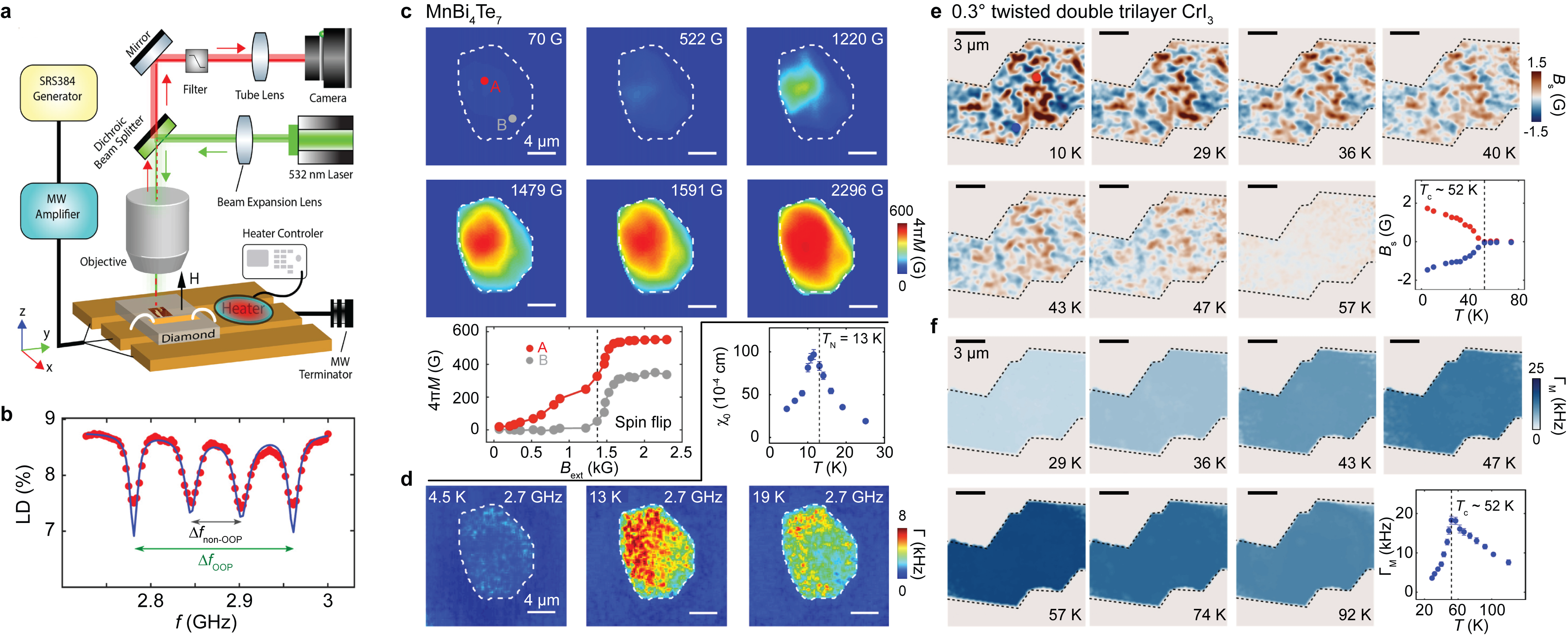}\caption{\small Examples of using wide-field NV magnetometry to image 2D vdW magnetism. (a) Sketch for the setup of wide-field NV magnetometry; (b) Characteristic ODMR spectrum as a function of applied microwave frequency for a [111] oriented diamond; (c-d) Magnetic field dependence of the reconstructed magnetization images and the longitudinal magnetic fluctuation maps for few-layer MnBi$_4$Te$_7$; (e-f) Temperature dependence of the reconstructed magnetization images and the longitudinal magnetic fluctuation maps for a 0.3$^\mathrm{o}$ twisted double trilayer CrI$_3$. Figures (a),(b-d), and (e-f) are adapted from Refs. \cite{chen2023above},\cite{mclaughlin2022quantum} and \cite{huang2023revealing}, respectively.}
\label{wide-field NV}
\end{center}
\end{figure}

\subsubsection{Wide-field nitrogen-vacancy magnetometry to study two-dimensional magnetism}
Nitrogen-vacancy (NV) center in diamond is the nearest neighboring pair of one substitutional nitrogen and one lattice vacancy. The spin-triplet ground state ($S=1, m_s=0,\pm1$) of the negatively charged NV$^-$ state has robust quantum coherence and is the basis for quantum sensing of both static dc- and dynamic ac-magnetic fields from the neighboring environment. The static dc-magnetic field leads to an energy split between the two $m_s=\pm1$ states while the dynamic ac-magnetic field influences the decay rates from the prepared $m_s=0$ state to the $m_s=\pm1$ states, which can be probed by optically detected magnetic resonance (ODMR) measurements over the NV centers \cite{degen2017quantum}. In the recent few years, NV centers in diamond have been exploited to study 2D vdW magnets in the wide-field imaging mode, such as MnBi$_4$Te$_7$ \cite{mclaughlin2022quantum}, VI$_3$ \cite{broadway2020imaging}, Fe$_x$GeTe$_2$ \cite{chen2022revealing,chen2023above}, and twisted double trilayer CrI$_3$ \cite{huang2023revealing}. For a general experiment, exfoliated 2D vdW magnet flakes are transferred onto an oriented diamond chip with NV centers implanted. A 532nm laser illuminates the sample area to optically excite the NV centers, and a wide-field imaging system collects the photoluminescence of the NV centers, whereas the microwave is delivered into the NV centers through a patterned waveguide. The sketch of the setup is shown in Figure \ref{wide-field NV}a.

Taking MnBi$_4$Te$_7$ as an example \cite{mclaughlin2022quantum}, a standard ODMR spectrum as a function of sweeping microwave frequency is shown in Figure \ref{wide-field NV}b, where the diamond is oriented along the [111] direction, i.e., the out-of-plane (OOP) direction. Using the frequency split in the ODMR spectrum at each pixel of the wide-field photoluminescence map, the magnetization map of the Mn$_4$Te$_7$ can be reconstructed and imaged as a function of an external magnetic field ($B_\mathrm{ext}$) that spans across the spin-flip transition in MnBi$_4$Te$_7$ (Figure \ref{wide-field NV}c). Because of the layered AFM state that compensates the magnetization in MnBi$_4$Te$_7$, the net magnetization in the low field regime ($B_\mathrm{ext} <1000$ G) shows vanishingly small net magnetization. Using the NV relaxometry at each pixel, the longitudinal magnetic fluctuation map of zero-magnetization MnBi$_4$Te$_7$ can be achieved as a function of temperature ($T$) that vary across the magnetic phase transition (Figure \ref{wide-field NV}d). A clear divergence at $T_N$ = 13 K can be seen in the magnetic susceptibility that is proportional to the spin fluctuations. 

Wide-field NV magnetometry has further been used to image moir\'e magnetic domains that extend over multiple moir\'e supercells in twisted double trilayer CrI$_3$ \cite{huang2023revealing}. Low twist angle twisted double trilayer CrI$_3$ have monoclinic stacked and rhombohedral stacked regions within each moir\'e supercell corressponding to the AFM and the FM coupling, respectively. The AFM-coupled region results in fully compensated zero magnetization whereas the FM-coupled region leads to non-compensated finite magnetization. Moreover, for the non-compensated magnetization, there are two degenerate domain states that are related by the time-reversal operation, i.e., magnetization along $+\hat{z}$ and $-\hat{z}$ directions, and therefore, there are extended moir\'e magnetic domains that are expected to be at length scales beyond several moir\'e wavelengths. Indeed under zero-field cooling, the reconstructed magnetization map shows two types of domains with opposite magnetization at a length scale of 1--2 $\mu$m (Figure \ref{wide-field NV}e). Simultaneously, the spin fluctuation map is homogeneous, but shows a peak at $\sim$52 K (Figure \ref{wide-field NV}f), which is 7 K higher than the layered AFM onset temperature \cite{huang2017layer} but 9K lower than the bulk FM critical temperature \cite{mcguire2015coupling}. Further experiments are needed to address this change in the critical temperature in twisted double trilayer CrI$_3$.

\begin{figure}[th]
\begin{center}
\includegraphics[width=1.0\textwidth]{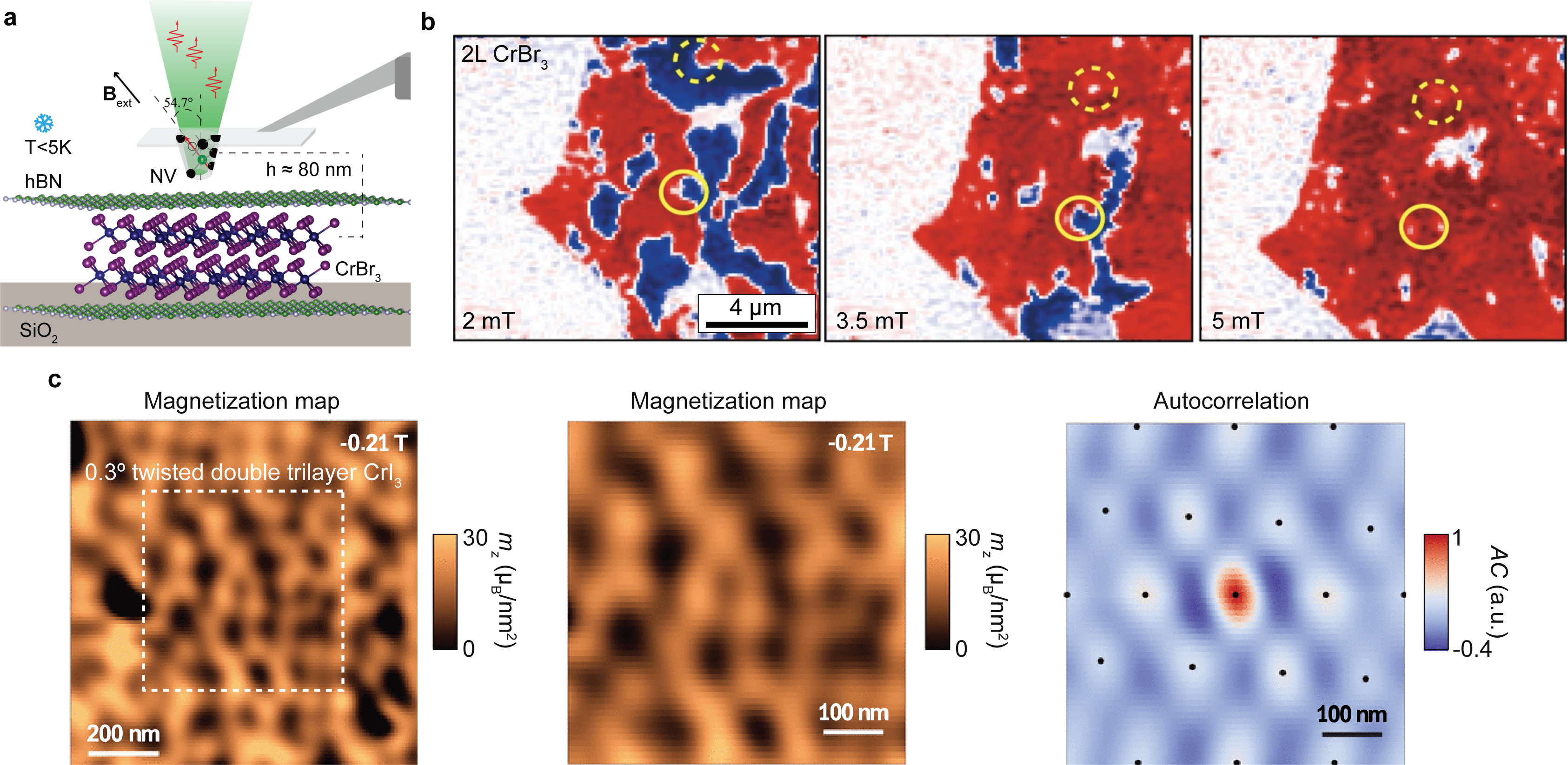}\caption{\small Examples of applying scanning NV magnetometry in the research of 2D vdW magnets. (a) Illustration of the scanning NV magnetometry apparatus; (b) Magnetic field dependence of the reconstructed magnetization maps probed by scanning NV magnetometry for a bilayer CrBr$_3$ flake; (c) Reconstructed magnetization maps and its autocorrelation map for 0.3$^\mathrm{o}$ twisted double trilayer CrI$_3$. Figures (a-b) and (c) are adapted from Refs. \cite{sun2021magnetic} and \cite{song2021direct}, respectively.}
\label{scanning NV}
\end{center}
\end{figure}

\subsubsection{Scanning nitrogen-vacancy magnetometry to image two-dimensional magnetism}
Scanning NV magnetometry employs a single NV center in diamond as a sensor to local static dc- and dynamic ac-magnetic field (Figure \ref{scanning NV}a), and can achieve a spatial resolution of $\sim$50 nm \cite{song2021direct}, an order of magnitude better than the optical diffraction-limited wide-field NV magnetometry. Recently, scanning NV magnetometry has been used to probe magnetization, localized defects, and magnetic domains in 2D vdW magnets, with examples including CrI$_3$ \cite{thiel2019probing}, CrBr$_3$ \cite{sun2021magnetic}, CrTe$_2$ \cite{fabre2021characterization}, and twisted double bilayer and double trilayer CrI$_3$ \cite{song2021direct}.

Scanning NV magnetometry shows $\sim\mu$m scale magnetic domains in bilayer CrBr$_3$ under a small external field of 2 mT (Figure \ref{scanning NV}b). Nearly equal populations of two types of domains with magnetizations along the $+\hat{z}$ and $-\hat{z}$ have been observed in the nanoscale magnetization map. By applying an external magnetic field of 5mT, a nearly uniform magnetization map can be visualized (Figure \ref{scanning NV}b). More interestingly, by tracking the magnetization maps across the magnetic hysteresis loop under an external magnetic field, pinned magnetic domains and domain boundaries have been captured throughout the hysteresis loop, which is found to be caused by local sites of defects.

In addition to 2D vdW magnetic atomic crystals, scanning NV magnetometry has also been employed to examine moir\'e magnetism in twisted double trilayer CrI$_3$. A low twist angle of $\sim$0.3$^\mathrm{o}$ is chosen to ensure that the moir\'e wavelength is longer than the scanning NV magnetometry spatial resolution. Indeed, the reconstructed magnetization map of the $\sim$0.3$^\mathrm{o}$ twisted double trilayer CrI$_3$ shows a spatial modulation at the moir\'e wavelength, with the modulation varying from finite magnetization in the FM coupled region to zero magnetization in the AFM coupled region within a moir\'e supercell (Figure \ref{scanning NV}c). 

\subsubsection{Magnetic force microscopy to map two-dimensional magnetism}
Magnetic force microscopy relies on magnetic interactions, such as magnetic dipole-dipole interaction, between a sharp magnetized tip and a measured magnetic sample, to achieve visualization of the magnetic structure at the surface of the sample. The typical spatial resolution by magnetic force microscopy is $\sim$50 nm. Recently, magnetic force microscopy has been used to image magnetic domains and domain walls in 2D vdW magnets, such as CrI$_3$ \cite{niu2019coexistence}, CrSBr \cite{rizzo2022visualizing}, and VSe$_2$ \cite{ci2022thickness}. 

\begin{figure}[th]
\begin{center}
\includegraphics[width=1.0\textwidth]{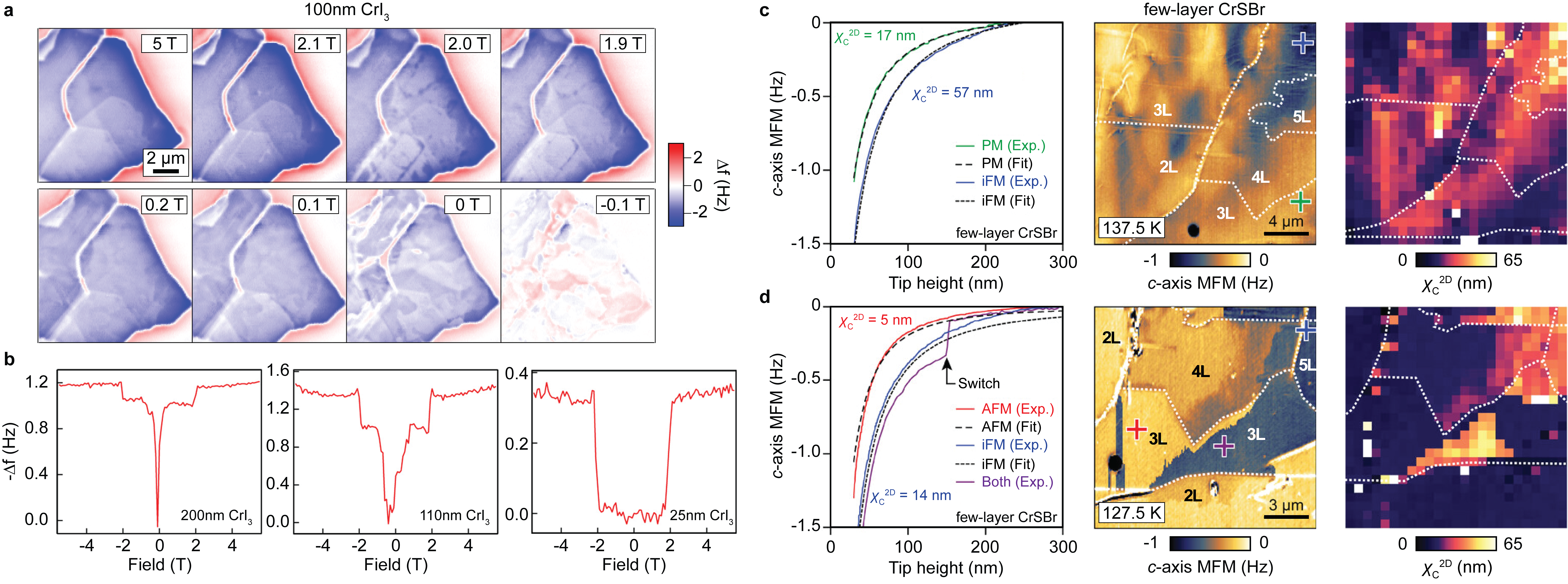}\caption{\small Examples of magnetic force microscopy used in imaging 2D vdW magnetism. (a) Magnetic field dependence of magnetic force microscopy images for a 200 nm-thick CrI$_3$ flake; (b) Thickness dependence of the integrated magnetic force microscopy signal for 200 nm-, 110 nm-, and 25 nm-thick CrI$_3$; (c-d) Measurements of the local sheet susceptibility, $\chi_c^{\mathrm{2D}}$, of few-layer CrSBr for the iFM phase at 137.5 K and the layered AFM phase at 127.5 K. Figures (a-b) and (c-d) are adapted from Refs. \cite{niu2019coexistence} and \cite{rizzo2022visualizing}, respectively.}
\label{MFM}
\end{center}
\end{figure}

Using the study of exfoliated CrI$_3$ as an example, magnetic force microscopy has been used to image the evolution of magnetic domains and domain walls as a function of the external magnetic field ($B$), with the magnetic map for a 200 nm thick CrI$_3$ at representative fields shown in Figure \ref{MFM}a \cite{niu2019coexistence}. At the high magnetic field, e.g., $B$ = 5 T, the spins are fully polarized, and nearly uniform magnetic force microscopy signals are observed across the entire flake. Reducing the magnetic field to be around 2T, within a tiny magnetic field window, 1.9 T < $B$ < 2.1 T, magnetic domains with a weak signal contrast show up, indicating the partial reversal of spins in the thick flake. Continued down to $B$ = 0.2 T, the magnetic force microscopy signal remains nearly uniform but slightly weaker than that at $B$ = 5 T, and finally experiences a dramatic decrease below $B$ = 0.2 T. A plot of magnetic force microscopy signal as a function of the external magnetic field is shown in Figure \ref{MFM}b, clearly showing a step at $B_c$ = 2 T and a dip near -0.1 T < $B$ < 0.1 T, corresponding to the spin-flip transition field in the layered AFM layers and the coercive field of the FM layers. This magnetic force microscopy study confirms the coexistence of layered AFM and FM orders in thick CrI$_3$ flakes (thicker than 25 nm), which is consistent with the findings from the magneto-Raman study of bulk CrI$_3$ \cite{li2020magnetic}. 

In addition to measuring magnetization, magnetic force microscopy can also be used to probe the local magnetic sheet susceptibility, $\chi^\mathrm{2D}_c$, by measuring and fitting the magnetic force microscopy signal, $\Delta f(z)$, depending on the sample-tip distance, $z$ \cite{rizzo2022visualizing}. By performing the $\Delta f(z)$ measurements and analysis for few-layer CrSBr at different magnetic states, paramagnetic phase (PM), incoherent c-axis ferromagnetic phase (iFM), and layered AFM phase (AFM), $\chi^\mathrm{2D}_c$ were found to be 17 nm for PM, 57 nm for iFM, and 5 nm for AFM (Figure \ref{MFM}c). Furthermore, the spatially dependent $\chi^\mathrm{2D}_c$ shown in Figure \ref{MFM}d can fully account for the layer- and position-resolved magnetic force microscopy signal contrast, for both the iFM phase and the AFM phase.  


\section{Outlook and challenges}
2D vdW magnetism is an emerging research topic that is expected to have profound impacts on both the fundamental science of phase transitions in the low dimensions and the practical applications in spintronics and microelectronics. As of now, six years after the experimental discovery of intrinsic 2D magnetism in 2016--2017, tremendous progress has been made in realizing, understanding, and controlling the magnetic ground states of 2D vdW FMs and AFMs with collinear spin textures. Yet, there are many more to explore and expect in the research of 2D magnetism. Below, we provide an outlook for this field in two directions -- novel magnetic phenomena and desired experimental probes, and also briefly comment on the key challenges as of now.

\noindent\textbf{Novel magnetic phenomena} 

Very recently, there have been revised classifications of magnets \cite{mazin2022altermagnetism}. Magnetic materials are firstly divided into two classes; collinear and noncollinear spin magnets. The collinear class can be further divided into commensurate and incommensurate ones, where the commensurate family further includes the crystal-symmetry compensated and the no crystal-symmetry compensated ones. Importantly, two new sub-groups of collinear magnets have been identified; altermagnetism in the crystal-symmetry compensated group \cite{vsmejkal2022emerging} and Luttinger compensated FMs for the no crystal-symmetry compensated group \cite{mazin2022altermagnetism}. In the context of this revised classification, 2D magnetism research thus far has primarily focused on conventional collinear magnetism. One direction for future fundamental research on 2D magnetism could be to realize unconventional magnetic phases in the 2D limit, such as 2D altermagnetism, 2D Luttinger compensated FM, and more types of noncollinear magnetism. 

Both the quantum and thermal fluctuations are expected to be significantly enhanced in the 2D limit. The research on 2D magnetism so far has treated the enhanced fluctuations as a foe to long-range magnetic orders and concentrated efforts in suppressing this effect. From a different perspective, such enhanced fluctuations can also be a friend that helps the emergence of new magnetic phases. Another research direction for future 2D magnetism could be to exploit strong quantum fluctuations in 2D to design and realize new quantum magnetic phases, such as quantum spin liquid \cite{du20182d,yang2023magnetic} and spin-induced nematicity \cite{fernandes2019intertwined}. 

In addition to the magnetic ground state, magnetic excitation -- including but not limited to magnon -- is another important component for 2D magnetism research. So far, mainly magnons at the Brillouin zone center, $\Gamma$ point, have been probed in a few limited 2D magnets, e.g., CrI$_3$ \cite{klein2018probing,kim2019evolution,li2020magnetic,cenker2021direct,zhang2020gate}, CrSBr \cite{bae2022exciton}, and FePS$_3$ \cite{lee2016ising,luo2023evidence}. In fact, there are topological magnons at the K points for honeycomb FMs \cite{pershoguba2018dirac} and at both K and $\Gamma$ points for kagome FMs \cite{yin2022topological}. Another future research direction could be to access and control such topological magnons in the 2D magnets. Moreover, nontrivial magnetic excitations, such as Majorana particles, spinons, etc., can exist in exotic magnetic systems if realized in the 2D form, which can be an additional future research direction.

\noindent\textbf{Desired experimental probes}

Probing 2D magnetism has primarily relied on optics, transport, and microscopy measurements. For the magnetic ground states, while optics can provide point symmetries and microscopy can reveal domain structures, a comprehensive picture still requires knowledge of translational symmetries, especially for the magnetic orders with finite momenta. For the magnetic excitations, although optics and transport can access magnons at limited momentum points, i.e., mostly Brillouin zone center and boundaries, a full determination of the spin Hamiltonian needs the complete momentum dependence of magnon energy, $\bold{E}(\Vec{k})$. To access the full Brillouin zone momenta, scattering techniques are needed, elastic ones for the magnetic ground states and inelastic ones for the magnetic excitation spectra. Due to the thinness and the micron-scale lateral dimension, resonant elastic and inelastic X-ray scatterings are needed to achieve sufficient signal-to-noise levels, whereas neutron and non-resonant X-ray scatterings may face critical challenges. So far, there have been very few resonant X-ray scattering experiments performed on 2D magnets \cite{discala2023dimensionality}. The application of resonant X-ray scattering to probing translational symmetries of magnetic orders and momentum-resolved spin wave spectra would be much appreciated in 2D magnetism research.

In addition to the momentum-space probe by resonant X-ray scattering, real-space imaging is also needed to resolve the detailed spin textures for incommensurate magnets and moir\'e magnets. Scanning NV magnetometry and magnetic force microscopy are two nanoscale imaging techniques applied in the research of 2D magnetism so far. However, the spatial resolution for state-of-the-art cryogenic scanning NV magnetometry and magnetic force microscopy are $\sim$50 nm \cite{song2021direct, rizzo2022visualizing}, which is not sufficient for resolving detailed spin textures in many incommensurate magnets and moir\'e magnets. Lorentz transmission electron microscopy is another nanoscale imaging tool with a spatial resolution of $\sim$1 nm, with the cryogenic temperature possibly reaching down to $\sim$10 K \cite{rajeswari2015filming}. Spin-polarized scanning tunneling microscopy is another option to achieve sub-nm spatial resolution at cryogenic temperatures \cite{bode2003spin}. Imaging complex spin textures in 2D magnets with Lorentz transmission electron microscopy and spin-polarized scanning tunneling microscopy will be highly desired in future research on 2D magnetism.

\noindent\textbf{Key chanllenges}

From the fundamental science perspective, the impact of disorders in 2D magnetic systems is expected to be more profound than in their 3D counterparts. It has been established that both the extrinsic and thermal disorders can degrade 2D emergent quantum phases arising from many-body interactions \cite{nie2014quenched}. While the thermal disorders can be quenched by lowering temperatures, the extrinsic disorders, such as atomic defects, stacking faults, moir\'e inhomogeneities, \textit{etc.}, are built in when the 2D magnetic systems are fabricated. Unfortunately, most of the 2D magnetic systems are extremely environment-sensitive, making extrinsic disorders inevitable and perhaps more common in 2D magnets than in graphene and TMDCs. It is a timely challenge of how to eliminate the extrinsic disorders in 2D magnetic systems, so that one can access the intrinsic physical properties of 2D magnetic systems.

From the practical application perspective, the stability, size, and critical temperature of 2D magnets cause major challenges. First, as mentioned above, the degradation of 2D magnets in an ambient environment necessarily requires novel protection schemes to isolate 2D magnets from air. At the research forefront, such protection is primarily achieved by the hBN encapsulation of $\mu$m-sized 2D magnets, which is, however, not feasible for scale-up applications due to its small sizes and time-consuming procedure. Efforts have been made to identify robust air-stable 2D magnets. So far, while the structural stability has been achieved in some 2D magnets \cite{liu2022three}, the magnetism stability for them remains under investigation. Second, as discussed in Section 3, large-scale film growth techniques, such as MBE and CVD, have not been widely applied in producing 2D magnetic films up to the wafer size. In addition, the physical properties of these large-scale magnetic films are much less explored and, therefore, less known. Finally, the critical temperatures for nearly all 2D magnets are at the cryogenic temperatures, significantly lower than the room temperature. The critical temperature for a 2D magnets is determined by the magnetic anisotropy whereas that for a 3D magnets is by the magnetic exchange coupling \cite{gong2019two}. It is typically the case that the magnetic anisotropy is much smaller than the magnetic exchange coupling. Other factors, including thermal and extrinsic disorders, further reduce the critical temperatures for 2D magnets. 

\section{Acknowledgement}
We thank Dr. Kai Sun for very helpful discussions. L.Z. and Y.A. acknowledge the support from the National Science Foundation under Award DMR-174774, the Air Force Office of Scientific Research under Award FA9550-21-1-0065 and the Alfred P. Sloan Foundation. X.G. and S.S acknowledge the support from the Office of Naval Research under Award N00014- 21-1-2770, and the Gordon and Betty Moore Foundation under Grant N031710. Z.S. acknowledges the support from the National Science Foundation under Award DMR-2103731.

\newpage

 \bibliographystyle{./elsarticle-num}
 \bibliography{./manuscript_arxiv.bib}

 \end{document}